\documentclass[submission, Phys]{SciPost}
\usepackage{subcaption}

\usepackage[utf8]{inputenc} 
\usepackage[T1]{fontenc} 	
\usepackage[english]{babel} 

\usepackage[bitstream-charter]{mathdesign}
\usepackage{geometry} 		
\usepackage{amsmath} 		
\usepackage{mathtools} 		
\usepackage{float} 			
\usepackage{graphicx} 		
\usepackage{tabularx} 		
\usepackage{booktabs} 		
\usepackage{color, xcolor} 	
\usepackage{pdfpages} 		
\usepackage{extarrows} 		
\usepackage{multirow} 		
\usepackage{multicol} 		
\usepackage{enumitem} 		
\usepackage{xspace} 		
\usepackage{stackrel} 		
\usepackage{tikz} 			
\usepackage{braket} 		
\usepackage{bm} 			
\usepackage{tensor} 		
\usepackage{slashed} 		
\usepackage[group-digits=false]{siunitx} 		
\usepackage{lastpage} 		
\usepackage{cite} 			
\usepackage[normalem]{ulem} 
\usepackage{fontawesome} 	
\usepackage{tocloft} 		
\usepackage{titlesec} 		
\usepackage{doi} 			
\usepackage{hyperref} 		
\usepackage[most]{tcolorbox} 					
\usepackage[nameinlink, capitalize]{cleveref} 	
\usepackage[nottoc, notlot, notlof]{tocbibind} 	
\usepackage[ruled, vlined]{algorithm2e} 		
\usepackage{makecell}
\usepackage[makeroom]{cancel}
\usepackage{feynmf}
\usepackage{orcidlink} 

\makeatletter
\def\BState{\State\hskip-\ALG@thistlm}
\makeatother

\makeatletter
\@ifundefined{pdfoutput}{}{\DeclareGraphicsRule{*}{mps}{*}{}}
\makeatother

\makeatletter
\DeclareRobustCommand*{\bfseries}{%
   \not@math@alphabet\bfseries\mathbf
   \fontseries\bfdefault\selectfont
   \boldmath
}
\makeatother

\hypersetup{
	pdftitle={XAI4QCDTaggers},
	pdfauthor={Vent et al.},
	colorlinks=true, 			
	linkcolor={red!50!black}, 	
	citecolor={blue!50!black}, 	
	urlcolor={blue!80!black} 	
} 

\DeclareSymbolFont{usualmathcal}{OMS}{cmsy}{m}{n}
\DeclareSymbolFontAlphabet{\mathcal}{usualmathcal}

\SetArgSty{textnormal}
\SetKwComment{Comment}{{\small\#}~}{}
\SetCommentSty{mycommfont}

\setitemize{itemsep=0pt, parsep=0pt} 				
\setenumerate{itemsep=0pt, parsep=0pt} 				
\setitemize{itemsep=2pt,topsep=2pt,parsep=0pt,partopsep=0pt,leftmargin=*}
\setenumerate{itemsep=0pt,topsep=2pt,parsep=0pt,partopsep=0pt,labelindent=3pt,leftmargin=*}
\usepackage{amsmath}
 
\usepackage{amsthm} 		
\theoremstyle{definition}

\newcommand{\dd}{\text{d}}

\definecolor{red_cb}{HTML}{e41a1c}
\definecolor{blue_cb}{HTML}{377eb8}
\definecolor{green_cb}{HTML}{4daf4a}
\definecolor{purple_cb}{HTML}{984ea3}
\definecolor{orange_cb}{HTML}{ff7f00}

\definecolor{EmeraldGreen}{HTML}{1ea78d}
\definecolor{EnglishRed}{HTML}{b02427}
\hypersetup{colorlinks=true,urlcolor=EmeraldGreen,citecolor=EmeraldGreen,linkcolor=EnglishRed}

\newcommand{\eg}{\text{e.g.}\;}
\newcommand{\ie}{\text{i.e.}\;}

\newcommand{\eqperiod}{\;.} 	

\newcommand{\mand}{\text{and}}

\newcommand{\Langle}{\bigl\langle}
\newcommand{\Rangle}{\bigr\rangle}
\newcommand{\XLangle}{\Bigl\langle}
\newcommand{\XRangle}{\Bigr\rangle}
\newcommand{\XXLangle}{\biggl\langle}
\newcommand{\XXRangle}{\biggr\rangle}

\newcommand{\qqquad}{\qquad\quad}
\newcommand{\qqqquad}{\qquad\qquad}

\newcommand\one{\leavevmode\hbox{\small1\normalsize\kern-.33em1}}

\newcommand{\loss}{\mathcal{L}} 	

\newcommand{\pythia}{\texttt{Pythia}\xspace}

\newcommand{\arXiv}[2][]{%
	\ifthenelse{\equal{#1}{}}%
	{\href{http://arxiv.org/abs/#2}{arXiv:#2}}%
	{\href{http://arxiv.org/abs/#2}{arXiv:#2~[#1]}}}

\def\slashchar#1{\setbox0=\hbox{$#1$}           
   \dimen0=\wd0                                 
   \setbox1=\hbox{/} \dimen1=\wd1               
   \ifdim\dimen0>\dimen1                        
      \rlap{\hbox to \dimen0{\hfil/\hfil}}      
      #1                                        
   \else                                        
      \rlap{\hbox to \dimen1{\hfil$#1$\hfil}}   
      /                                         
   \fi}

\newcommand{\tikznode}[2]{%
\ifmmode%
\tikz[remember picture,baseline=(#1.base),inner sep=0pt] \node (#1) {$#2$};%
\else
\tikz[remember picture,baseline=(#1.base),inner sep=0pt] \node (#1) {#2};%
\fi}

\def\mathswitchr#1{\relax\ifmmode{\mathrm{#1}}\else$\mathrm{#1}$\xspace\fi}
\def\mathswitch#1{\relax\ifmmode#1\else$#1$\xspace\fi}

\newcommand{\PZ}{\mathswitchr Z}

\newcommand{\Pg}{\mathswitchr g}

\newcommand{\Pd}{\mathswitchr d}
\newcommand{\Pu}{\mathswitchr u}
\newcommand{\Ps}{\mathswitchr s}

\graphicspath{{./figs/}}

\begin{document}

\begin{center}{\Large \textbf{
The Latent Information Geometry of Jet Classification
}}\end{center}

\begin{center}
Rebecca Maria Kuntz\orcidlink{0009-0006-0960-817X}\textsuperscript{1},
Tilman Plehn\orcidlink{0000-0001-5660-7790}\textsuperscript{2,3},
Björn Malte Schäfer\orcidlink{0000-0002-9453-5772}\textsuperscript{1,3},\\
Benedikt Schosser\orcidlink{0009-0007-8905-7749}\textsuperscript{1}, and
Sophia Vent\orcidlink{0009-0002-9106-3447}\textsuperscript{2}
\end{center}

\begin{center}
{\bf 1} Zentrum f\"ur Astronomie der Universit\"at Heidelberg, \\ Astronomisches Rechen-Institut, Germany \\
{\bf 2} Institut f\"ur Theoretische Physik, Universit\"at Heidelberg, Germany\\
{\bf 3} Interdisciplinary Center for Scientific Computing (IWR), Universit\"at Heidelberg, Germany
\end{center}

\begin{center}
\today
\end{center}


\section*{Abstract}
{\bf Latent representations are an important theme in modern machine learning. Any network training with the notion of locality introduces a latent geometry which we can analyze with the help of differential geometry, specifically information geometry. We introduce the main concepts needed to analyze learned latent geometries, specifically curvature and nonmetricities, and show how they can be used for decoder and classifier geometries. We then apply our new methods to understand the physics behind binary quark-gluon classification and three-fold fat jet tagging.}

\vspace{10pt}
\noindent\rule{\textwidth}{1pt}
\tableofcontents\thispagestyle{fancy}
\noindent\rule{\textwidth}{1pt}
\vspace{10pt}

\clearpage
\section{Introduction}
\label{sec:intro}

A leading theme of modern machine learning (ML) for fundamental physics~\cite{Plehn:2022ftl} is representation learning. Starting with exploiting structures or symmetries to increase performance, it leads us directly to explainability, which requires a mathematically sound and reliable framework in fundamental physics. Quantitatively understanding a complex network output or classifier decision allows us to learn about the underlying physics and strengthens confidence in the soundness of trained neural networks. If we manage to relate latent and feature spaces we can answer questions like: What physical features does the network rely on? Does the latent representation align with given physics features? How does such an alignment change with the network task?

We propose a new way to analyze latent representations based on the simple assumption that networks encode a task-specific correlations or similarity in a local structure. A variational autoencoder~\cite{kingma2022autoencodingvariationalbayes} realizes this through an explicit regularization, but self-attention~\cite{vaswani2023attentionneed} in transformers can be viewed in complete analogy. Given the local structure, we can study the geometry of the latent representation using differential geometry. Because network training can be viewed as statistical inference of a meaningful latent representation, we first turn concepts of General Relativity and its alternative formulations into the analogous information geometry. Starting with the Fisher information as a metric, we can study the latent geometry, curvature, nonmetricity, and geodesic distances, and in turn relate the learned representations to the physical feature space.

An interesting and relevant LHC task for this kind of study is modern jet tagging~\cite{deOliveira:2015xxd}, where quark-gluon tagging~\cite{Nilles:1980ys,Gallicchio:2012ez,Komiske:2016rsd,Cheng:2017rdo,Komiske:2018vkc,Kasieczka:2018lwf,Larkoski:2019nwj,Lee:2019ssx,Lee:2019cad,Kasieczka:2020nyd,Romero:2021qlf,Dreyer:2021hhr,Bright-Thonney:2022xkx,Bogatskiy:2023nnw,Athanasakos:2023fhq,Shen:2023ofd,Dolan:2023abg} is in many ways the most challenging application and affects a vast number of LHC analyses, typically separating quark jets produced in weak decays from QCD background jets. The theoretical background of quark-gluon is challenging, because quark and gluon jets are ill-defined beyond leading order in QCD~\cite{Gras:2017jty}. Furthermore, it strongly depends on the parton shower, hadronization, and detector effects~\cite{Gras:2017jty,Butter:2022xyj}. If we want to trust quark-gluon tagging in LHC analyses, we need to understand what it is based on. Once we do understand some of its intricacies, we can apply this insight to close the simulation gap from simulation-based training and improve the resilience of top taggers~\cite{Kasieczka:2017nvn,Macaluso:2018tck,Sahu:2024fzi, Larkoski:2024hfe, Woodward:2024dxb}, $W/Z$ taggers~\cite{Chen:2019uar,CMS:2020poo,Kim:2021gtv, Li:2025tsy}, and bottom/charm taggers~\cite{VanStroud:2023ggs,Hassan:2025yhp,ATLAS:2025rbr,ATLAS:2025dkv}.

Modern machine learning enables quark-gluon tagging through graph-based classification networks like the classic ParticleNet~\cite{Qu:2019gqs}. It represents modern taggers operating on a point cloud of low-level detector inputs or jet constituents. Advanced graph-based architectures, including advanced transformer implementations~\cite{Qu:2022mxj,He:2023cfc,Wu:2024thh,Brehmer:2024yqw,Spinner:2025prg,Esmail:2025kii,Petitjean:2025zjf},  have led to significant advances. To understand these graph-based networks we employ our information geometry methodology. While it is clear that low-dimensional Euclidean latent representations are not sufficient to encode all tagging-relevant information, the question is how this information is encoded. Curvature, nonmetricity and torsion offer three possibilities at least in principle. Going beyond binary classification, the separation of jets in terms of latent distance in the learned geometry becomes a key question.

In Sec.~\ref{sec:info_geo} we describe our information geometry methodology. In Sec.~\ref{sec:toy_geo} we apply it to a toy classification problem between digits, where we control the relevant geometric features. We show how four new scalars for information geometry capture much of the relevant information about the learned decoder and classifier geometries. In Sec.~\ref{sec:qg} we then apply these concepts to quark-gluon tagging and relate the learned geometries to the known jet shapes. Finally, we extend our framework to 3-label classification of top vs $Z$ vs $q/g$ jets in Sec.~\ref{sec:top} and link the latent space geometries to simple observables like the N-subjettiness and the jet mass. We provide an outlook and a detailed development of the theory in a series of appendices.

\clearpage
\section{Information geometry methodology}
\label{sec:info_geo}

Information geometry applies differential geometry to decision making~\cite{Amari:2016mig, Nielsen:2022mfg}. It is based on the fact that a parametric family of probability distributions,
\begin{align}
    \mathcal{P} = \{p_\theta :\, \theta \in \Theta \}
    \qqquad p:\Theta\rightarrow\mathcal{P(\mathcal{X})}\; ,
\end{align}
constitutes a topological space and thereby a notion of closeness. Here, $\Theta \subseteq \mathbb{R}^n$ is the parameter space and $\mathcal{X}$ the (sample) data space~\cite{Nielsen:2022mfg}. This topological space $\mathcal{P}$ becomes a statistical manifold $\mathcal{M}$, where the model parameters $\theta$ constitute a local coordinate chart, \ie a map to Euclidean space. A reparametrization of the statistical model then corresponds to a coordinate change on the manifold, and geometric invariants correspond to the key properties of the statistical model which are invariant under reparametrization~\cite{Nielsen:2022mfg}. In our study, such a statistical model will be a classifier or a decoder network.

\subsubsection*{Metric space}

A metric tensor is a symmetric positive-definite bilinear form which allows us to measure infinitesimal distances on the manifold. If we identify it with the Fisher information~\cite{Fisher:1922},
\begin{align}
    \tensor{g}{_i_j}(\theta) \equiv \tensor{F}{_i_j}(\theta) 
    = \int \dd \mu(x) \,p_\theta(x) \partial_i \log p_\theta (x) \, \partial_j \log p_\theta (x)
    \qquad \text{with} \qquad 
    \partial_i \equiv \frac{\partial}{\partial \theta^i} \; ,
    \label{eq:fisher}
\end{align}
the statistical manifold becomes Riemannian with a measure $\dd \mu$ over the data space. The Fisher information is the unique Riemannian metric invariant under sufficient statistics~\cite{Amari:2000, Nielsen:2022mfg}. The relation of the Fisher information metric to the quality of a potential optimal measurement is illustrated by the Cram\'er-Rao bound. For an unbiased measurement, the uncertainty of the measurement is given by the covariance matrix at $\theta$. The Cram\'er-Rao bound states that this covariance matrix is always larger than the inverse Fisher information. This bound can be used to compute the optimal outcome of multi-dimensional LHC precision measurements~\cite{Brehmer:2016nyr,Brehmer:2017lrt}.

We can construct a differential structure on the statistical manifold. It globally connects different tangent spaces $T_p\mathcal{M}$, where the connection allows us to differentiate one vector field $X \in \Gamma(T\mathcal{M})$ with respect to another $Y \in \Gamma(T\mathcal{M})$ as $\nabla_Y X$. In such an induced coordinate frame, the $k$-th component of the covariant derivative is related to the connection coefficients $\Gamma$ as  
\begin{align}
    (\nabla_Y X)^k = Y^i \left( \frac{\partial X^k}{\partial Y^i} + \tensor{\Gamma}{^k_i_j} X^j\right)\; .
\label{eq:def:connection}
\end{align}
The unique connection of Riemannian geometry, the Levi-Civita-connection (LC) $\tensor[^{\text{LC}}]{\nabla}{}$, is characterized by two properties: it is torsion-free and metric compatible. Therefore, the LC-connection is free to express invariant information as scalar curvature. However, the LC-connection alone is unable to fully capture the geometry of decision making. We need a different way to construct such a geometry, which in general has LC-curvature, nonmetricity, but no torsion~\cite{Nielsen:2020, Nielsen:2022mfg}.

\subsubsection*{Divergence-induced geometry} 

A divergence on the statistical manifold $D:\mathcal{M}\times\mathcal{M} \rightarrow \mathbb{R}_{0+}$ allow us to construct an information geometry~\cite{Eguchi:1983},
\begin{alignat}{7}
    &\text{metric tensor} &\qqquad 
    \tensor[^D]{g}{_{ij}} &= -\partial_i \partial_{j^\prime} D[p_\theta, p_{\theta^\prime}]\Bigg|_{\theta = \theta^\prime} \notag \\[4mm]
    &\text{connection coefficient} &\qqquad 
    \tensor[^D]{\Gamma}{_{ijk}} &= -\partial_i \partial_j \partial_{k^\prime} D[p_\theta, p_{\theta^\prime}]\Bigg|_{\theta = \theta^\prime} 
    \label{eq:divergence_geometry} \\[4mm]
    &\text{dual connection coefficient}  &\qqquad 
    \tensor[^{D^*}]{\Gamma}{_{ijk}} &= -\partial_k \partial_{i^\prime} \partial_{j^\prime} D[p_\theta, p_{\theta^\prime}]\Bigg|_{\theta = \theta^\prime}
    \notag     &\; . \notag
\end{alignat}
The asymmetry $D[p_\theta, p_{\theta^\prime}] \neq D[p_{\theta^\prime}, p_\theta]$ expresses the innate duality of information geometry. It is manifest in the existence of two connections $\tensor[^D]{\nabla}{},  \tensor[^{D^\star}]{\nabla}{}$, defined in terms of the respective connection coefficients in Eq.\eqref{eq:def:connection}, which are torsion-free, not metric-compatible, but conjugate to each other: only together they covariantly preserve the metric tensor, in this case the Fisher information~\cite{Efron:1975, Amari:1982, Amari:2016, Jost:2017, Nielsen:2022mfg},
\begin{align}
 \frac{1}{2}\left(\tensor[^D]{\nabla}{} + \tensor[^{D^\star}]{\nabla}{}\right)\tensor[^D]{F}{} = \tensor[^{\text{LC}}]{\nabla}{} \tensor[^D]{F}{} = 0
 \qquad \text{whereas} \qquad  \tensor[ ^{D, D^\star}]{\nabla}{} \tensor[^D]{F}{} \neq 0 \; \label{eq:conjugate_connections}.
\end{align}

\subsubsection*{Amari $\alpha$-geometry and Amari-Chentsov tensor} 

One particular choice of divergence is the $\alpha$-divergence~\cite{Amari:1980tis},
\begin{align}
    D_\alpha[p_\theta, p_{\theta^\prime}] = \frac{4}{1-\alpha^2}\left( 1 - \int \dd x\, p_\theta(x)^{(1-\alpha)/2} p_{\theta^\prime}(x)^{(1+\alpha)/2} \right) \; .
\end{align}
The information geometry objects in Eq.\eqref{eq:divergence_geometry} then read
\begin{alignat}{7}
    \tensor[^{(\alpha)}]{g}{_{ij}} &= \tensor{F}{_{ij}} \notag \\
    \tensor[^{(\pm\alpha)}]{\Gamma}{_{ijk}} &= \tensor[^{\text{LC}}]{\Gamma}{_{ijk}} \mp \frac{\alpha}{2} C_{ijk} 
    \quad \text{with} &\quad 
    \tensor[^{\text{LC}}]{\Gamma}{_{ijk}} &= \frac{1}{2} \left( \partial_k \tensor{F}{_{ij}} + \partial_j \tensor{F}{_{ik}} - \partial_i \tensor{F}{_{jk}} \right) \notag \\
     && C_{ijk} &= \int \dd \mu(x) \, p_\theta \, \partial_i \log p_\theta \, \partial_j \log p_\theta \, \partial_k \log p_\theta \; .
\label{eq:alpha_connection}
\end{alignat}
Here, $C_{ijk}$ is the symmetric rank-(0,3) Amari-Chentsov tensor (ACT) or skewness tensor. In analogy to the Fisher information, the ACT is the unique rank-3 tensor invariant under sufficient statistics~\cite{Nielsen:2020, Nielsen:2022mfg}.

\subsubsection*{KL-divergence geometry} 

In the limit $\alpha\rightarrow \pm 1$, one can show that the $\alpha$-divergence becomes the forward (inverse) KL-divergence~\cite{Amari:2016}
\begin{align}
      D_{\text{KL}}[p_\theta, p_{\theta^\prime}] 
      = \int \dd\mu(x)\; p_\theta(x) \, \log \frac{p_\theta(x)}{p_{\theta^\prime}(x)} \; ,
\end{align}
for which we can confirm that the Fisher metric in Eq.\eqref{eq:fisher} follows from Eq.\eqref{eq:divergence_geometry} 
\begin{align}
-\partial_i \partial_{j^\prime}  D_{\text{KL}}[p_\theta, p_{\theta^\prime}] \Big|_{\theta = \theta^\prime} 
    &= \partial_i \int \dd\mu(x) \; p_\theta(x) \;\partial_{j^\prime} \log p_{\theta^\prime}(x) \Big|_{\theta = \theta^\prime} \notag \\
    &= \int \dd\mu(x) \;  p_\theta(x) \;\partial_i \log p_\theta(x) \,\partial_{j} \log p_{\theta}(x)
    = F_{ij}(\theta) \; .
\end{align}
Accordingly, the connection coefficients in Eq.\eqref{eq:alpha_connection} become
\begin{align}
     \tensor[^{( \pm 1)}]{\Gamma}{_{ijk}} = \tensor[^{\text{LC}}]{\Gamma}{_{ijk}} \mp \frac{1}{2} C_{ijk} 
     \qquad \Leftrightarrow \qquad 
    C_{ijk} &= \tensor[^{(-1)}]{\Gamma}{_{ijk}} - \tensor[^{(+1)}]{\Gamma}{_{ijk}} \notag \\
    &\equiv \tensor[^{D_{\text{KL}}}]{\Gamma}{_{ijk}}
    - \tensor[^{D_{\text{KL}}^*}]{\Gamma}{_{ijk}}\; .
     \label{eq:kl_connection}
\end{align}
We can compute the ACT from the $(\pm 1)$-connection coefficients. From a statistical point of view, it quantifies the skewness of the likelihood and arises as
\begin{align}
     \tensor[^{(-1)}]{\Gamma}{_{ijk}} -& \tensor[^{(+1)}]{\Gamma}{_{ijk}}  
     =  -\partial_i \partial_j \partial_{k^\prime} D_{\text{KL}}[p_\theta, p_{\theta^\prime}] + \partial_k \partial_{i^\prime} \partial_{j^\prime} D_{\text{KL}}[p_\theta, p_{\theta^\prime}] \Bigg|_{\theta = \theta^\prime} \notag\\
     =& \partial_i \int \dd\mu(x)\; p_\theta(x) \; \partial_j \log p_\theta(x) \partial_{k^\prime}\log p_{\theta^\prime}(x) 
     - \partial_k \int\dd\mu(x) \; p_{\theta}(x) \; \partial_{i^\prime} \partial_{j^\prime} \log p_{\theta^\prime}(x) \Bigg|_{\theta = \theta^\prime} \notag \\
     =& \int \dd\mu(x) \; p_\theta(x) \; \partial_i\log p_\theta(x) \partial_j \log p_\theta(x) \partial_{k^\prime} \log p_{\theta^\prime}(x)  \notag \\
    &+ \int \dd\mu(x) \; p_\theta(x) \; \partial_{k^\prime} \log p_{\theta^\prime}(x) \; \partial_i \partial_j \log p_\theta(x) \notag \\
    &- \int\dd\mu(x) \; p_{\theta}(x) \; \partial_k \log p_\theta(x) \;\partial_{i^\prime} \partial_{j^\prime} \log p_{\theta^\prime}(x) \Big) \Bigg|_{\theta = \theta^\prime} \notag
    \\
    =& \int\dd\mu(x)\; p_\theta(x) \;\partial_i\log p_\theta(x) \partial_j\log p_\theta(x) \partial_k\log p_\theta(x) \; ,
\end{align}
which gives the ACT in Eq.\eqref{eq:alpha_connection}. 

\subsubsection*{Riemann curvature and Ricci scalar} 

\begin{figure}[t]
    \centering
    \includegraphics[width=0.8\linewidth]{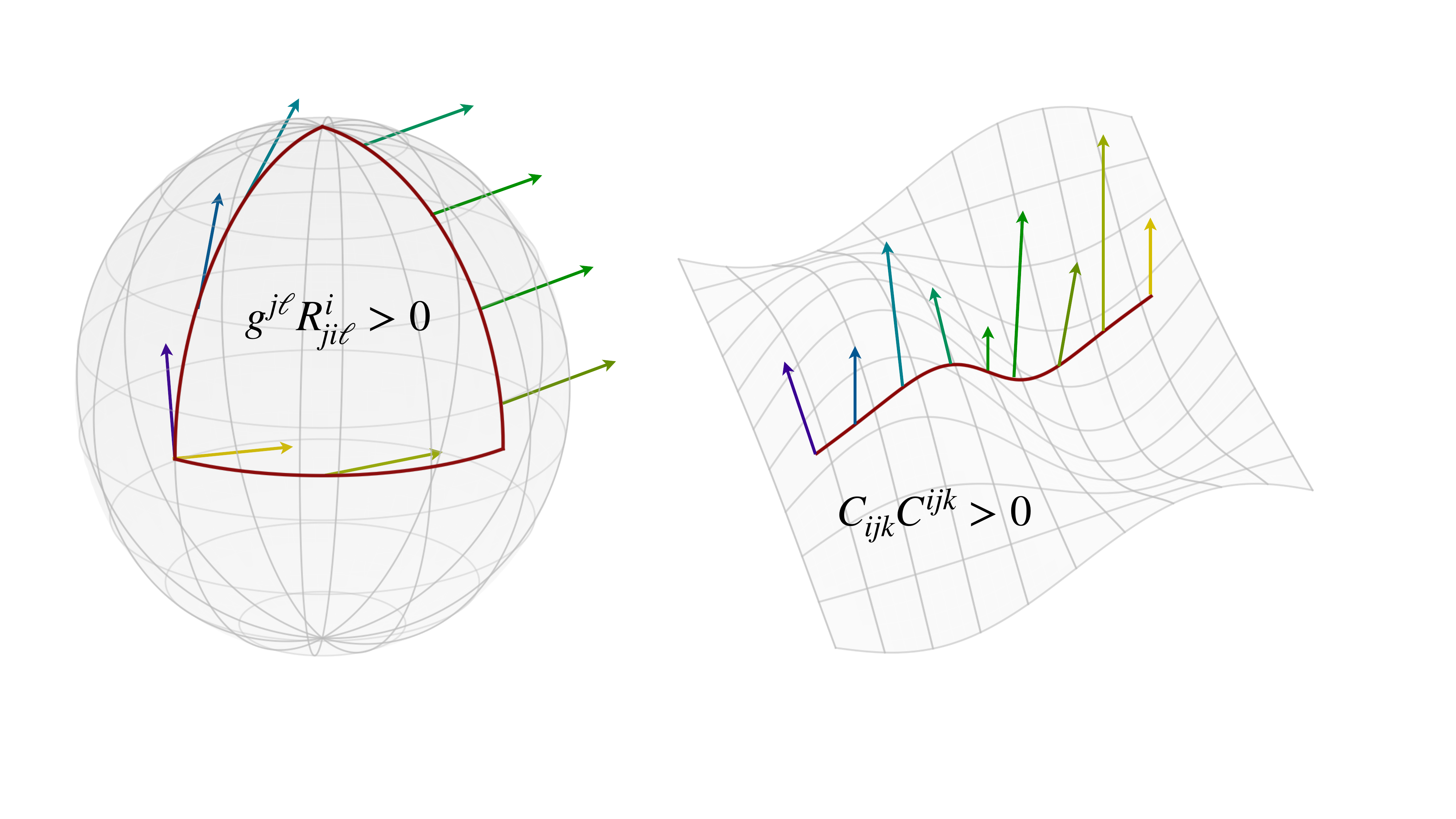}
     \vspace{-1.2cm}
    \caption{Left: Riemann curvature changes the orientation of a vector when it is parallel-transported in a loop. Right: the Amari-Chentsov tensor causes length and angular defects of vectors under parallel transport.}
    \label{fig:tensor_illustration}
\end{figure}

The left side of Fig.~\ref{fig:tensor_illustration} illustrates the effects of positive scalar curvature on the example of the 2-sphere: when a vector is parallel-transported in a closed loop on this manifold, a mismatch between the orientations of the initial and the final vectors emerges. This angular defect of the geometry is quantified by the Ricci curvature scalar $R$. As an essential property of the manifold it follows directly from the Riemann curvature tensor
\begin{align}
     \tensor{R}{^i_{j k \ell }} = \partial_k \tensor{\Gamma}{^i_{\ell j}} - \partial_\ell  \tensor{\Gamma}{^i_{kj}} + \tensor{\Gamma}{^i_{k\lambda}} \tensor{\Gamma}{^\lambda_{\ell  j}} - 
  \tensor{\Gamma}{^i_{\ell \lambda}} \tensor{\Gamma}{^\lambda_{jk}}\; ,
\end{align}
given a connection $\nabla$ and the corresponding coefficients. For the $(\pm 1)$-geometry, the Ricci scalars are then defined as \cite{Nielsen:2020}
\begin{align}
 R_{(\pm1)} = \tensor[^{(\pm 1)}]{R}{^i_{ j i \ell}} g^{j\ell} \; .
\end{align}

\subsubsection*{Curvature, nonmetricity, and the ACT}

There are three main ways to encode information in a connection: curvature, nonmetricity, and torsion. In standard General Relativity, the LC-connection $\tensor[^{\text{LC}}]{\nabla}{}$ is assumed to map out curvature, encoded in the Riemann tensor, and to covariantly preserve the metric, $\nabla_{\text{LC}} \,g =0$. Nonmetricity is an important concept in Weyl geometry formulations of gravity, which instead treat the geometry as flat and give up the assumption of metricity, $\nabla g \neq 0$. In this case, a nonmetricity tensor $Q_{ijk} \equiv \nabla_i g_{jk}$ quantifies geometry~\cite{Iarley:2015, Jaerv:2018}. 

In information geometry, nonmetricity appears because the $(\pm1)$-connections of the KL-divergence do not separately preserve the metric tensor, as we can see in Eq.\eqref{eq:conjugate_connections}. The nonmetricity tensor for the $(\pm 1)$-connection is the ACT~\cite{Udriste:2014, Amari:2016}, 
\begin{align}
 \tensor[^{(\pm 1)}]{\nabla}{_i} F_{jk} = \pm C_{ijk} \; .
\end{align}
The right panel of Fig.~\ref{fig:tensor_illustration} illustrates the geometric effect of  nonmetricity: a non-vanishing ACT changes a vector's length and direction under parallel-transport along a path on the statistical manifold. This leads to an an additional stretch, compression, or a shear of statistical distances. It is helpful to split the ACT into a trace and a part which is traceless in the last two indices,
\begin{align}
    C_{ijk}= \frac{1}{2} \tau_i\, F_{jk} + \tilde{C}_{ijk}
    \qquad \text{with} \qquad 
    \text{tr}_F \tilde{C} = 0 \; .
    \label{eq:act_split}
\end{align}
The trace $\tau_i$ is the Chebyshev (co-)vector field~\cite{Matsuzoe:2006}. It is the information geometry analogue of the Weyl field in gravity theories and as such dictates the failure to preserve lengths on the statistical manifold~\cite{Iarley:2015}. 

\begin{figure}[t]
    \centering
    \includegraphics[width=0.8\linewidth]{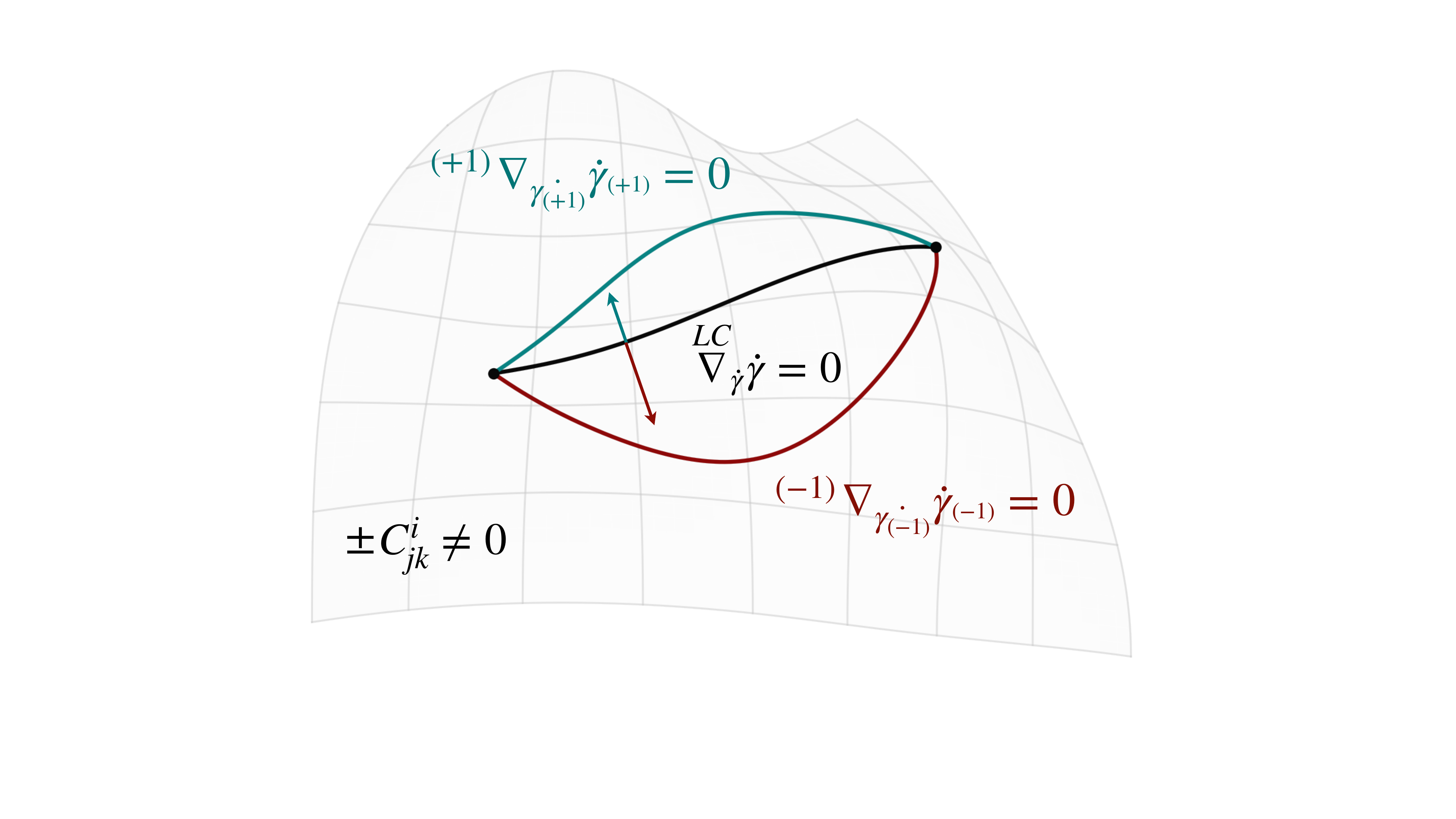}
    \vspace{-1cm}
    \caption{Illustration of parallel transport with the $(\pm 1)$- and LC-connection, respectively. Adapted from Ref.~\cite{Nielsen:2022mfg}.}
    \label{fig:dual_pt_illustration}
\end{figure}

\subsubsection*{Geodesics, autoparallels, and latent distance} 

To globally relate different points in the latent space and interpolate between them, we need to generalize the concept of straight lines and distances in flat, Euclidean space to a curved manifold. Given a connection, autoparallels are curves $\gamma: I\subset \mathbb{R}\to \mathcal{M}$, whose tangent vector $\dot{\gamma}\in\Gamma(T\mathcal{M})$ remains parallel to itself when transported along the curve. In local coordinates, autoparallel curves fulfill the ODE
\begin{align}
    \ddot{\gamma}^k(t) + \tensor{\Gamma}{^k_i_j} \dot{\gamma}^i(t) \dot{\gamma}^j(t) = 0 \; .
    \label{eq:geodesic}
\end{align}
If the connection coefficients in this ODE are taken from the LC-connection, the curves $\gamma$ are also length-minimizing geodesics, \ie the straightest and shortest paths. 
For the $(\pm 1)$-connections the ODE instead becomes
\begin{align}
    \ddot{\gamma}^k(t) + \left(\tensor[^{\text{LC}}]{\Gamma}{^k_i_j} \mp \frac{1}{2}\tensor{C}{^k_i_j} \right) \dot{\gamma}^i(t) \dot{\gamma}^j(t) = 0\; ,\label{eq:dualgeodesic}
\end{align}
where the ACT pushes the dual autoparallels ($\alpha = \pm 1$) away from the LC-geodesics ($\alpha = 0$)~\cite{Nielsen:2020, Nielsen:2022mfg}. This deviation of the dual autoparallels from the length-minimizing paths is illustrated in Fig.~\ref{fig:dual_pt_illustration}. 

As a measure of dissimilarity between two latent points, we therefore define the statistical Fisher-Rao distance along a path $\gamma$ on the manifold, for instance, a geodesic or autoparallel~\cite{Nielsen:2022mfg},
\begin{align}
    \rho_F(z_i, z_f) = \int_0^1  \dd t \; \sqrt{F_{ab}(\gamma(t))\; \dot{\gamma}^a(t) \dot{\gamma}^b(t)}
    \qquad \text{for} \qquad 
    \gamma: & t \in [0,1]\to z \in \mathbb{R}^2 \notag \\
    z_i &= \gamma(0),\; z_f = \gamma(1) \; .
    \label{eq:fisher_rao_distance}
\end{align}

\clearpage
\section{Decoder-classifier geometries and nonmetricity scalars}
\label{sec:toy_geo}

\begin{figure}[b!]
    \centering
    \includegraphics[width= 0.73\linewidth]{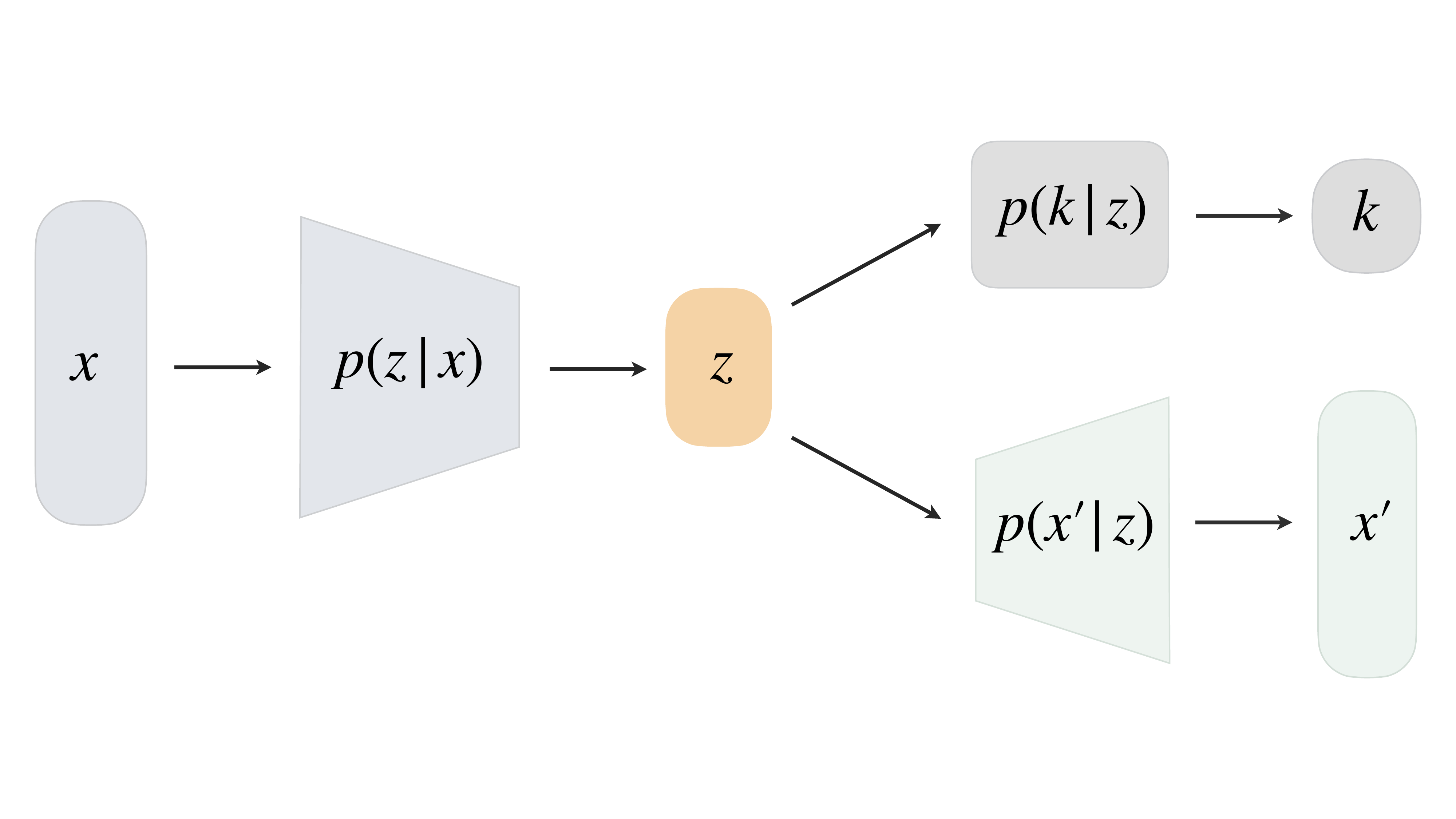}
    \vspace{-0.4cm}
    \caption{Conceptual overview of our network setup. The decoder and classifier are both attached to the latent space $z$ and each induce a geometry on it.}
    \label{fig:vae_classifier}
\end{figure}

As an illustration, we apply information geometry to the latent space of a variational autoencoder with an additional classifier head, shown in Fig.~\ref{fig:vae_classifier}. As the full data information is too complex for the latent space $z \in \mathbb{R}^2$, the network has to construct latent differential geometries, using curvature and nonmetricity, to encode information relevant for the decoding for the classification: Effectively, this corresponds to a highly nonlinear embedding of the data into the latent manifold. After introducing the classifier and decoder geometries, we go through the geometric quantities for analyzing the latent space alongside their statistical significance and illustrate them for subsets of the MNIST dataset. Crucially, we propose new scalars for information geometry and discuss their meaning.

\subsubsection*{Classifier Geometry}

A multi-class probability in terms of the network parameters $\phi$ is given by the softmax function
\begin{align}
    p_\phi(k | z) = \frac{e^{\eta_{\phi,k}(z)}}{\sum_{l}e^{\eta_{\phi,l}(z)}}
    \qqqquad k,l = 1~...~K \; .\label{eq:softmax}
\end{align}
For the discrete probability distribution, we keep the expectation values in the Fisher metric and the $\alpha$-connection
\begin{alignat}{7}
    \tensor[]{F}{_i_j}(z) 
    &= \XLangle s_i(r) \, s_j(r) \XRangle_{r\sim p_\phi(r| z)} 
    \qquad &\text{with} \quad 
    s_i(r) &= \partial_i \log p_\phi(r | z) \notag \\
    \tensor[]{\Gamma}{^i_j_k} 
    &= \tensor[]{F}{^{il}}
    \XXLangle s_l(r) t_{jk}(r) + \frac{1-\alpha}{2} s_l(r)s_j(r)s_k(r) \XXRangle_{r\sim p_\phi(r| z)} \\
    &= \tensor[]{F}{^{il}} \left( \XLangle s_l(r) t_{jk}(r)\XRangle_{r\sim p_\phi(r| z)}  + \frac{1-\alpha}{2} C_{ljk} \right)
    \qquad &\text{with} \quad 
    t_{jk}(r) &= \partial_j \partial_k \log p_\phi(r | z)\; . \notag 
\end{alignat}
Here, the nonmetricity or ACT can be finite.

\subsubsection*{Decoder geometry}

The decoder is parametrized by a Gaussian with learned mean,
\begin{align}
    p_\phi (x^\prime | z) = \mathcal{N}(\mu_\phi(z), \sigma_\phi)\; .
\end{align}
The variance does not depend on $z$. This Gaussian decoder yields simple analytic expressions for the Fisher metric, the connection coefficients, and the ACT,
\begin{align}
    \tensor[]{F}{_{ij}}(z) 
    &=\frac{1}{\sigma_\phi^2} \, \partial_i \mu_\phi(z) \, \partial_j \mu_\phi(z) \notag \\
    \tensor[^{\text{LC}}]{\Gamma}{^k_i_j}(z) 
    &= \frac{1}{\sigma_\phi^2 } \, \tensor[]{F}{^k^r} \, \partial_r  \mu_\phi(z) \, \partial_i \partial_j  \mu_\phi(z) \notag \\
    \tensor[]{C}{_i_j_k} 
    &= 0   
    \qqqquad \qqquad \text{with} \qqqquad \partial_i = \frac{\partial}{\partial z^i}
\end{align}
and the inverse metric $F^{ij}$. With the vanishing ACT the decoder geometry is Riemannian. For the rest of the paper we omit the network index $\phi$ whenever possible.

\subsubsection*{Fisher Frobenius norm and ellipses} 

The classifier and decoder each construct a latent coordinate system $z$ and a Fisher information or metric tensor. A straightforward  measure for the magnitude of $F_{ij}$ in the Euclidean sense is the Frobenius norm
\begin{align}
 \left\| F_{ij} \right\| = \sqrt{\sum_{ij} F_{ij}F_{ij} } \; .
\end{align}    
It is not coordinate-invariant, but a direction-averaged summary of how stretched or
compressed distances are in the Fisher information metric compared to a Euclidean metric in the coordinate chart selected by the network during training. Larger Fisher distances imply that two instances or likelihoods are easier to distinguish in latent space.

\begin{figure}[t]
    \includegraphics[width=0.45\textwidth]{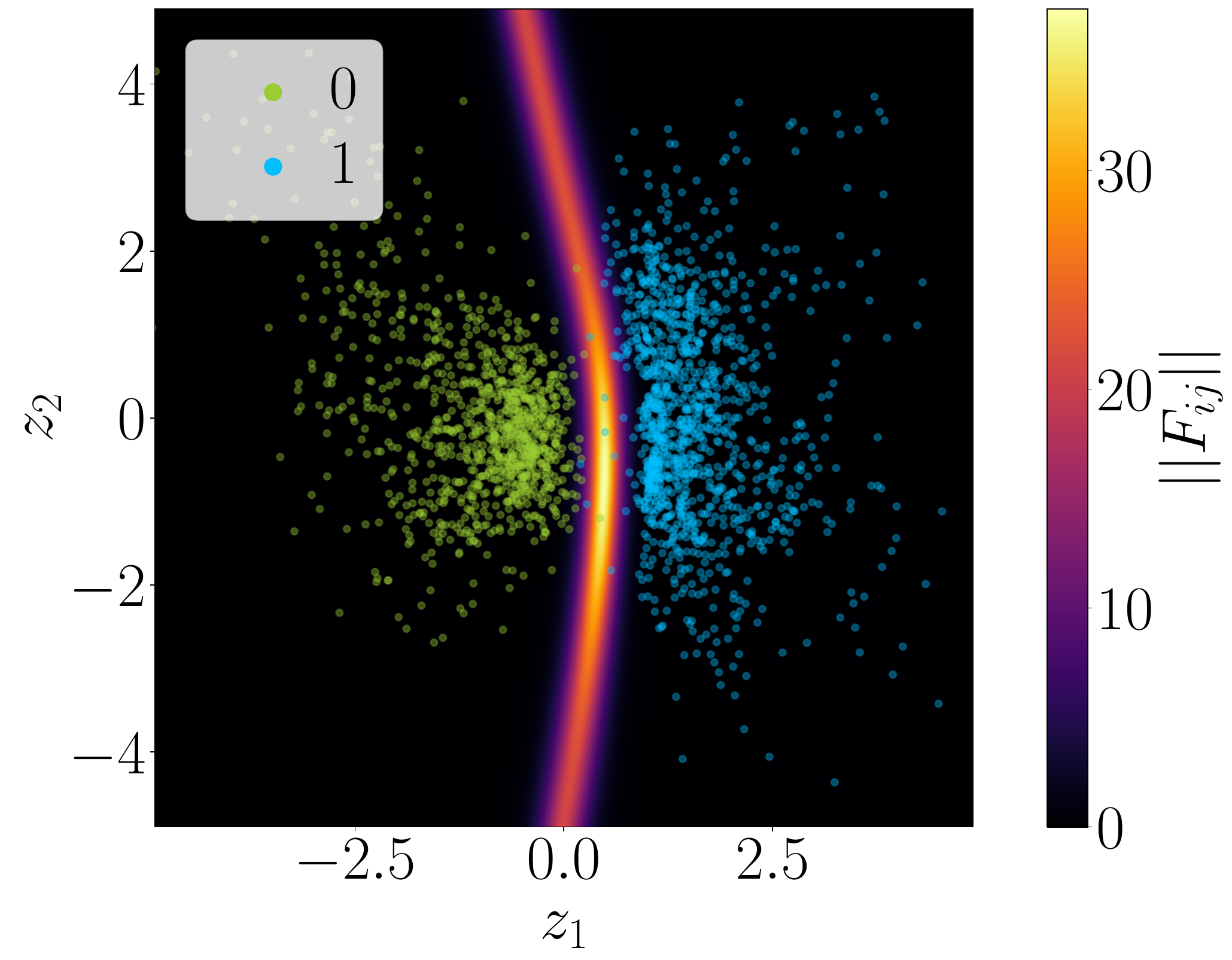}
    \includegraphics[width=0.45\textwidth]{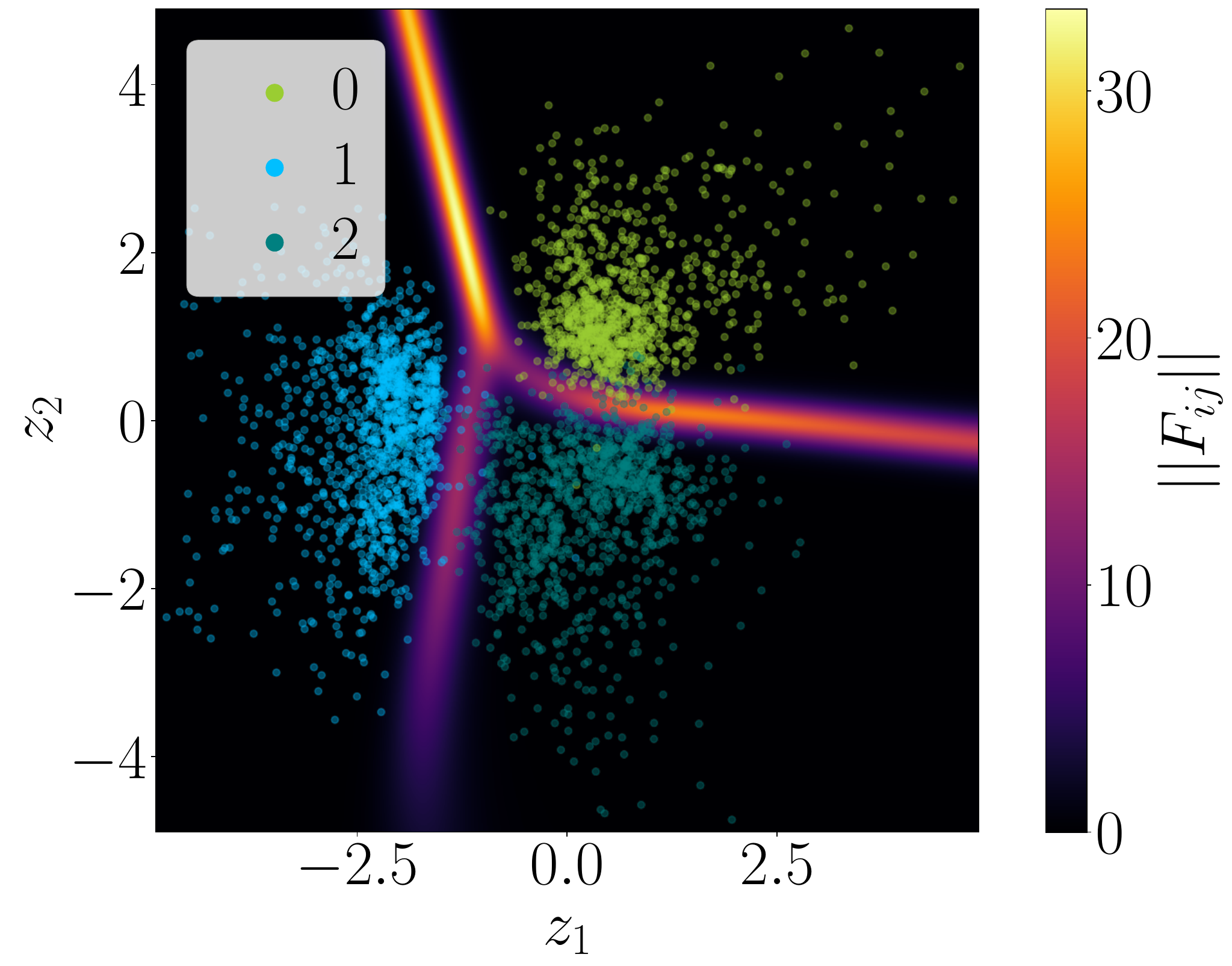} 
    \hfill \\
    \includegraphics[width=0.45\textwidth]{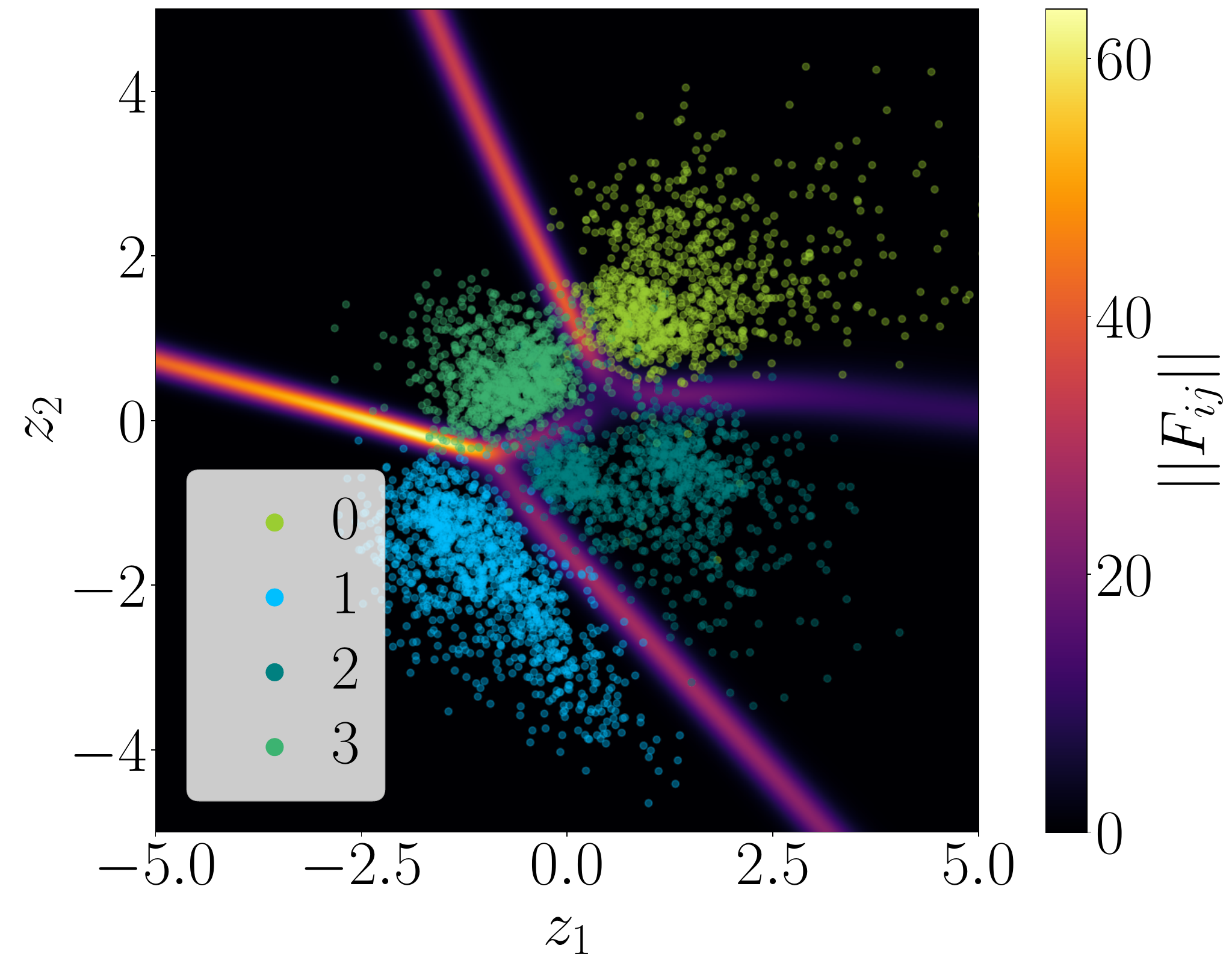}
    \includegraphics[width=0.52\textwidth]{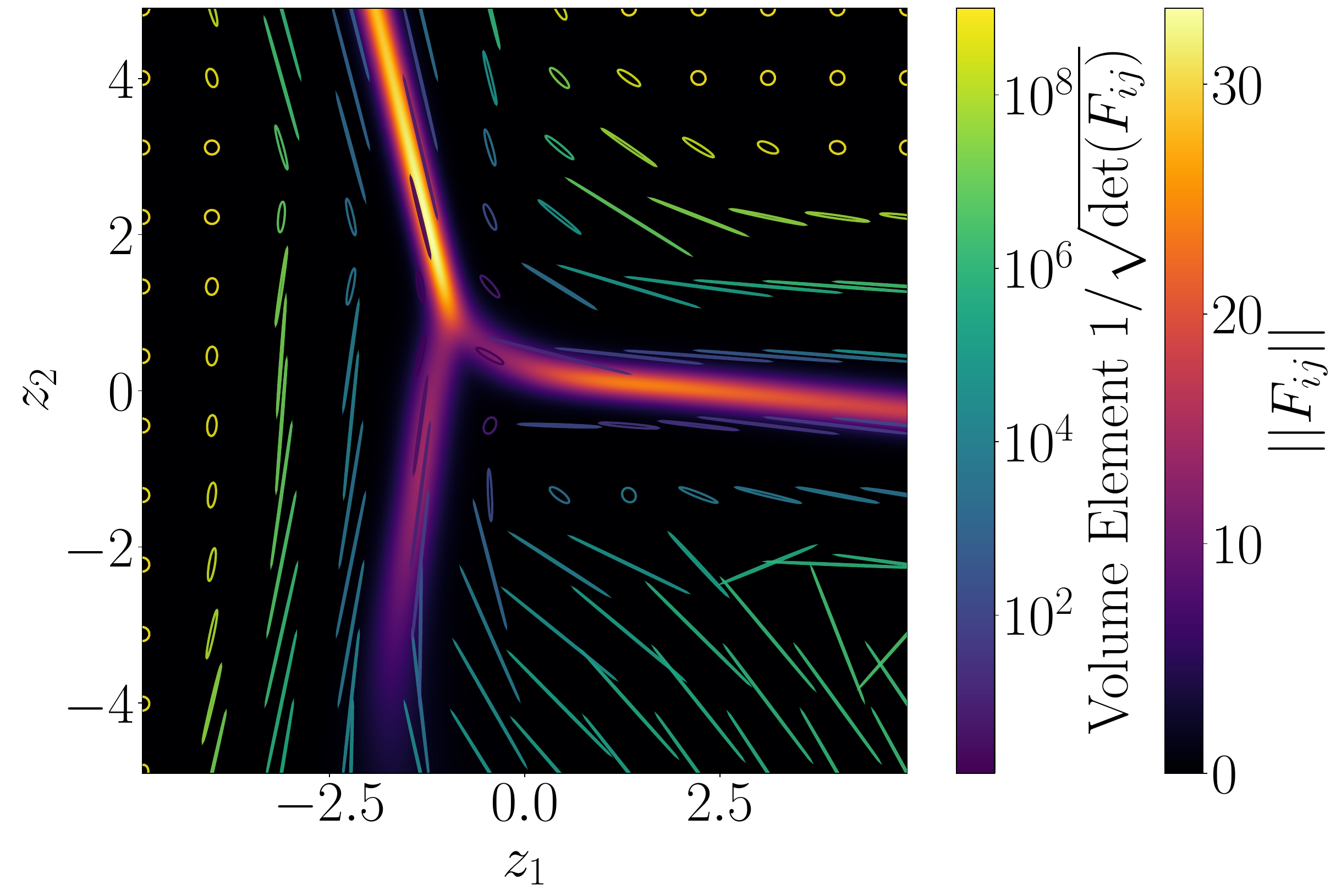}
    \caption{Upper: test data and Frobenius norm of the Fisher information for a binary classifier (left) and a three-label classification (right). Lower: test data and Frobenius norm of the Fisher information for four labels (left) and Fisher ellipses for the three-label case (right).}
    \label{fig:frobenius_ellipses}
\end{figure}

In Fig.~\ref{fig:frobenius_ellipses} we first see that the Frobenius norm traces decision boundaries in the latent classifier space. Because the Fisher information is the curvature of the log-likelihood surface, a large Frobenius norm signals high sensitivity of the classifier output to changes in $z$. At a decision boundary the predicted label changes most rapidly. In contrast, the Frobenius norm is small within the classes, where the classifier output hardly changes. While the Frobenius norm measures the discriminative power over the latent space, the encoded local information is averaged over all directions.

We resolve the directional stretch or compression through Fisher ellipses in the lower right panel of Fig.~\ref{fig:frobenius_ellipses}. They are aligned with the eigenvectors of the Fisher information, and the inverse eigenvalue scales the corresponding axis. The eigenvalues tell us how much classification power a latent direction carries, so Fisher ellipses should align with decision boundaries. For our graphic representation, the ellipse areas are encoded in a color map, while their shown area is normalized. The information loss from the directional averaging of the Frobenius norm is indeed small for elongated ellipses close to the decision boundaries.

\subsubsection*{Fisher directional derivative}

The eigenvector corresponding to the largest Fisher information eigenvalue marks the direction in which the network output is most sensitive to a latent space perturbation. For a classifier, we can test how the output features change along this direction. Given the leading Fisher information eigenvector $\upsilon$ of the classifier, the directional derivative of the decoder, learning $\mu(z)$,
\begin{align}
    D_\upsilon \mu(z) = \partial_\lambda \mu(z + \lambda\upsilon) \Bigg\rvert_{\lambda=0}
    \label{eq:directional_derivative}
\end{align}
locally quantifies how much the output features change along the most decisive direction.

\begin{figure}[t]
  \includegraphics[width=0.44\textwidth]{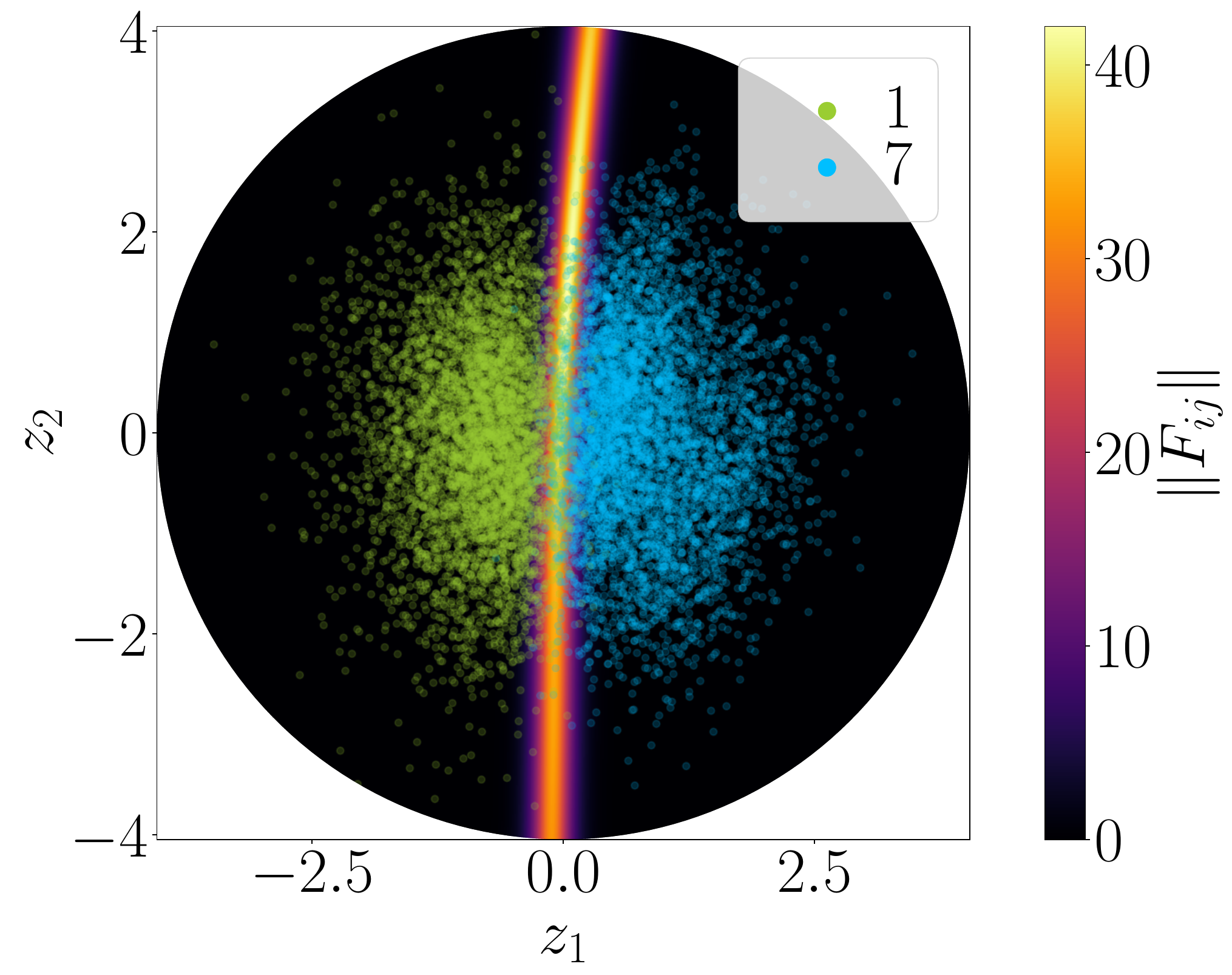}
  \hfill
  \includegraphics[width=0.54\textwidth]{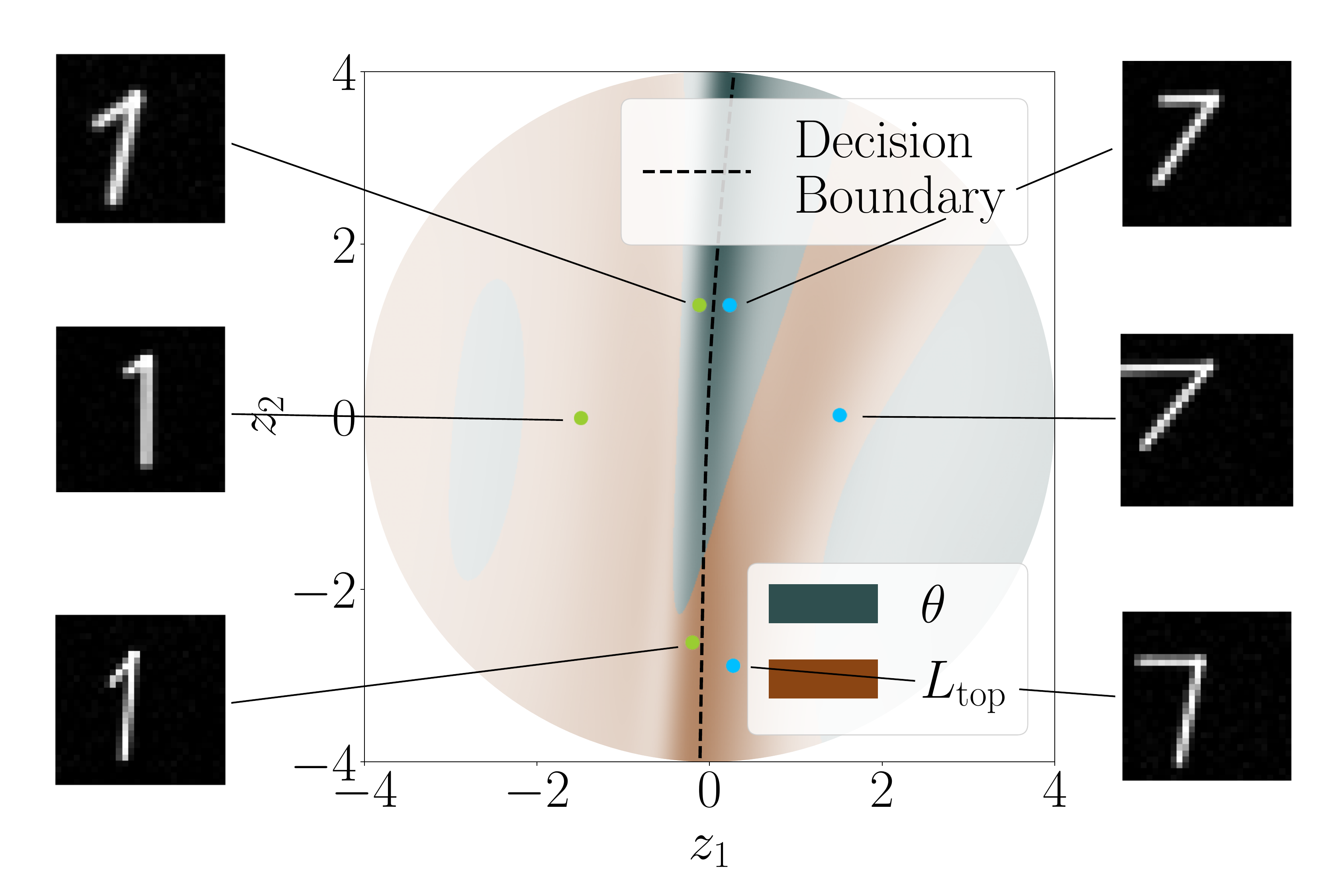}
  \caption{Left: test data and Frobenius norm for the simplified classification of 1 vs. 7. Right: feature based on the size of the directional derivative of the decoder in the direction of the dominant eigenvector of the classifier metric.} 
  \label{fig:mnist_directional_derivative}
\end{figure}

The left panel of Fig.~\ref{fig:mnist_directional_derivative} shows the Frobenius norm and test data over the latent space. Here, and for all upcoming plots, we whiten the out of distribution region of the latent space, which results in a circular plot. For this example, we do not use the original MNIST data, but a toy version described in App.~\ref{sec:datagenerationmnist}. It consists of the digits 1 and 7 and each number is characterized by two features, the length of the top line $L_\text{top}$ and the rotation angle of the vertical line $\theta$. The two numbers are separated with minimal overlap. In the right panel we show the Fisher directional derivative. For two features, it is also two-dimensional, and we color-code the feature with the larger component. The actual size of the directional derivative is encoded as increased transparency for smaller values. Because the features are standardized, the magnitudes of the derivatives are comparable. 

First, we see that the directional derivative is largest around the decision boundary, \ie the features change most rapidly there. Along the upper part of the decision boundary, the rotation angle of the vertical line is the dominant feature. This means that for numbers where the classifier is slightly uncertain, changing the angle is most effective to push it towards the other class. The upper examples of 1 and 7 illustrate this point. At the bottom of the latent space, the length of the vertical line has the larger impact on the classification. This reflects the geometry of this toy example, where 1 and 7 can be separated in different ways. 

Away from the decision boundary the directional derivative for the two features has to be interpreted with care. For almost all of the 7-region, decreasing the top length moves the classifier output more towards a 1. This only implies that the top length is the feature that changes most when we move slightly towards the decision boundary. However, a very tilted 7 eventually needs to be rotated to become a 1 when we move classes.

\subsubsection*{Geometry alignment}

\begin{figure}[t]
    \centering
  \includegraphics[width=0.49\textwidth]{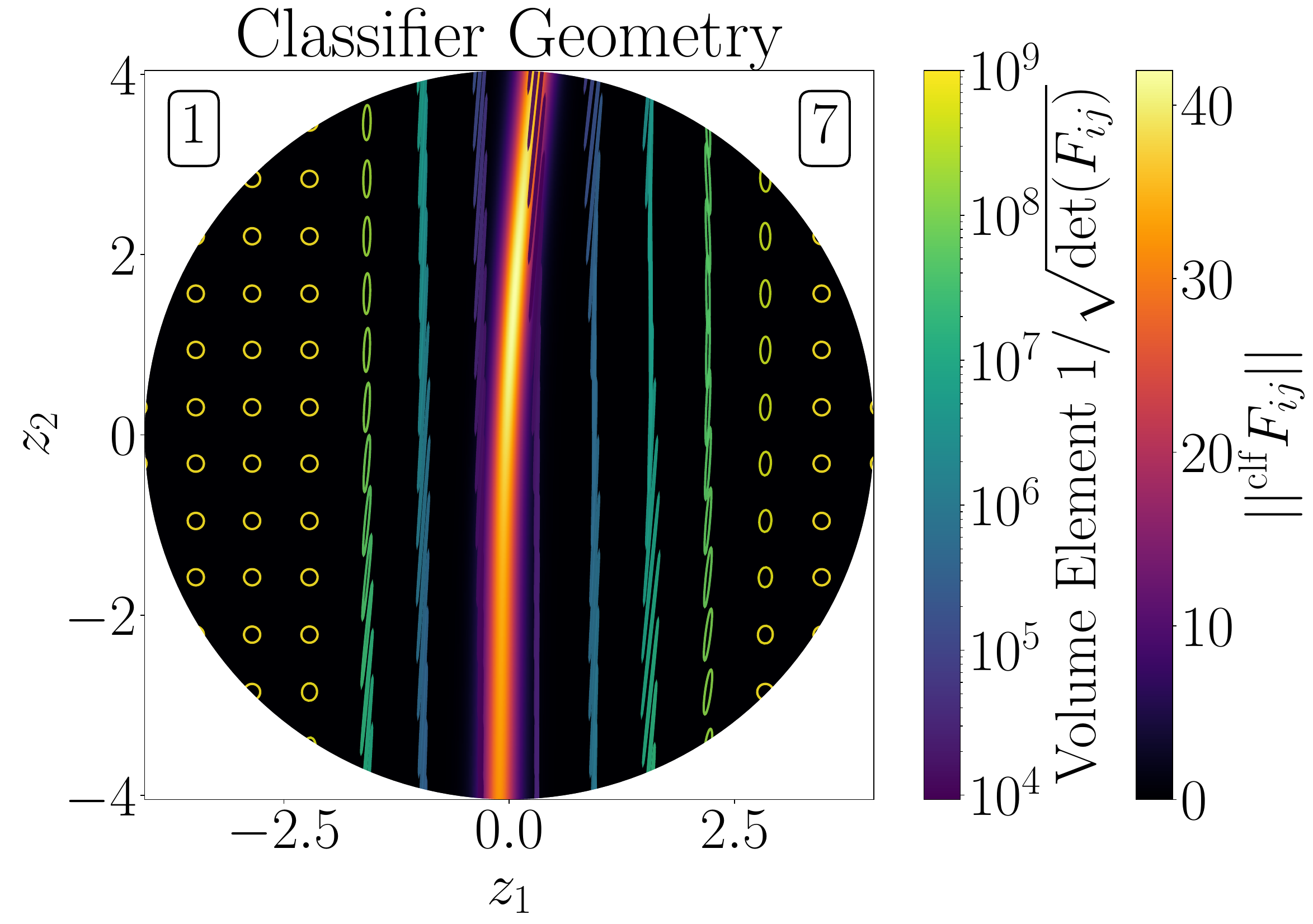}
  \hfill
  \includegraphics[width=0.49\textwidth]{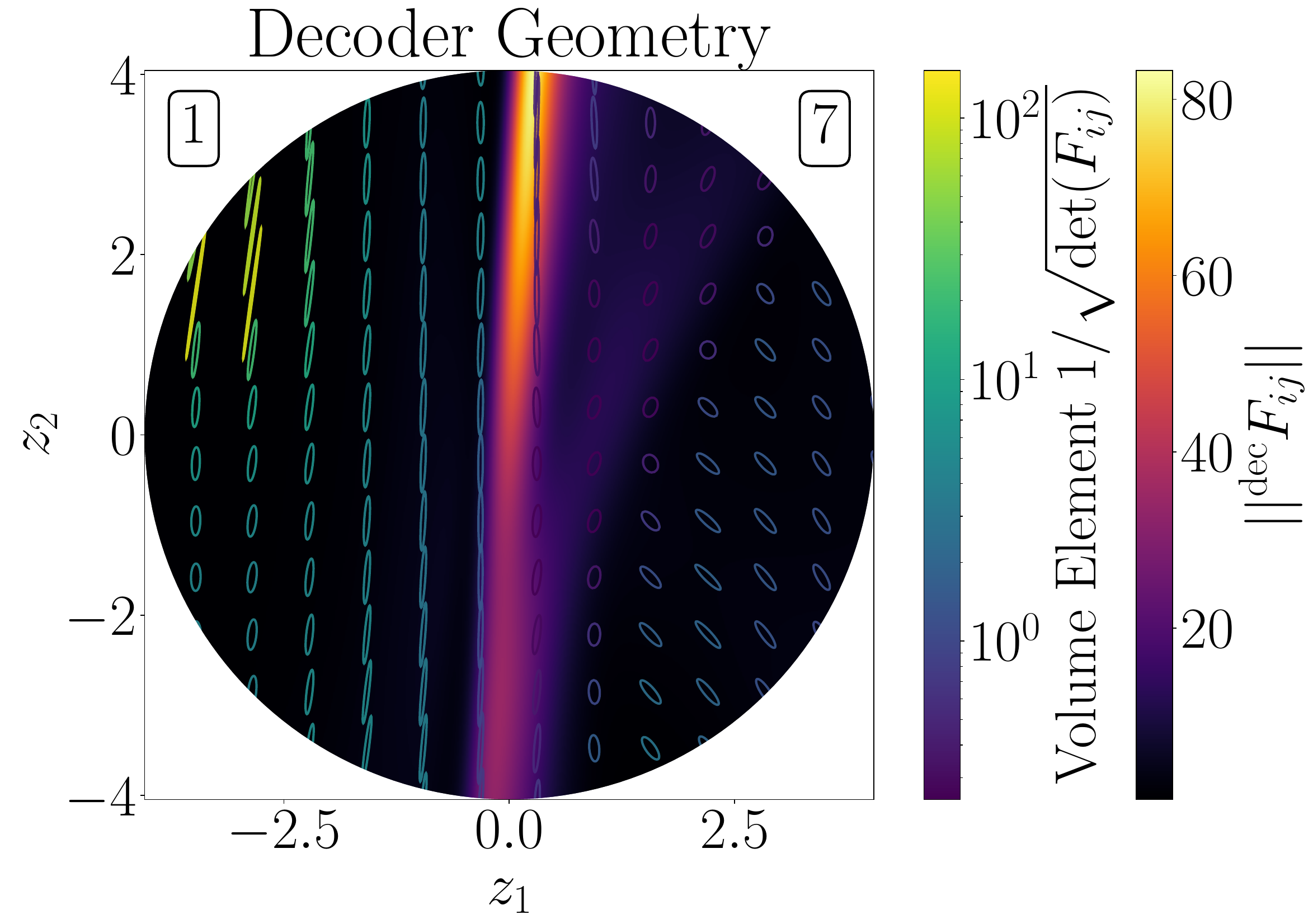}  
  \includegraphics[width=0.45\textwidth]{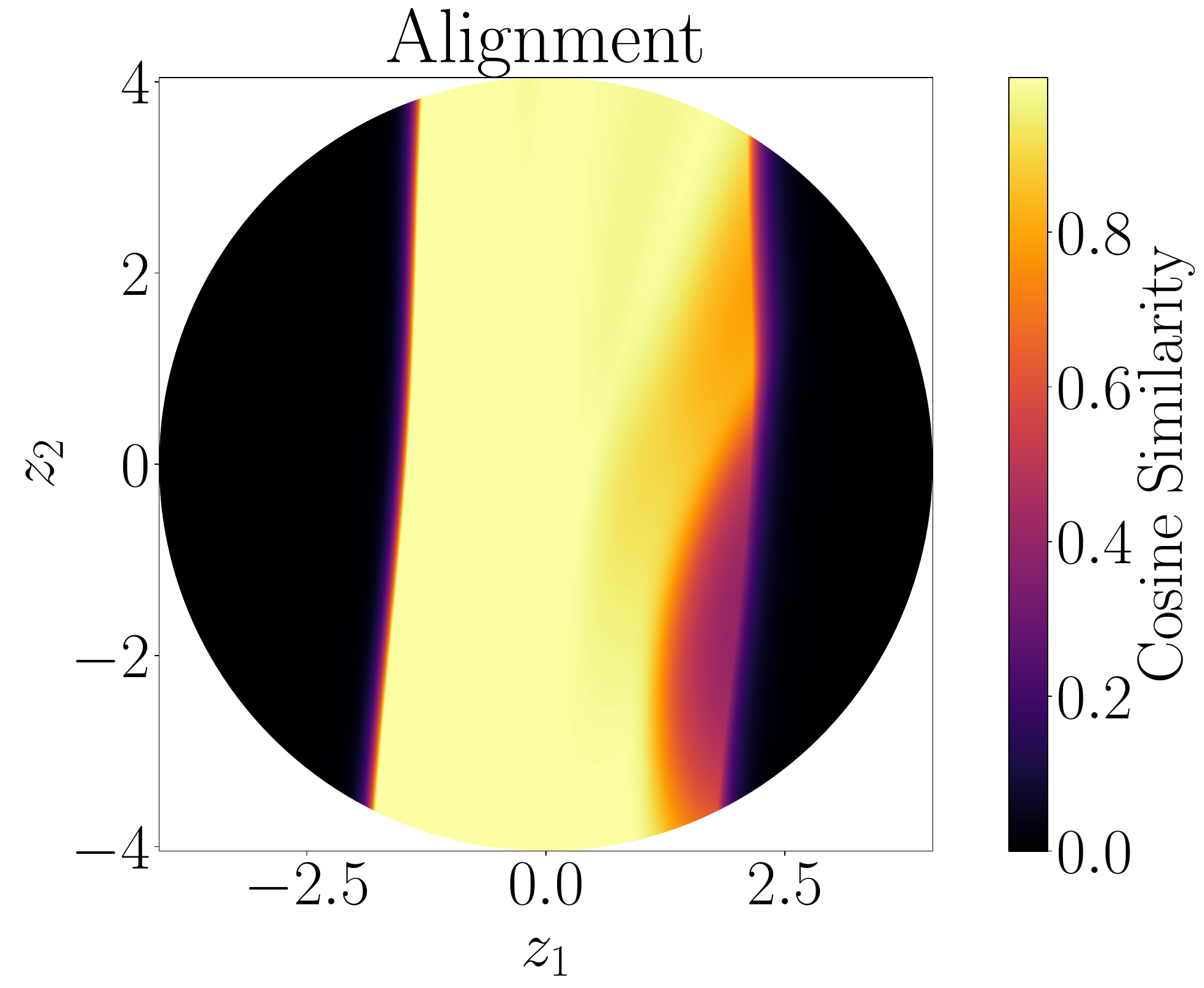}
  \caption{Left: Classifier geometry with the Frobenius norm and ellipses. Center: Cosine similarity as alignment measure between the classifier and decoder geometry. Right: Decoder geometry with the Frobenius norm and ellipses.} 
  \label{fig:mnist_alignment}
\end{figure}

To assess the quality of a set of features for a specific classification task, we measure the alignment between the two geometries with a cosine similarity between the two metric tensors,
\begin{align}
    \text{align}(z) =\frac{\sum_{i,j} \, {}^{\text{dec}}F_{ij}(z)\;{}^{\text{clf}}F_{ij}(z)}{ \left\| {}^{\text{dec}}F_{ij} \right\|\;\left\| {}^{\text{clf}}F_{ij} \right\|}\; .
\end{align}
This is not an invariant scalar and only tests whether the orientation entailed in the two metric tensors is the same. The induced geometry from the decoder becomes interpretable once the reconstructed features are well understood. In the 1 vs. 7 example, the features represent physical characteristics of the digits, while before, when the full image is reconstructed, interpreting how the value of each pixel influences the space is difficult. The decoder geometry is much simpler, as it is built on a Gaussian distribution, which does not permit any skewness, $C_{ijk}=0$. 

In Fig.~\ref{fig:mnist_alignment} we show the Frobenius norms for the classifier and the decoder and the cosine similarity for the two spaces for the 1 vs. 7 binary classification. The near-perfect alignment close to the decision boundary implies that the the decoder is sensitive in the same direction as the classifier, so the chosen features are a good basis for the classification. It fails far away from the decision boundary, where the certain classifier is insensitive to small changes in the latent space, and the decoder still reconstructs the features. If the two spaces are aligned at the decision boundary, the chosen features are sufficient and do not carry unnecessary information for the classification. 

\subsubsection*{Geodesics, autoparallels, and latent distance} 

\begin{figure}[t]
    \centering
    \includegraphics[width=0.9\linewidth]{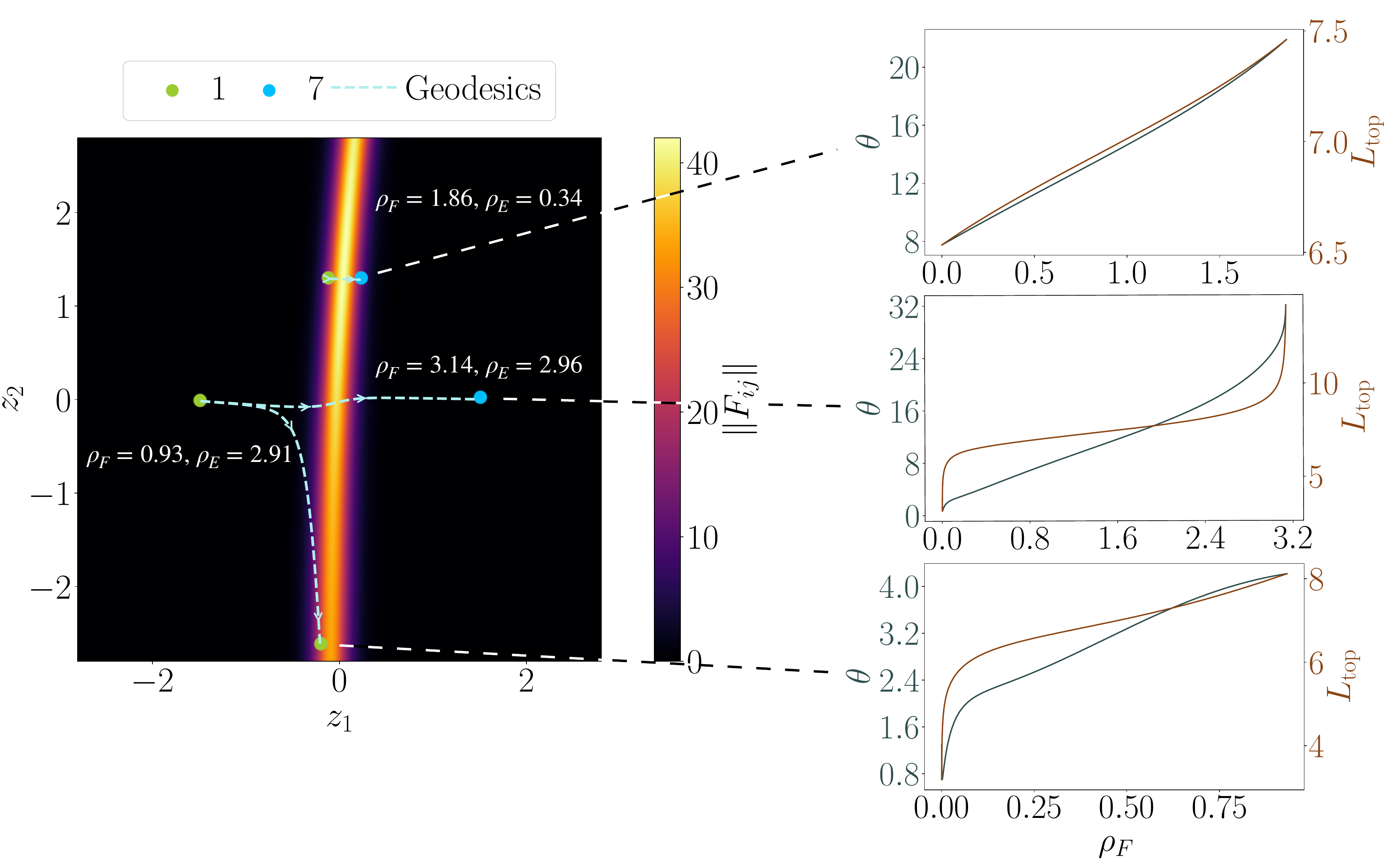}
    \caption{Simplified 1 vs. 7 classification task. Left: three LC-geodesics with their Fisher-Rao distance $\rho_F$ and the Euclidean distance $\rho_E$ superimposed on the Frobenius norm. Right: reconstructed rotation angle and top length as a function of the Fisher-Rao distance along the geodesics.}
    \label{fig:frobeniusgeodesics17}
\end{figure}

We illustrate Fisher-Rao distances for the 1 vs 7 classification in Fig.~\ref{fig:frobeniusgeodesics17}, where we show three LC-geodesics. The upper geodesic is limited to just the boundary region. All three clearly differ from Euclidean straight lines in $\mathbb{R}^2$, as does the statistical distance from the Euclidean distance $\rho_E$. Whenever we traverse regions of high Frobenius norm, the Fisher-Rao distance increases much faster than the Euclidean distance.

The right panels in Fig.~\ref{fig:frobeniusgeodesics17} show the reconstructed features, rotation angle and top length, along each classifier geodesic. For the first geodesic, the features change approximately linearly with $\rho_F$ from the classifier. If we instead measure the features along the decoder geodesics, they grow linearly by definition. Linear change of a feature along a classifier geodesic thus suggests an agreement between the classifier and decoder metric around the decision boundary, which confirms that the given feature is precisely what pushes events across the boundary. We confirm this by comparing with Fig.~\ref{fig:mnist_directional_derivative}, including the fact that the most relevant feature for the boundary crossing is the angle in this region.  

For the second geodesic, the Frobenius norm in the black region is small and the Fisher-Rao distance grows more slowly than $\rho_E$ at first. Conversely, the top length feature changes rapidly, indicating that the decoder and classifier metric define vastly different geometries. Closer to the decision boundary we find a region of linear growth of the rotation angle with the classifier distance $\rho_F$. Like for the first geodesic, the angle is the leading classification feature. Fig.~\ref{fig:mnist_directional_derivative} confirms this the top length is the dominant feature as one moves towards the decision boundary and is then surpassed by the rotation angle, which ultimately changes the classification. Beyond the boundary, the top length dominates again. 

The third geodesic never crosses the decision boundary, \ie the classification never changes and only becomes less certain. Since we never cross regions of high Frobenius norm, the flat Euclidean distance is significantly larger than the Fisher-Rao distance. The reconstruction in Fig.~\ref{fig:frobeniusgeodesics17} shows an initially rapid change in top length, which is subsequently replaced by a period of more linear growth. Hence, the certainty in the classification drops at a rate similar to that with which the top length changes. This implies that the most influential feature for classification in this latent region is the top length, which is confirmed by Fig.~\ref{fig:mnist_directional_derivative}.   

\subsubsection*{Interpretable coordinate systems}

Because the likelihoods of our classifier form an exponential family~\cite{Amari:2016, Nielsen:2020, Nielsen:2022mfg}, their geometry is dually flat, meaning the Ricci scalars of both $(\pm 1)$-connections vanish everywhere. For that reason, there exist two privileged coordinate frames in which the $(\pm 1)$-autoparallels appear as straight lines~\cite{Nielsen:2020, Amari:2016}. These so-called natural and expectation coordinates have a straightforward statistical interpretation in that they correspond to log-likelihood ratios or likelihoods~\cite{Nielsen:2020}. 

For discrete likelihoods $p(k|z)$ over classes $k$, an exponential family $\mathcal{P}_e$ is defined as 
\begin{align}
    \mathcal{P}_e = \left\{ p_z \,|\,  p(k|z) = h(k) \cdot \exp{\left(z^\top T(k) - A(z)\right)}\right\}\; ,
\end{align}
with $h(k)$ a function of the class $k$, $T(k)$ a sufficient statistic of the class counts, and $A(z)$ ensuring normalization~\cite{Amari:2016}. The corresponding natural coordinate system $\lambda \in \mathbb{R}^{K-1}$ is 
\begin{align}
    \lambda_k = \log \frac{p(k|z)}{p(k_0|z)} 
    \qquad \text{for} \qquad 
    k_0 \in K,\; k_0 \neq k 
    \; .
\end{align}
In these coordinates, the categorical likelihoods from Eq.\eqref{eq:softmax} become
\begin{align}
    p_{\phi}(k| z) 
    \; \mapsto\;   
    p_{\phi}(k|\lambda) = \frac{\exp \sum_{i=1}^{K-1} ( \lambda_i T_i(k)) }{\sum_{l=1}^{K}\exp \sum_{i=1}^{K-1} \lambda_i (T_i(l))}\; ,
\end{align}
with $T(k)$ a one-hot encoding. 
\begin{enumerate}
\item The $(+1)$-autoparallels appear as straight lines in $\lambda$ by definition, they are log linear interpolations between two distributions $p^{(0)}$ and $p^{(1)}$ at $\lambda^{(0,1)}$,
\begin{align}
    \lambda^{(t)} &= (1-t)\lambda^{(0)} + t \lambda^{(1)} \qqqquad t\in[0,1] \notag \\
    \Rightarrow \qquad     
    \log p^{(t)}(k) &= (1-t)\log p^{(0)}(k) + t \log p^{(1)}(k) - \log Z(t) \notag \\
    &\qquad \text{with} \qquad 
    Z(t) = \sum_{i=1}^{K}\left[p^{(0)}(i) \right]^{1-t} \; \left[ p^{(1)}(i)\right]^{t} \; .
\end{align}
\item Conversely, $(-1)$-autoparallels are straight lines in the expectation coordinates
\begin{align}
    \eta_i &= \XLangle T_i(k) \XRangle_{k\sim p(k|z)}\; ,
\end{align}
and for the one-hot sufficient statistic we obtain $\eta_i = p(i|z)$. They are a linear mixture of the start and end distribution,
\begin{align}
    \eta^{(t)} &= (1-t)\eta^{(0)} + t \eta^{(1)} \qqqquad t\in[0,1] \notag \\
    p^{(t)} &= (1-t)p^{(0)} + t p^{(1)}\,.
    \label{eq:mixture_autoparallel}
\end{align}
\end{enumerate}
Depending on which learned autoparallel is straight in these coordinate frames, we can infer whether the model exhibits a log-likelihood ratio interpolation or a linear (mixture) interpolation.

\begin{figure}[t!]
    \includegraphics[width=\textwidth]{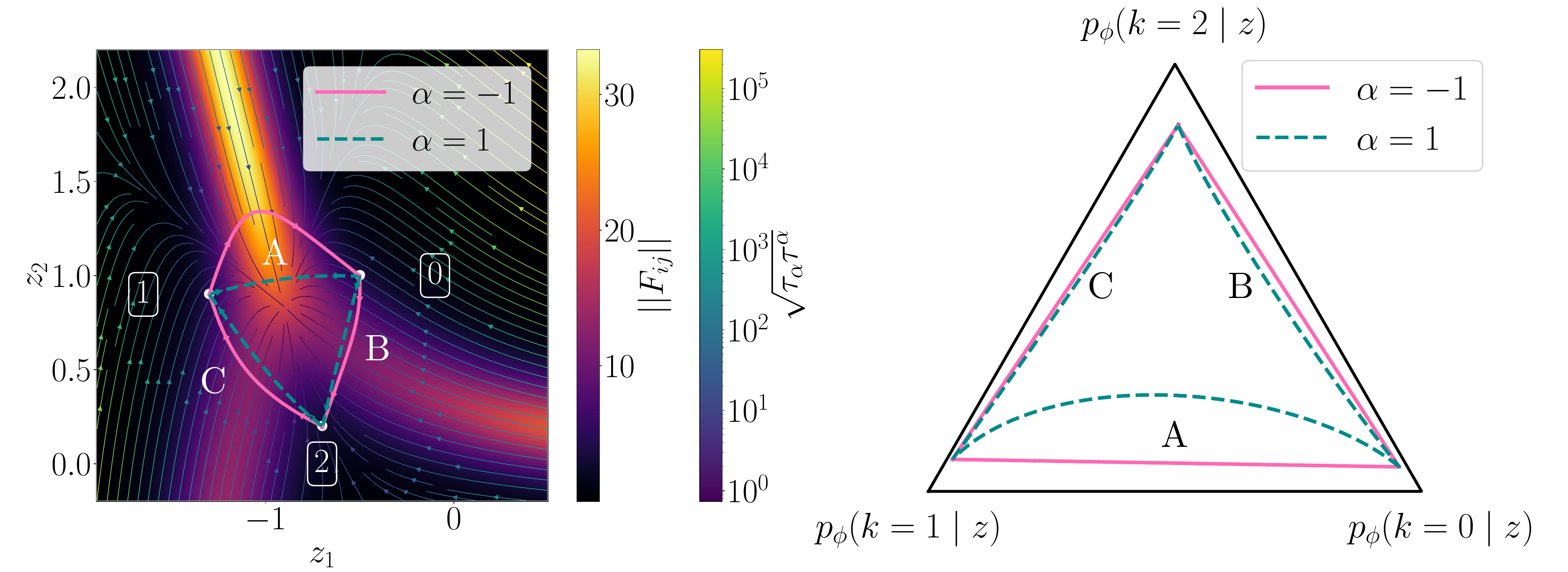}
    \caption{Left: Frobenius Norm and Chebyshev field with color coded norm for the three digit dataset. Three autoparallels connecting three test digits for the two $\alpha$ connections.  Right: Paths on the simplex along the autoparallels on the left.}
    \label{fig:mnist_simplex}
\end{figure}

In Fig.~\ref{fig:mnist_simplex} we show the Chebyshev field over the latent space for the three digit classification from the right panels of Fig.~\ref{fig:frobenius_ellipses}. We also show $(\pm 1)$-autoparallels connecting test points in the three class regions. The right panel shows the output of the classifier along these autoparallels. We observe a preference for natural coordinates from the almost straight latent curves from the $(+1)$-connection. In contrast, the $(-1)$ connection enforces a curved latent path, but straight lines on the simplex through the linear mixture of probabilities, Eq.~\eqref{eq:mixture_autoparallel}. This indicates that the classifier metric prefers the natural coordinates over the expectation coordinates.

The natural coordinates, representing the log likelihood ratio, are curved on the simplex, but the deviation from a straight line in $z$ between the digits 0 and 2 and 2 and 1, respectively, is minor. The $(-1)$-autoparallels have to be straight lines on the simplex, so if we want to use them to connect two points of high certainty, we cannot go through the region of the third class. However, in our case, the direct path in latent space from 0 to 1 involves a section of non-zero probability for 2. This pattern is also visible in the direction and norm of the Chebyshev field, which can be viewed as a force pushing the $(+1)$-autoparallel away from the $(-1)$-autoparallel. This force is strongest at the decision boundary between 0 and 1, affected by a metric that is larger where the decision is clearer. Explicitly, we can infer something about how the digits are connected. Turning a 0 into 1 prefers a phase of non-zero probability for a 2. However, the shift from 0 to 2 and from 1 to 2, does not require such an intermediate step. 

\subsubsection*{Novel nonmetricity scalars for information geometry} 

In general, statistical manifolds have both curvature and nonmetricity but, as shown in Eq.\eqref{eq:ricciscalars}, this is not to the case for our flat but non-metric dual classifier geometry. This motivates us to propose novel informative scalars for information geometry. These are different from the magnitude of nonmetricity discussed in Ref.~\cite{Hirano:2023}, which refers to the constant Amari-$\alpha$ scaling in Eq.\eqref{eq:alpha_connection}.

The ACT is the nonmetricity tensor of the $(+1)$-geometry, $\tensor[^{(+1)}]{\nabla}{_i} F_{jk} = C_{ijk}$~\cite{Udriste:2014, Amari:2016}, where a skewness in the likelihood implies more complex distributions than Gaussians. Geometrically, the nonmetricity tensor quantifies the non-preservation of lengths and angles on a manifold~\cite{Iarley:2015}. A finite ACT thereby signals a deviation from Riemannian geometry towards higher geometric complexity. Similarly to nonmetricity scalars in gravity theories, we propose different ACT contractions at quadratic order to access this information,
\begin{alignat}{7}
  C_1 &= C_{ijk}C^{ijk} &\qqqquad
  C_2 &= \tau_i \tau^i \notag \\
  C_3 &= \tilde{C}_{ijk} \tilde{C}^{ijk} &\qqqquad 
  C_4 &= \frac{C_{ij k} C^{ij k}-\tau_i \tau^i}{4} \; . 
  \label{eq:nonmetricity_scalars}
\end{alignat}
\begin{enumerate} 
\item $C_1$ is the complete contraction of the fully symmetric ACT. In affine differential geometry, its analogue is the Fubini-Pick invariant, where the Fubini-Pick form corresponds to our ACT~\cite{Alekseevsky:2024}.

\item $C_2$ is the contraction of the ACT trace or Chebyshev field.

\item $C_3$ encodes the (non-symmetric) traceless portion of the ACT. Vanishing $C_3$ indicates an additional conformal symmetry of the dual geometry. To see that, we look at the Chebyshev field $\tau^i$. In gravity, the corresponding Weyl field is eliminated locally by a conformal transformation, $g \mapsto e^{-\Phi/n}\cdot g$. This holds for the statistical manifold as well, because the Chebyshev field can be written as the gradient of a scalar potential, $\tau_i =\partial_i \Phi$. The potential $\Phi$ then defines the proper conformal factor $e^{-\Phi/n}$~\cite{Iarley:2015}. As the dual curvature scalars vanish for the classifier, they are unchanged by this transformation. The LC-curvature scalar changes under conformal transformations, with details discussed in App.~\ref{app:infogeo}.

This implies that if $C_3 = 0$ locally, a conformal transformation locally eliminates the skewness in the likelihoods and the lowest-order irremovable deviation from Gaussianity appears at fourth order. Conversely, $C_3 \ne 0$ is a local statement about the model complexity, indicating an irremovable asymmetry. 

\begin{figure}[t]
    \includegraphics[width=0.44\textwidth]{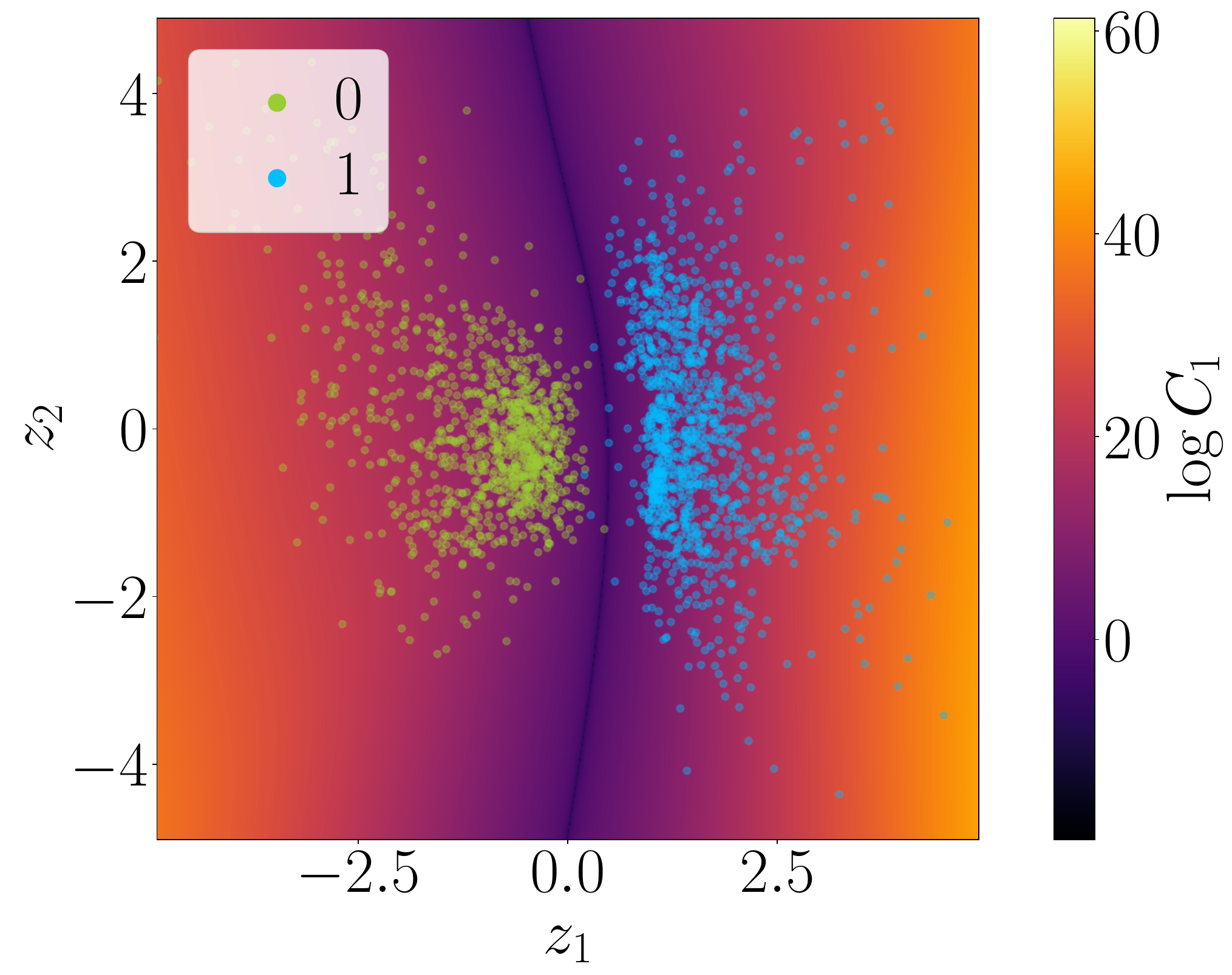}
    \includegraphics[width=0.53\textwidth]{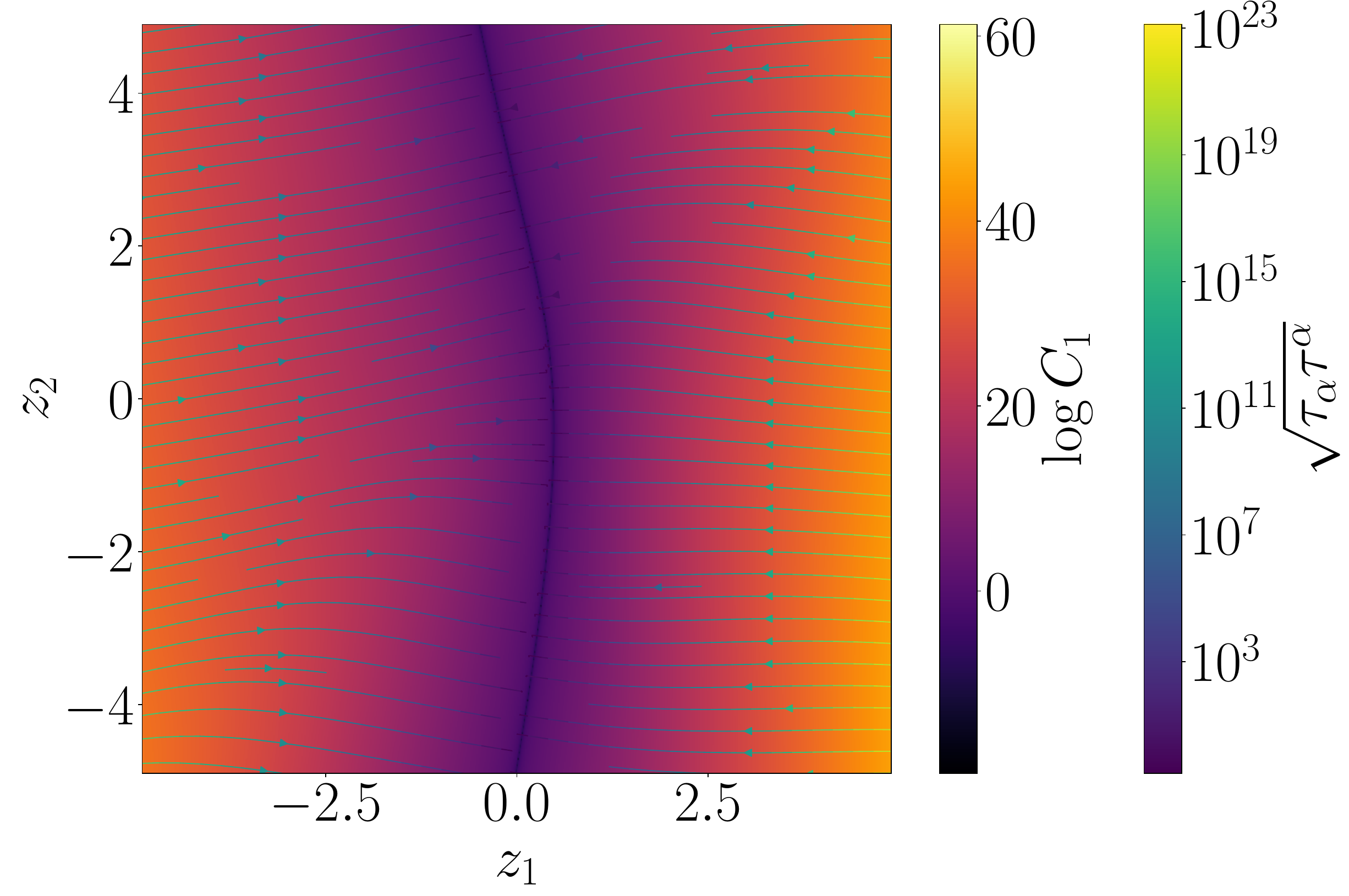}
    \caption{Binary classification: Left: test data and the scalar map $C_1$. Right: flow lines of the Chebyshev field $\tau$ superimposed on $C_1$.}
    \label{fig:mnist_act_twoclass}
\end{figure}

\item $C_4$ is motivated by gravity theories, where a unique nonmetricity scalar $Q$ ensures the equivalence of predictions from alternative formulations to those of General Relativity~\cite{Iarley:2015, Jaerv:2018},
\begin{align}
  Q = -\frac{1}{4}Q_{ijk}Q^{ijk} + \frac{1}{2}Q_{ijk}Q^{kji} + \frac{1}{4}\tensor{Q}{_i^j_j}\tensor{Q}{^{ik}_k} -\frac{1}{2}\tensor{Q}{_i^j_j}\tensor{Q}{_k^{ki}}
  \qquad \text{with} \qquad
  Q_{ijk} = \nabla_i g_{jk}
\end{align}
We compute these contractions for the ACT to obtain $C_4$ in Eq.\eqref{eq:nonmetricity_scalars}.
\end{enumerate}

In Fig.~\ref{fig:mnist_act_twoclass} we find that $C_1$ sharply traces the decision boundary between two classes on the classifier manifold. The same is true for $C_{2,3}$, while $C_4$ vanishes identically. Intuitively, all $C_{1,2,3}$ measure the asymmetry of the likelihoods, in our case evaluated for the symmetric setup of two classes. Note that all Ricci scalars vanish for binary classification and an exponential family of likelihoods. Hence, $C_{1,2,3}$ are the simplest informative, coordinate-invariant scalars for a binary classifier geometry. We also see that for binary classification the flow of the Chebyshev field $\tau$ is orthogonal to the decision boundary, which means that the flow of $\tau$ signifies the quickest path towards a change of the class label.

At the decision boundary, the skewness vanishes, the geometry becomes Riemannian, and likelihoods are most uninformative. Away from the boundary, the classifier encodes its information in a non-metric geometry with additional stretching and shearing of distances. A binary classifier geometry is effectively one-dimensional, $p_1 = 1-p_0$, and can be represented by one line orthogonally crossing the decision boundary. Only the reconstruction depends on both dimensions of the latent space. The degeneracy means that the classifier metric does not have full rank, which obstructs the construction of the inverse metric. For this reason, the pseudoinverse of the metric tensor is used. 

\begin{figure}[b!]
    \includegraphics[width=0.44\textwidth]{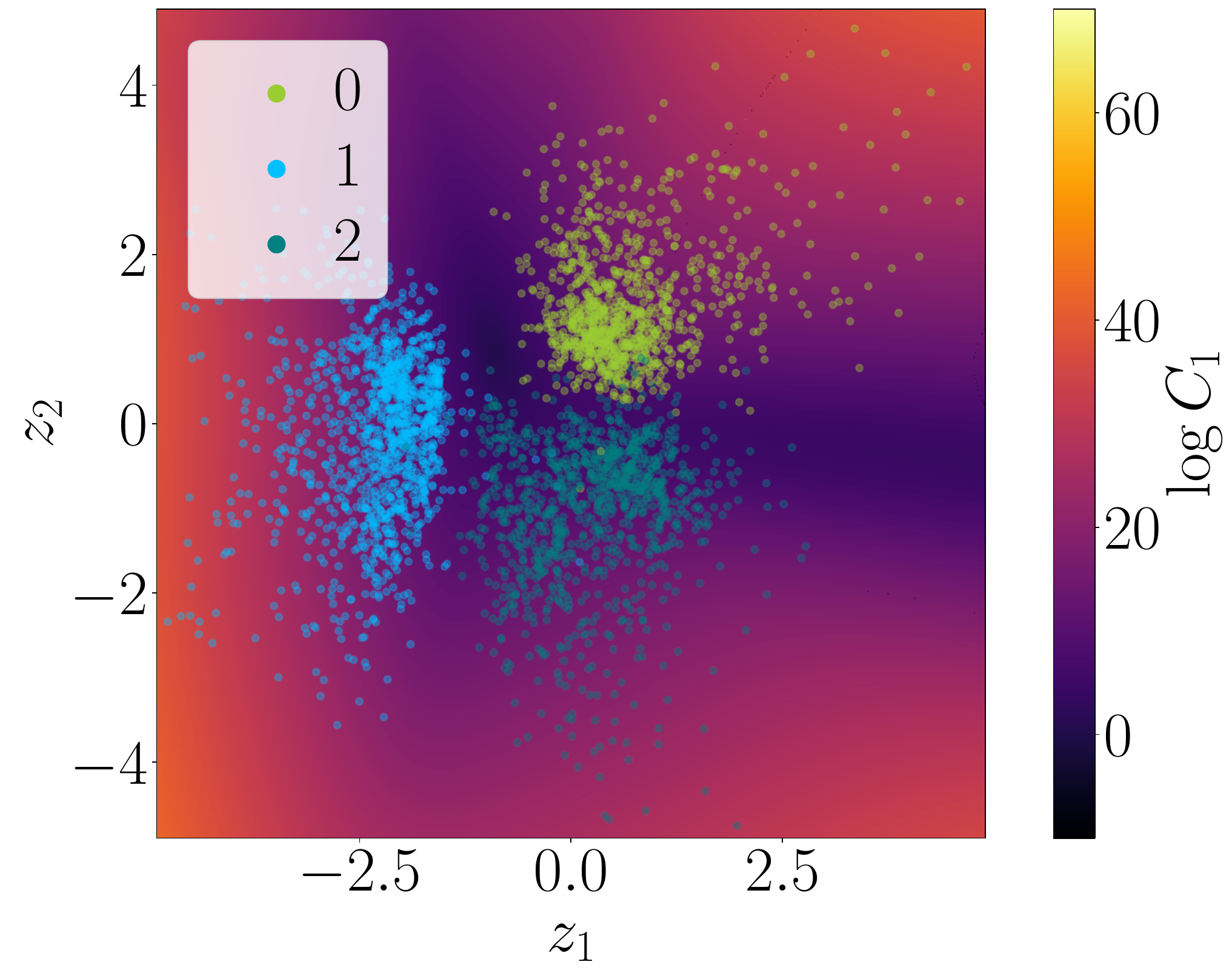}
\includegraphics[width=0.44\textwidth]{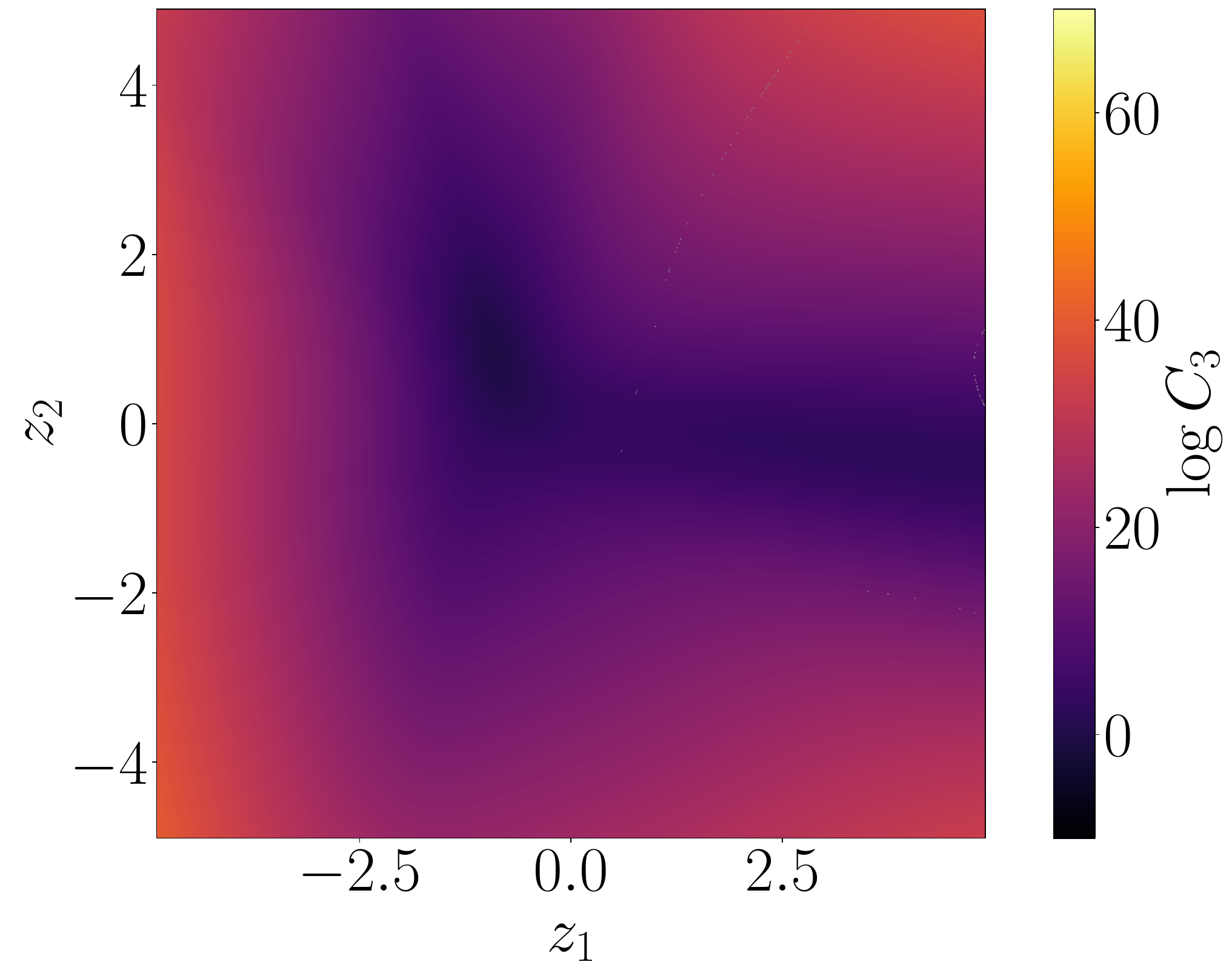}
    \caption{Three-label classification. Left: test data and the scalar map $C_1$. Right: traceless scalar $C_3$.}
    \label{fig:threeclass_off_trace}
\end{figure}

Moving to three-label classification, in Fig.~\ref{fig:threeclass_off_trace} we see that the fully contracted skewness $C_1$ is no longer a sharp tracer of the decision boundary, as it is obscured by an asymmetric pull from the third class. Nevertheless, the decision boundary stays imprinted in $C_1$ and $C_3$, while $C_4=0$ remains. As discussed above, $C_3\neq 0$ geometrically signals higher model complexity. Within the class regions, no conformal rescaling will eliminate the non-Gaussianities and lead to a Riemannian geometry, even to third order only. 

Altogether, we have seen that the classifier encodes invariant information in the scalars $C_{1,2,3}$. Unlike the Frobenius norm, our new scalars are diffeomorphism-invariant and invariant under sufficient statistics transformations of the data. Statistically, $C_{1,2,3}$ trace the decision boundary and are diagnostics of model redundancies.

\subsubsection*{Ricci scalars} 

In differential geometry, curvature scalars are important tools to analyze a manifold. In information geometry, the dual curvature scalars of an exponential family vanish, $R_{(+1)} = 0 =  R_{(-1)}$, and our classifier likelihoods form an exponential family \cite{Udriste:2014,Nielsen:2020}. The only curvature scalar that might not vanish is the LC-Ricci scalar $R_{\text{LC}}$. Here, the relevance of $C_4$ for statistical manifolds becomes clear, as $R_{\text{LC}}$ differs from the dual Ricci scalars by the nonmetricity scalar $C_4$,
\begin{align}
  R_{\text{LC}} = R_{(+1)} + C_4  = R_{(-1)}+ C_4 \equiv C_4
  \qqquad \Leftrightarrow \qqquad 
  R_{(+1)} = 0 =  R_{(-1)} \;,
  \label{eq:ricciscalars}
\end{align}
as shown in App.~\ref{app:infogeo}. We use this identity to compute $R_{\text{LC}}$ without the need for second derivatives of the network. Because we empirically find $C_4=0$ for binary and three-label classification, but $C_{1,2,3} \ne 0$, we conclude that the classifier expresses informative geometry in the nonmetricity tensor ACT, rather than in the curvature. Trivially, for a metric geometry \ie $C_{ijk}=0$, all curvature scalars coincide $R_{(+1)} = R_{\text{LC}} = R_{(-1)}$ and would not vanish in general.  

\subsubsection*{Fisher determinant} 

\begin{figure}[t]
    \includegraphics[width=0.45\textwidth]{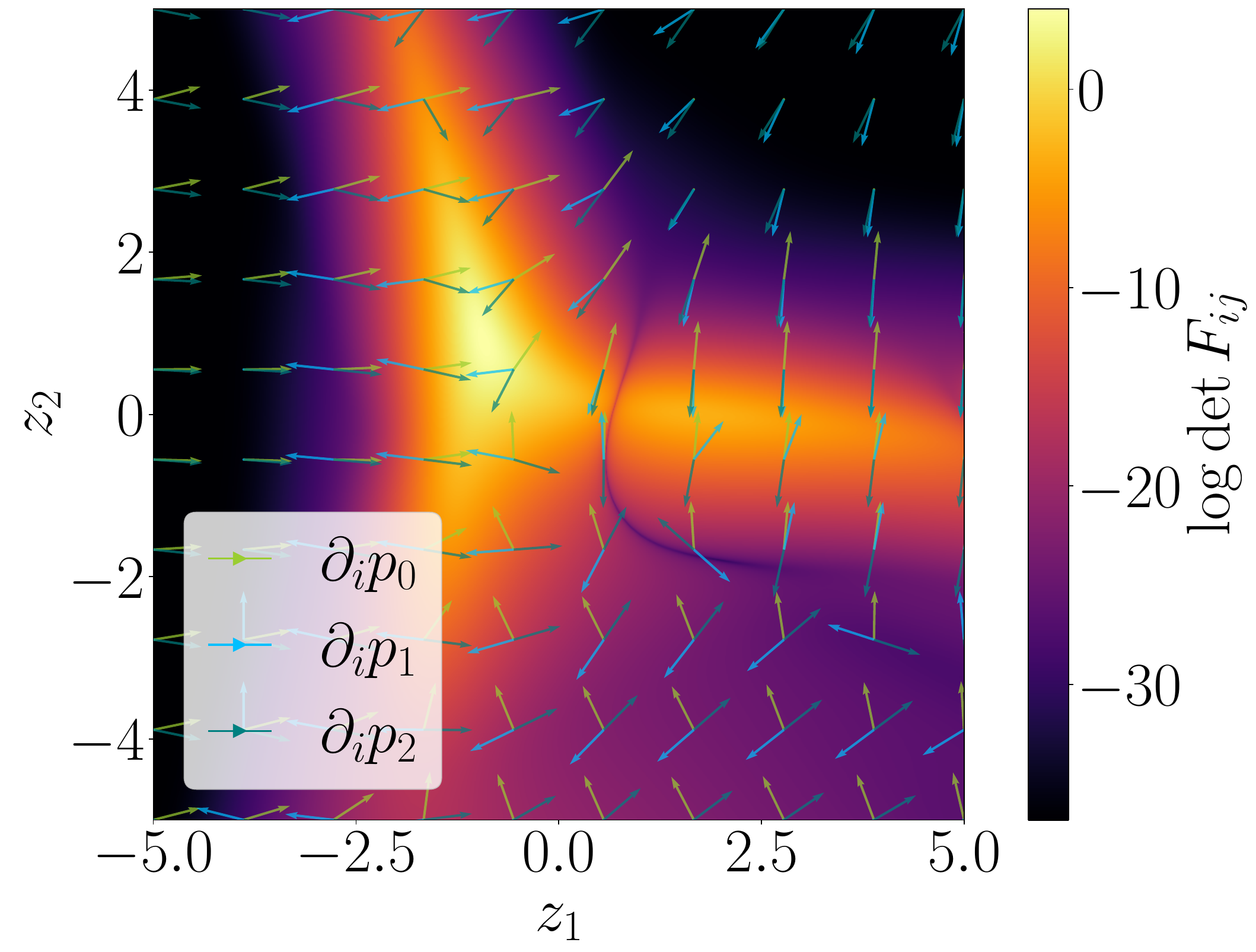}
    \hfill
    \includegraphics[width=0.52\textwidth]{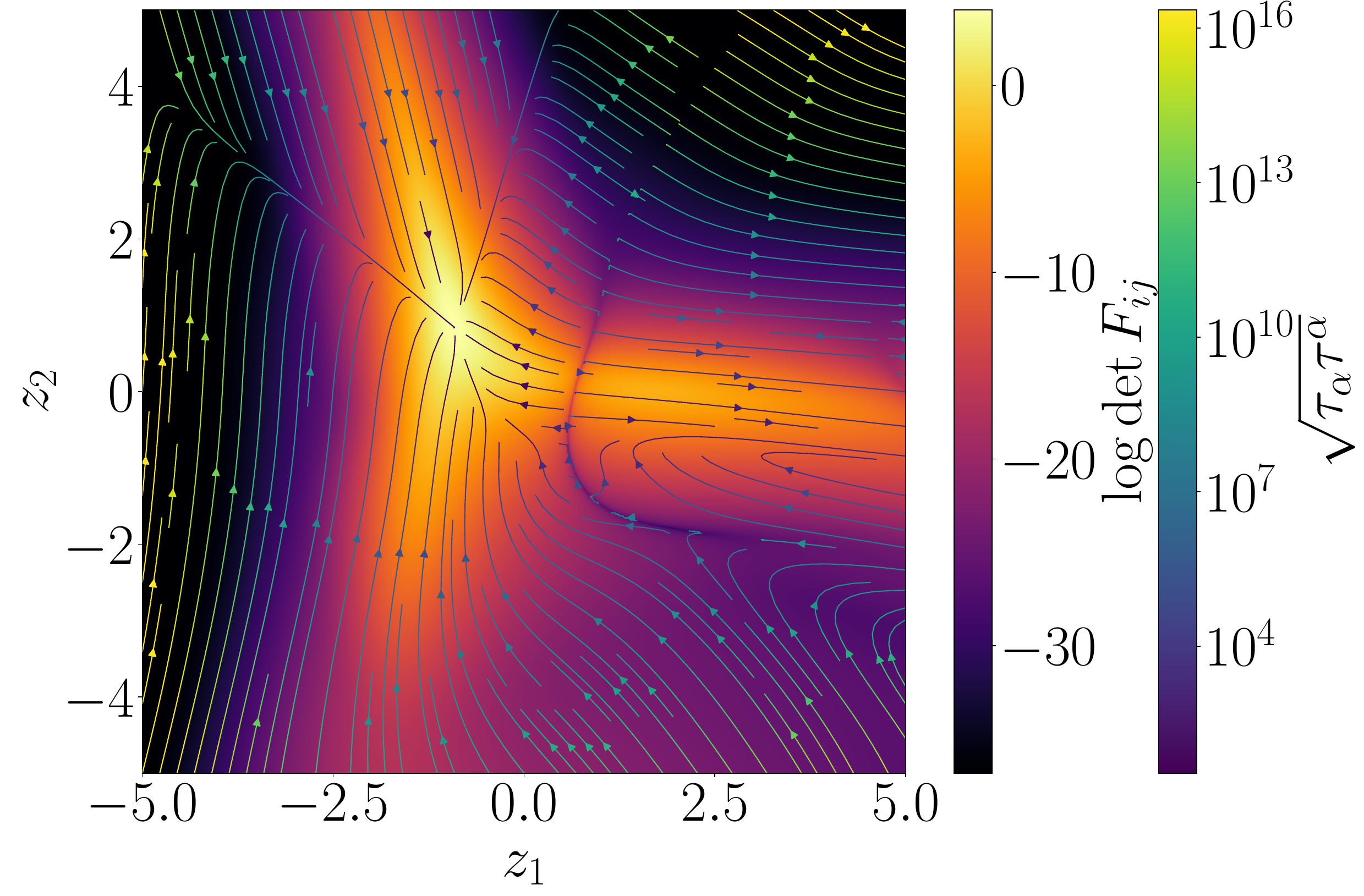}
    \caption{Three-label classification. Left: Fisher determinant map with a degeneracy line superimposed with the column vectors of the Jacobian. Right: Fisher determinant map superimposed with the flow lines of the Chebyshev field.}
    \label{fig:logdet_Chebyshev_jacobian}
\end{figure}

The metric determinant of the Fisher information is useful for analyzing the classifier geometry. On the one hand, $\det F$ vanishes if the metric is singular and its inverse is not defined, signaling a degeneracy in our representation of the geometry. For effectively one-dimensional binary classification, this happens over the entire latent space. 

For three classes, we show in Fig.~\ref{fig:logdet_Chebyshev_jacobian} that $\det F \neq 0$ almost everywhere in the boundary region, confirming that the geometry is well-represented in the latent coordinates. The decision boundary is clearly visible. Across the boundary, we observe a line of infinitesimal width with $\det F =0$, indicating a one-dimensional sub-manifold from an inherent degeneracy. The left panel also shows its origin: the Jacobian used to compute our classifier metric does not have full rank there and the direction in which the probability for being in class 1,2,3 (modulo the sign) changes is the same for all classes. Consequently, statistical models on this line could be fully specified by one parameter only rather than two. 

For the exponential family of classifier likelihoods, the potential $\Phi$ that gives the Chebychev field ($\tau_i = \partial_i \Phi$) is precisely $\Phi = \ln \det F$~\cite{Takeuchi:2005, Zhang:2014, Jiang:2020}. We confirm this using vector field divergences in App.~\ref{app:infogeo}. The right panel of  Fig.~\ref{fig:logdet_Chebyshev_jacobian}, which depicts the Chebyshev field superimposed onto the metric determinant, supports this point. From $\Phi = \ln\det F$, we get the proper conformal factor, which gives a local Gaussianization to third order given that $C_3$ vanishes. Interestingly, this factor is identical to Jeffrey's prior $\sqrt{\det F}$ (see App.~\ref{app:infogeo}). To put it differently, whenever  $C_3=0$ we can conformally rescale the metric tensor with Jeffrey's prior $F \mapsto \sqrt{\det F}\cdot F$ such that our latent geometry locally becomes Riemannian. Figure~\ref{fig:threeclass_off_trace} presents an instance where this is possible: in a limited region on the boundary, $C_3$ vanishes. The likelihoods in such regions only deviate from Gaussians beyond third order at most. Of course, this does not hold on the degenerate line. 

\subsubsection*{Scalars in decoder-classifier geometries} 

When constructing a low-dimensional representation of their information, the classifier and decoder networks encode surplus information in quantities that are accessible to differential geometry. We can interpret these geometric properties and understand their significance for the classifier and the decoder.

The networks express part of their information in statistical distances using the Fisher information metric. In this metric, instances deemed dissimilar appear farther apart. In the classifier geometry, the decision boundary is a region of vast distance between the respective classes, while for the decoder different reconstructions are placed apart. The combination of statistical distances in the classifier geometry with derivatives of the decoder gives the dominant feature for a classification decision. 

To globally connect different regions on the manifold, we define a connection characterized by torsion, curvature, and nonmetricity. In general, the decoder and classifier can store information in these invariant properties of the connection that govern the movement of paths across the manifold. While all connections in information geometry are torsion-free by definition, the networks can generally express information in the form of LC-curvature. In spite of this, we find that the classifier does not --- its scalar curvature always vanishes. To confirm this, we introduce a scalar $C_4$, which coincides with the LC-curvature for our likelihood family and indeed vanishes. 

In light of this, we emphasize the importance of nonmetricity for statistical manifolds, where our networks differ at a fundamental level. Only the classifier can construct a nonmetricity tensor, the ACT, which measures the failure of the connection to preserve inner products. This gives an additional distance stretch or compression. To access this geometric information, we propose several new scalars, in line with nonmetricity scalars used in alternative formulations of General Relativity. The nonmetricity scalars $C_{1,2,3}$ express invariant properties of the classifier, show the decision boundary, and convey information about the model complexity. 

\clearpage
\section{Quark-gluon tagging}
\label{sec:qg}

Modern jet taggers follow a common three-stage structure: $(i)$ a point-cloud feature extractor embeds the constituent information, typically using a graph-based architecture or a transformer; $(ii)$ an aggregation layer reduces these constituent embeddings to a fixed-length jet embedding; $(iii)$ a classifier head maps the jet embedding to class probabilities. The aggregation step compresses the jet information while retaining the information needed for classification. The resulting jet embedding serves as input to a VAE, for which we analyze the statistical geometry of the latent space. We emphasize that the novelty of this analysis is not a new quark–gluon observable, but the use of the new information-geometric scalars from Sec.~\ref{sec:toy_geo} as invariant diagnostics of the learned classifier geometry and its alignment with known QCD radiation patterns.

For binary classification, we focus on quark–gluon classification as the experimentally and theoretically most challenging and interesting LHC application. Related information-geometric approaches to QCD have recently been explored \cite{Assi:2025ibi}. Beyond leading order, the notion of a quark or gluon jet is not infrared and collinear safe and therefore not well defined in perturbative QCD. As the simplest ways to define quark and gluon jets meaningfully, we use a Monte Carlo definition, where the jet label is assigned from the initiating parton in the simulated leading order hard process. We use the standard benchmark quark-gluon dataset generated with \pythia~8.226~\cite{Sjostrand:2014zea,Komiske:2018cqr, komiske_2019_3164691}, using default tunes and parton shower settings for the processes
\begin{align}
q\bar{q} \to \PZ(\to \nu\bar{\nu}) + \Pg 
\qquad \mand \qquad 
q\Pg \to \PZ(\to \nu\bar{\nu}) + (\Pu\Pd\Ps) \eqperiod
\label{eq:def_procs}
\end{align}

Our baseline graph network is ParticleNet Lite. It maps each jet to a 64-dimensional jet-level embedding. Following the established convention, the network learns the quark signal label 1 and the gluon background label 0. We then train a VAE to map the 64-dimensional embeddings to a two-dimensional latent space. This representation maintains competitive classification performance and enables a stable reconstruction of physically interpretable observables~\cite{Vent:2025ddm}. It also allows us to easily visualize the  geometric properties. For the orientation of our latent representation, we break the rotational symmetry by aligning the average latent coordinates of gluon jets and quarks jets with the $z_1$-axis and assign the negative side to gluon jets. 

For the physics feature reconstruction we again employ a VAE, parametrized by a normal distribution with a learned local mean, and a learned global variance. Allowing local variance leads to overfitting, illustrated through the emergence of geometric structure, which differ in every re-training. This setup gives us access to interpretable features, while still utilizing the generality of the input for the classification task.

\subsubsection*{From likelihood ratio to decision boundary}

\begin{figure}[t]
    \includegraphics[width=0.49\textwidth]{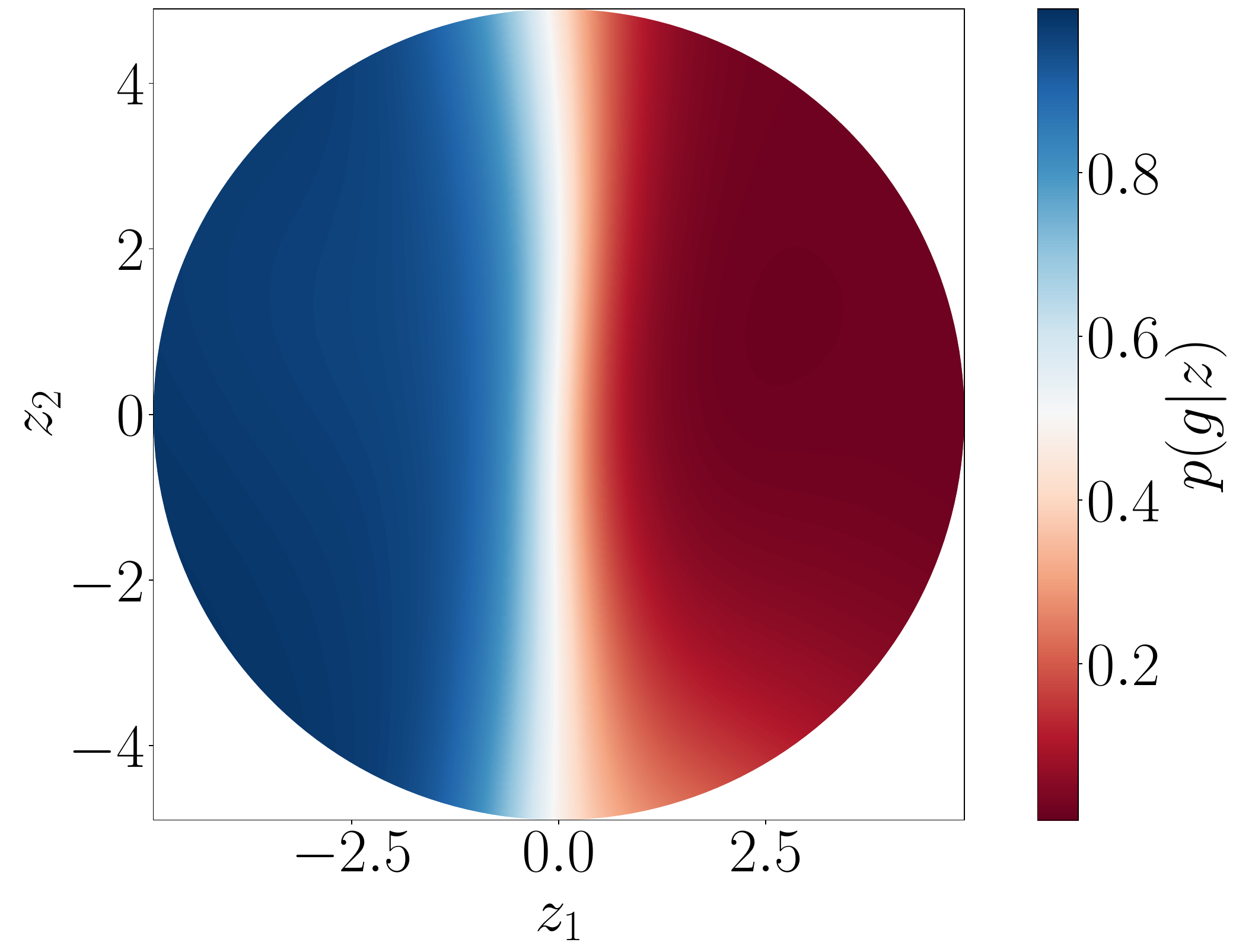}
    \includegraphics[width=0.488\textwidth]{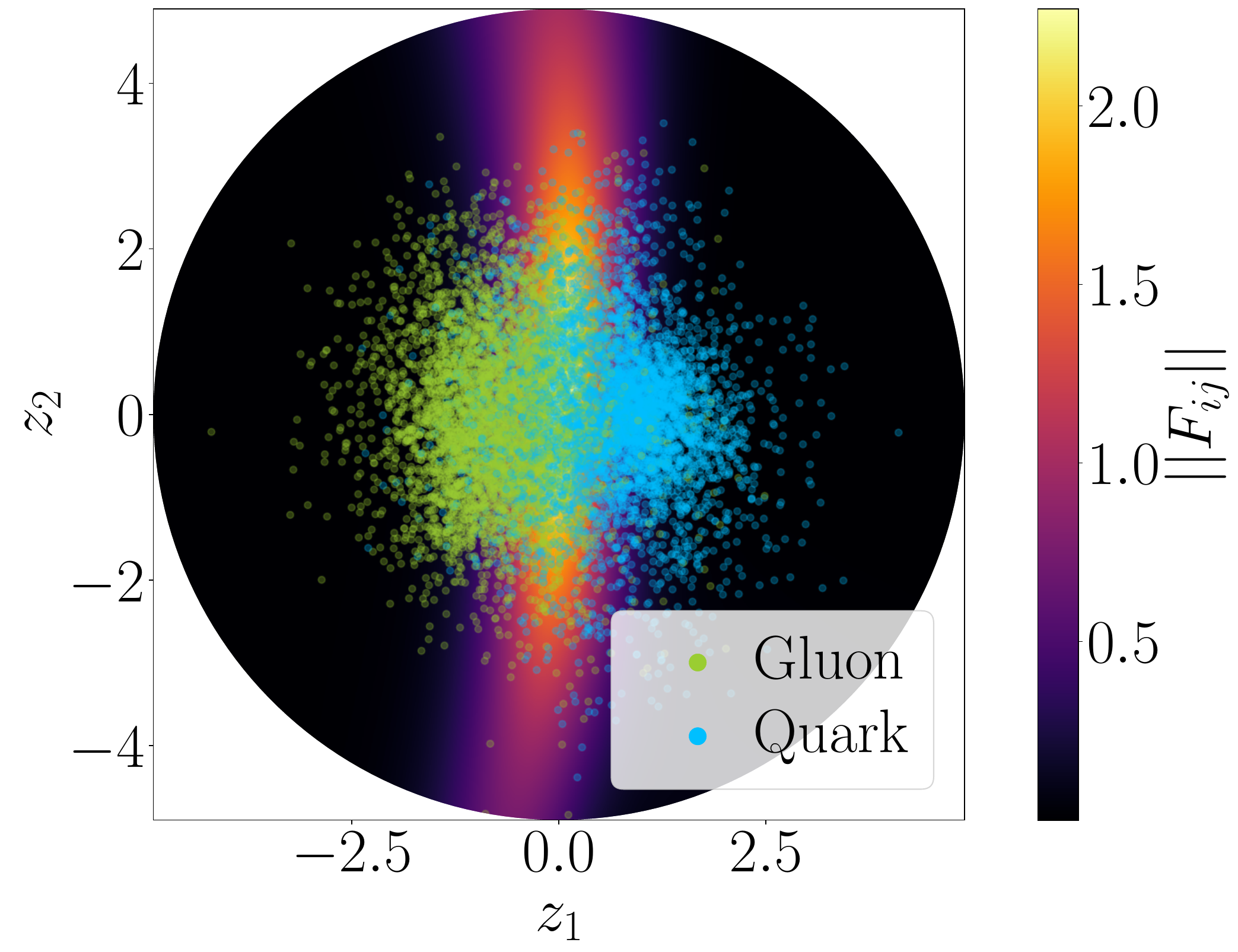} 
    \includegraphics[width=0.505\textwidth]{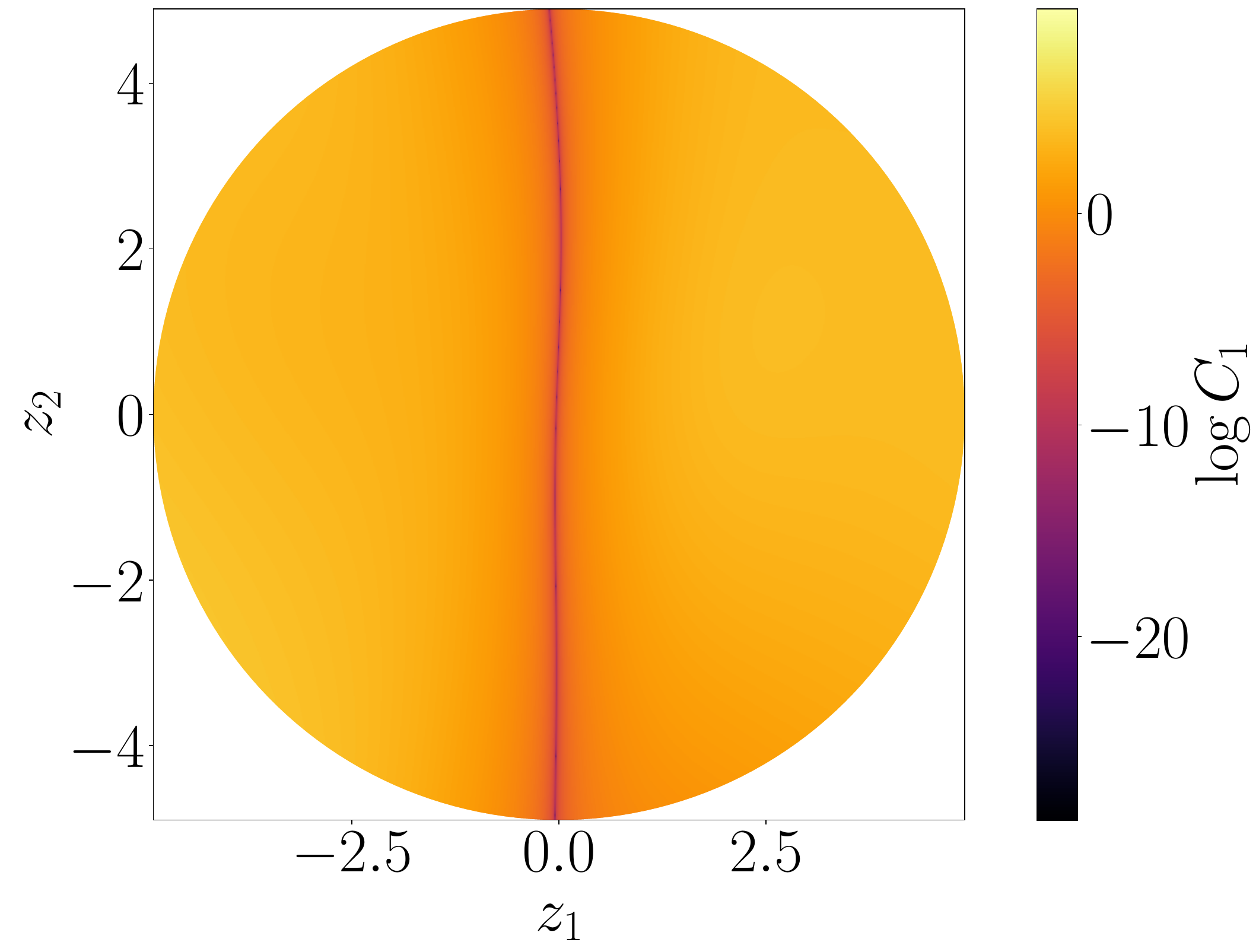}   
    \includegraphics[width=0.49\textwidth]{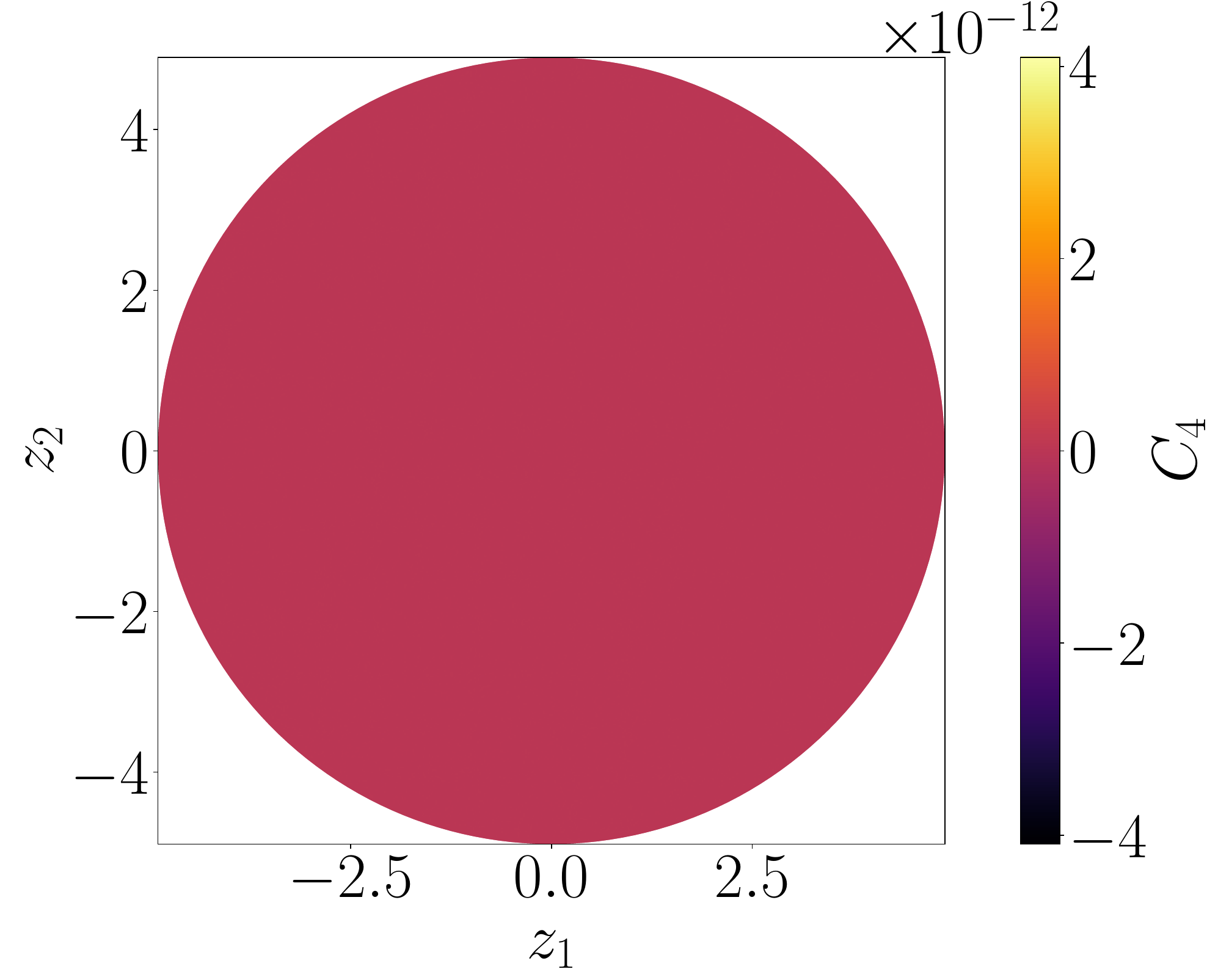} 
    \caption{Decision boundary of the quark-gluon classifier. Upper left: empirical evaluation of the classifier probability, Upper right: Frobenius norm. Lower left: ACT scalar $C_1$, Lower right: ACT scalar $C_4$.} 
    \label{fig:act_decision_tracer}
\end{figure}

Given the large datasets available in LHC physics, we usually follow the Neyman–Pearson lemma and think of jet classification in terms of a continuous likelihood ratio. A threshold on the class probabilities, for example $p= 1/2$, defines the decision boundary shown in Fig.~\ref{fig:act_decision_tracer}. At this boundary, the two hypotheses contribute equally, and the local statistical geometry is approximately symmetric. 

We then compare the gluon probability map to the Frobenius norm and the scalar $C_1$ defined in Eq.\eqref{eq:nonmetricity_scalars}. All three quantities trace the decision boundary, where the classifier output is most sensitive to small latent perturbations. While the Frobenius norm peaks where the classifier output varies most rapidly, $C_1$ tracks the imbalance between the two hypotheses, so it is minimal near the decision boundary and increases toward the interiors of the latent quark and gluon regions. In that sense, $C_1$ resolves class-dominant regions directly and the decision boundary only indirectly. Finally, $C_4$ is flat across the latent space, indicating that information is encoded in nonmetricity and not in curvature.

\begin{figure}[t]
    \centering
  \includegraphics[width=0.49\textwidth]{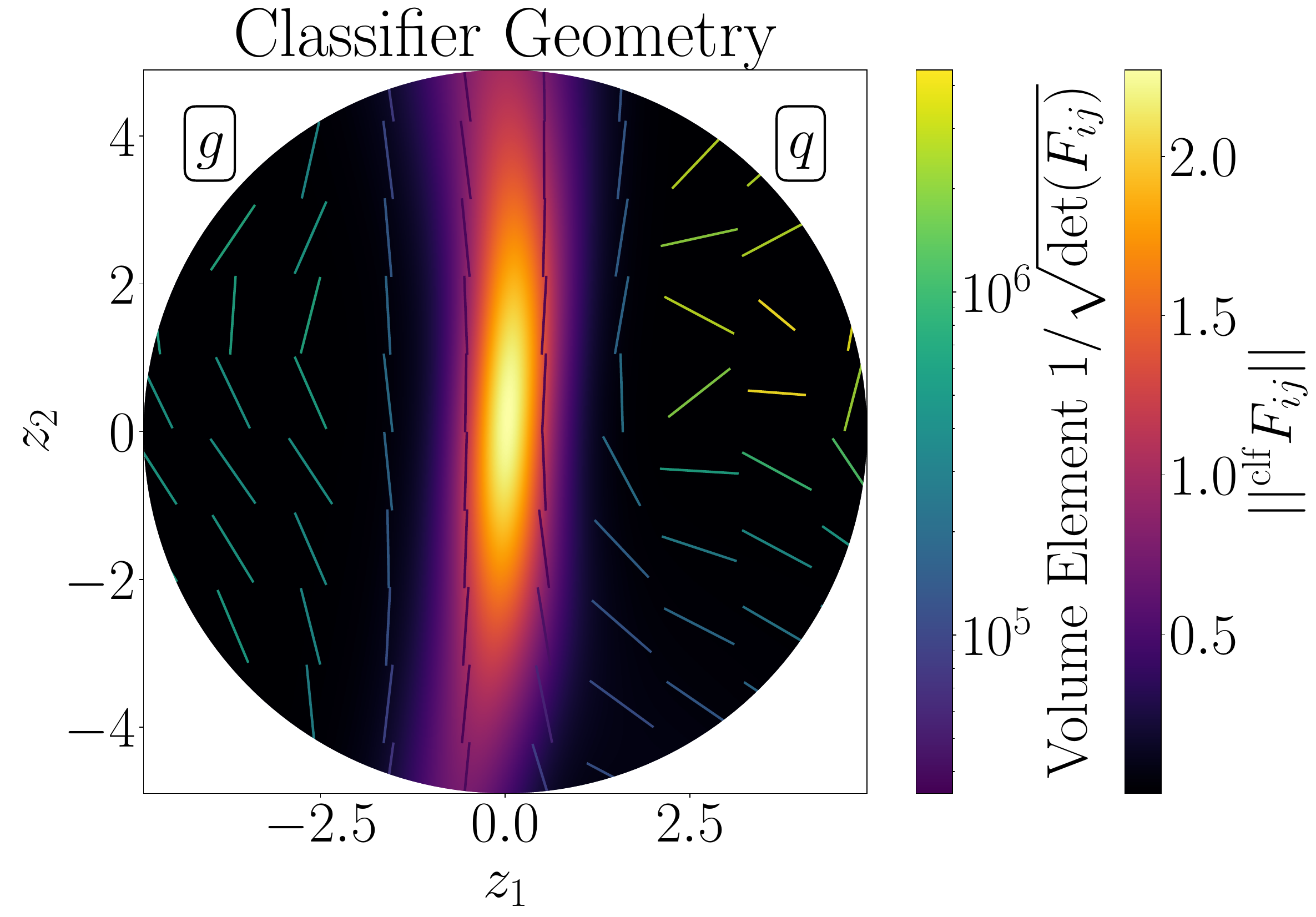}
  \hfill
  \includegraphics[width=0.49\textwidth]{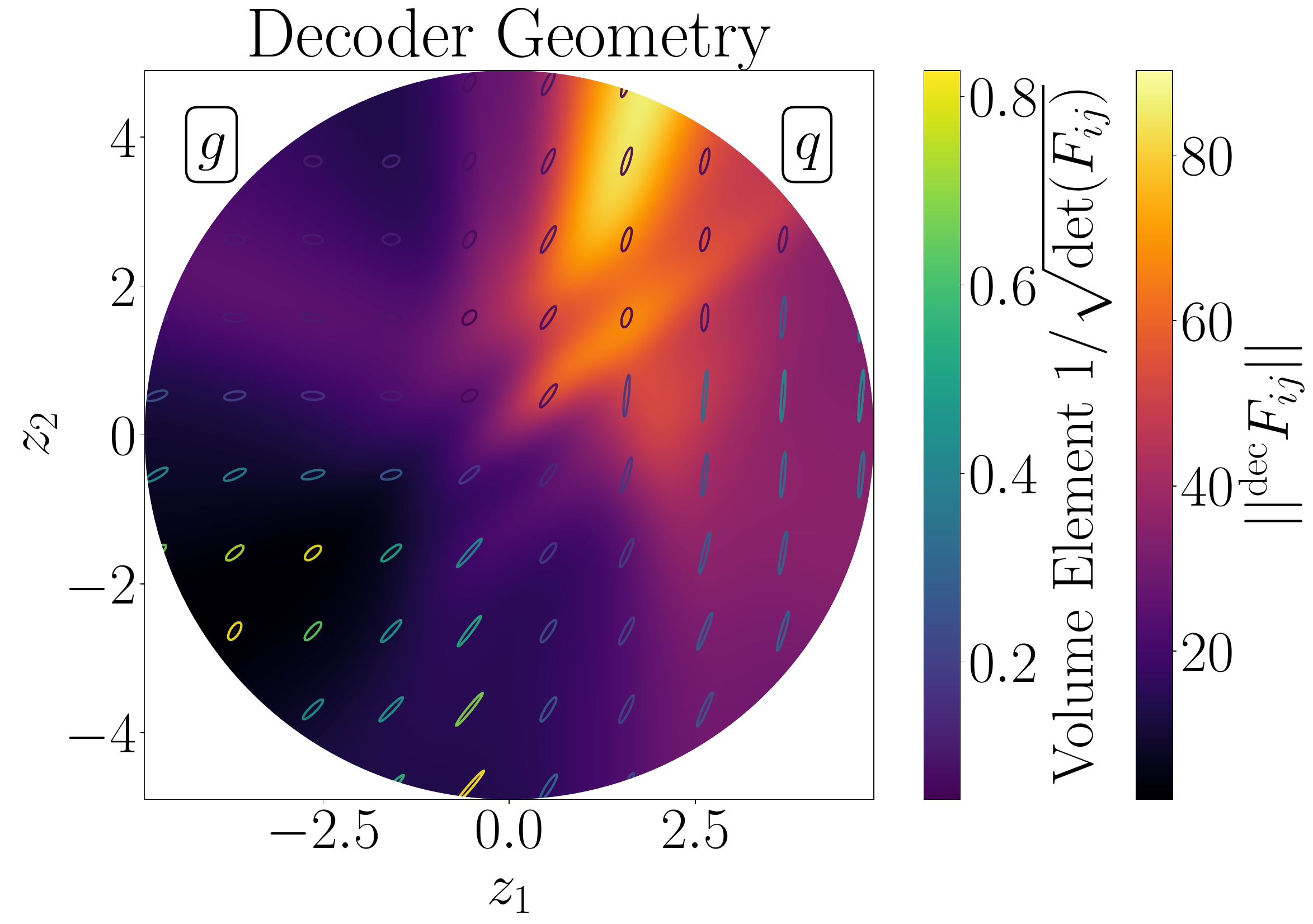}  
  \includegraphics[width=0.45\textwidth]{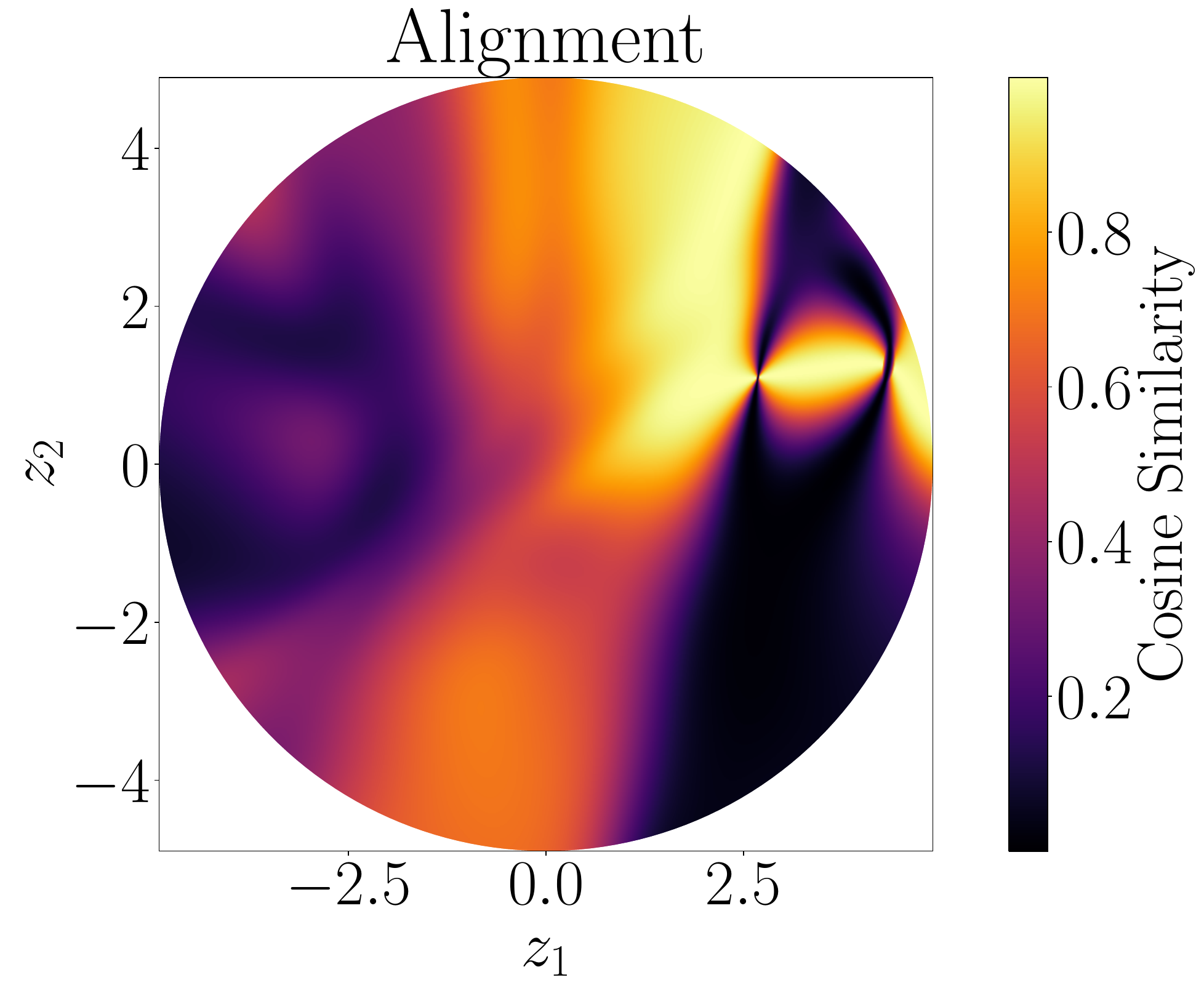}
  \caption{Left: Classifier geometry with the Frobenius norm and ellipses. Center: Cosine similarity as alignment measure between the classifier and decoder geometry. Right: Decoder geometry with the Frobenius norm and ellipses.} 
  \label{fig:qg_alignment}
\end{figure}

Next, we show the Fisher ellipses for the classifier and decoder metrics in Fig.~\ref{fig:qg_alignment}. As for the toy problem, the classification ellipses are elongated and parallel to the decision boundary, reflecting the one-dimensional classification structure. Large eigenvalues in the Fisher metric correspond to minor semi-axis. The decoder geometry is actually two-dimensional and has a non-trivial structure. In the lower panel of Fig.~\ref{fig:qg_alignment} the large cosine similarity shows that the decoder and classifier derivatives align over large latent space regions, implying that the network has identified meaningful features for the classification that coincide with the reconstructed physics observables~\cite{Vent:2025ddm}.

\subsubsection*{Feature reconstruction}

Given the information geometry results on the latent structure of quark-gluon tagging, we need to build a relation to observable physics features~\cite{Vent:2025ddm}, 
\begin{align}
    \Big\{ \; n_\text{PF},\, w_\text{PF}, \, p_T D,\, C_{0.2}\,,\, \epsilon,\, m_\text{jet},\,x_\text{max}, \, S_\text{frag} \; \Big\}.
\end{align}
Here, $n_\text{PF}$ measures particle-flow object multiplicity (radiation activity),  $w_\text{PF}$ ~\cite{Gallicchio:2010dq,Gallicchio:2011xq} the radial energy spread, $p_TD$~\cite{CMS:2012rth} the transverse-momentum dispersion, $C_{0.2}$ a two-point energy correlator~\cite{Larkoski:2013eya}  probing the angular structure,  $\epsilon$ the jet shape anisotropy or ellipticity~\cite{Brandt:1964sa}, $m_\text{jet}$ the jet mass, $x_\text{max}$~\cite{Pumplin:1991kc} the leading momentum fraction, and $S_\text{frag}$~\cite{Neill:2018uqw} the fragmentation entropy. Energy Flow Polynomials~\cite{Komiske:2017aww} (EFPs) form a complete basis of IRC-safe observables, but are a challenge to interpretability, so we show the corresponding results in App.~\ref{EFP_appendix}. 

In perturbative QCD, the probability that a parton of type $i$ emits no resolved radiation between scales $Q$ and $Q_0$ is given by the Sudakov form factor~\cite{Plehn:2015dqa}
\begin{align}
\Delta_i(Q,Q_0)= \exp\!\left[- \int_{Q_0}^{Q}\frac{\dd k}{k}\int \dd z\,\frac{\alpha_s(k)}{\pi}\,C_i\,P_{i}(z)\right] \; ,
\end{align}
where $P_i$ denotes the DGLAP splitting kernels with the color factors $C_F = 4/3$ for quarks and $C_A = 3$ for gluons. The latter radiate more strongly, their Sudakov no-emission probability is smaller, and jets characterized by little resolved radiation are typically quark-initiated.

Another difference in the QCD evolution arises because a quark can radiate collinear and soft gluons, while a gluon can radiate another gluon or split into a $q \bar q$ pair. As gluons do not preserve their identity, there will be purely quark-like regions associated with suppressed radiation, but no purely gluon-like regions.

\begin{figure}[t]
    \centering
    \includegraphics[width=\linewidth]{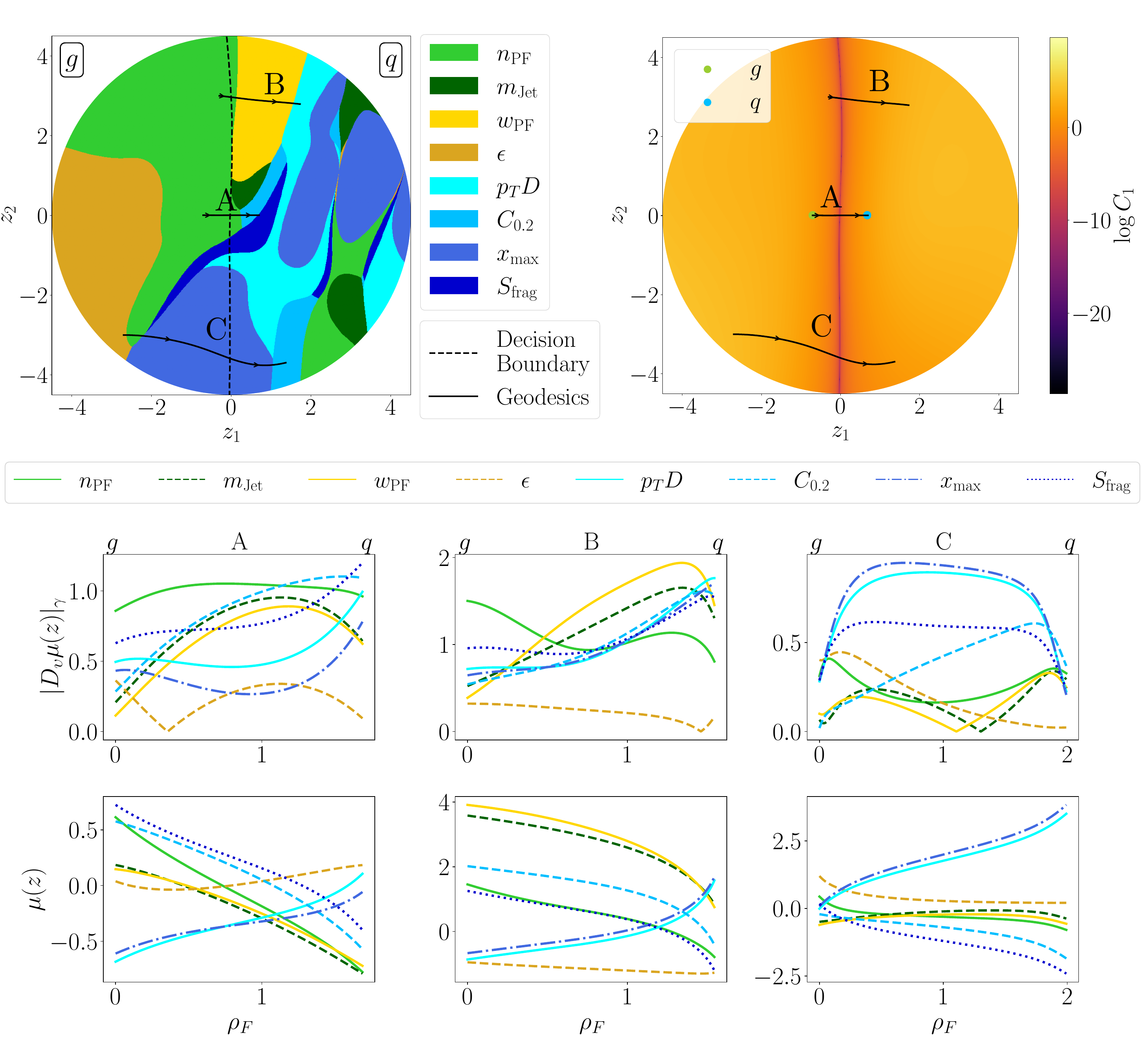}
    \caption{Quark-gluon classification, three exemplary geodesics (A,B,C), dominant derivatives and feature value in Fisher-direction reconstructed along each geodesic.}
    \label{fig:gq_geodesics_derivatives}
\end{figure}

The Fisher directional derivative in Eq.\eqref{eq:directional_derivative} allows us to determine where the classifier output is sensitive to a given feature. In 
the left panel of Fig.~\ref{fig:gq_geodesics_derivatives} we show which feature is most effective in changing the classifier output in the direction of the other class, \eg turning a specific quark jet into a gluon jet. Clusters of similar jets in the latent representation share effective features and lead to regions with one dominating feature. The dominant observable separating quark and gluon jets is the multiplicity. Beyond that, we see that the latent space is split into radiation spread (yellow) and fragmentation (blue) regimes.

From the above physics argument we expect that moving from the gluon region towards the quark region is primarily associated with constituent multiplicity and radiation activity. In Fig.~\ref{fig:gq_geodesics_derivatives} we follow this kind of pattern through classifier geodesics from the gluon region to the quark region. Geodesic A connects an average gluon with an average quark, where the average jet is determined by their latent space coordinates. For the geodesics B and C we probe different dominant directional derivatives regions. 

In the sub-panels of Fig.~\ref{fig:gq_geodesics_derivatives} we show the directional derivative and the value of the normalized feature. Along the stable geodesic A the constituent multiplicity evolves approximately linearly with the Fisher–Rao distance, indicated by the nearly constant derivative. Several other features display a similarly smooth behavior. This is consistent with the expected radiation hierarchy between quarks and gluons and compatible with approximate Casimir scaling~\cite{Larkoski:2014pca}.

Geodesics B and C probe interesting lower-density regions, where the Fisher derivative is driven by only a few features. Along geodesic B, the dominant contribution aligns with $w_{PF}$, which decreases steadily, corresponding to jets with higher multiplicity and a broader radiation pattern. Geodesic C explores a rare regime, where the classifier relies on hard fragmentation patterns.

\clearpage
\section{Three-class jet tagging}
\label{sec:top}

After establishing the geometric structure for binary quark--gluon tagging, we extend the analysis to three classes~\cite{qu_2022_6619768}:
\begin{enumerate} 
\item quark and gluon jets;
\item $Z \to q \bar{q}$ jets;
\item $t \to b q \bar{q}'$ jets. 
\end{enumerate} 
For the network we use the full ParticleNet, pre-trained on the JetClass dataset, and again extract the 256-dimensional latent space for the three classes. As for the quark-gluon classification, these embeddings serve as input to a VAE with a two-dimensional latent space. 

To illustrate the relation of the latent geometry to meaningful observables, we focus on $N$-subjettiness ratios and the jet mass~\cite{Stewart:2010tn,Thaler:2010tr},
\begin{align}
\tau_N^{(\beta)} =
\min_{\{\hat{n}_k\}}
\frac{1}{d_0}
\sum_{i \in \text{jet}} p_{T,i}
\min_{k=1,\dots,N}
\left( \Delta R_{ik} \right)^{\beta} ,
\end{align}
where $p_{T,i}$ denotes the transverse momentum of constituent $i$, and $\Delta R_{ik}$ is the angular distance between the constituent $i$ and the axis $\hat{n}_k$ and $d_0$ a normalization factor. These axes are chosen to minimize the expression. The exponent $\beta=1$ describes broadening, while $\beta=2$ emphasizes collinear radiation. $N$-subjettiness also has a geometric interpretation where it measures the optimal transport cost of re-arranging a jet energy distribution into an $N$-particle configuration~\cite{Komiske:2020qhg}. In practice, we benefit most from the ratios
\begin{align}
\tau_{21} = \frac{\tau_2}{\tau_1}
\qqquad \text{and} \qqquad 
\tau_{32} = \frac{\tau_3}{\tau_2} \;
\end{align}
which describe the shift from one ($q/g$) to two ($Z$) to three ($t$) prongs or decay products.

\subsubsection*{Decision boundaries}

\begin{figure}[b!]
    \includegraphics[width=0.51\textwidth]{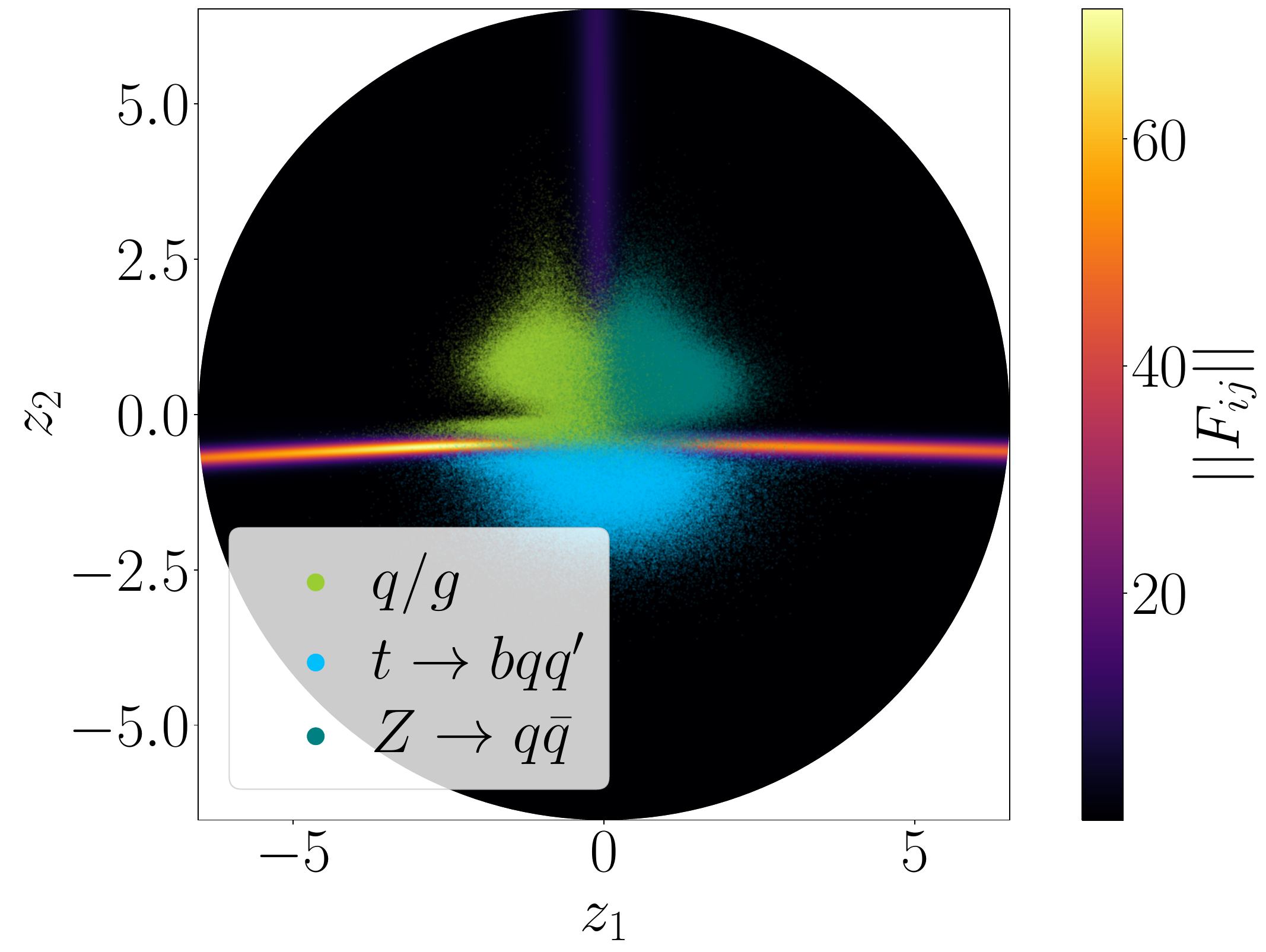}
    \hfill
    \includegraphics[width=0.48\textwidth]{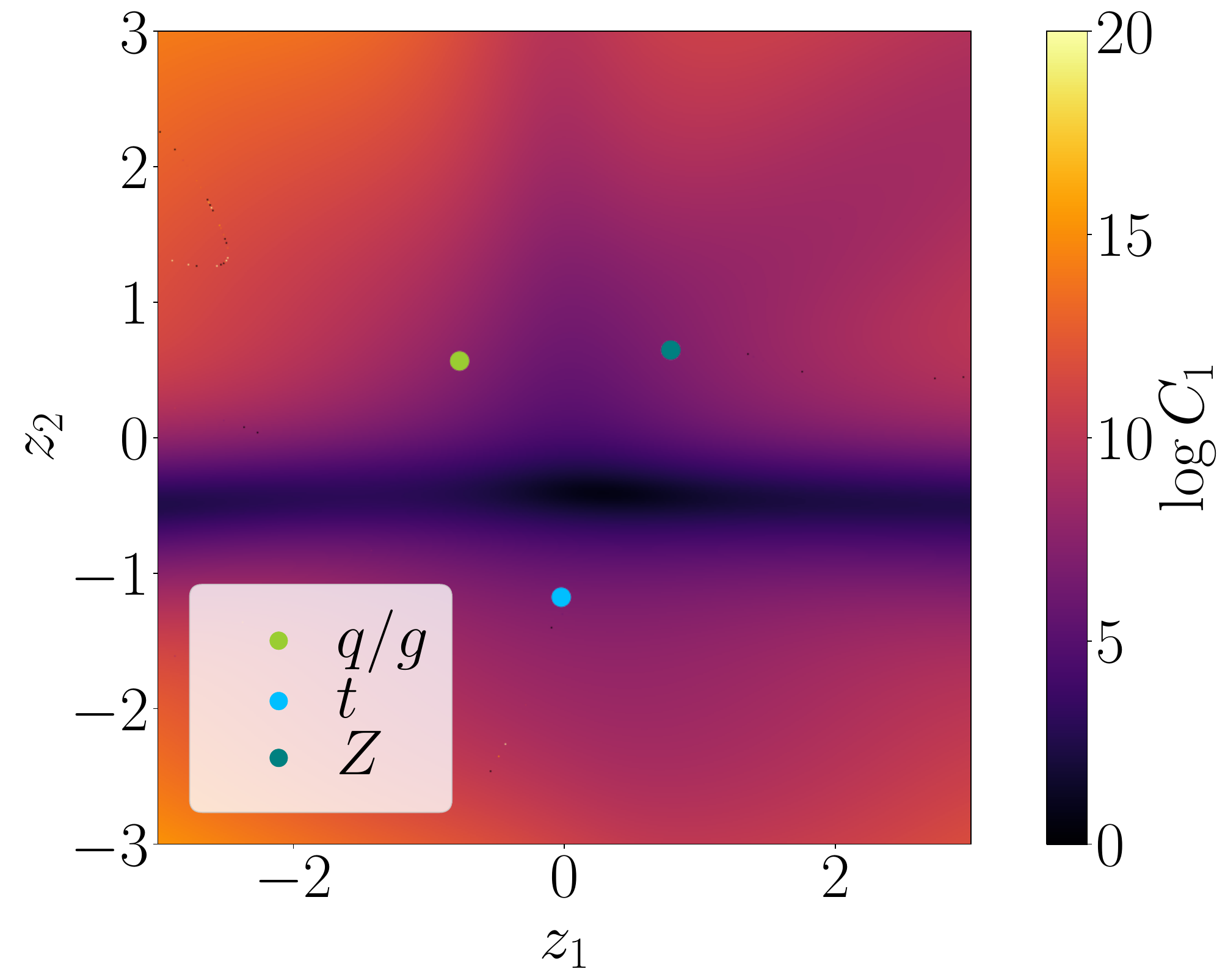}
    \caption{Left: Frobenius norm and test data. Right: fully contracted ACT and average jets.}
    \label{fig:jet_class_standard}
\end{figure}

In Fig.~\ref{fig:jet_class_standard} we show the Frobenius norm and the scalar $C_1$ for the three-label classification. While the Euclidean latent embedding looks different for every training, for example the angle between the decision boundaries changes, we have confirmed that the learned geometry remains the same. The Frobenius norm captures distances in the learned latent metric and shows that small latent shifts close to the decision boundary have a large effect on the classification outcome. The top jets are separated from the $Z$ and $q/g$ jets by a much larger Frobenius norm than the $Z$ and $q/g$ jets from each other. In the right panels we see that the skewness is not at all symmetric. The strong pull from the top jets leads to $\log C_1 \sim 0$ at the decision boundary, while between the other two classes we find small but finite $\log C_1$.

\begin{figure}[t]
  \centering
  \includegraphics[height=5cm]{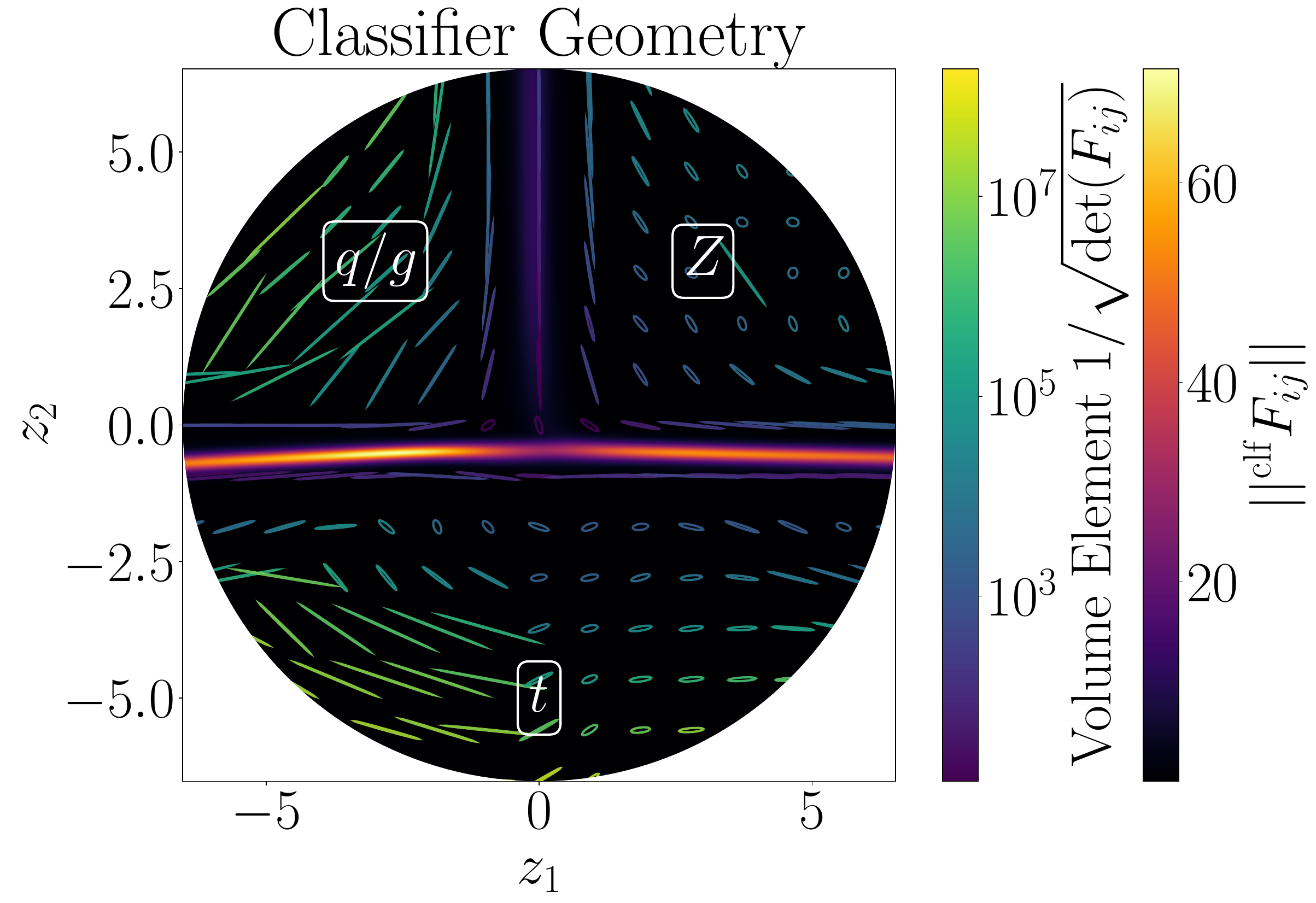}
  \includegraphics[height=5cm]{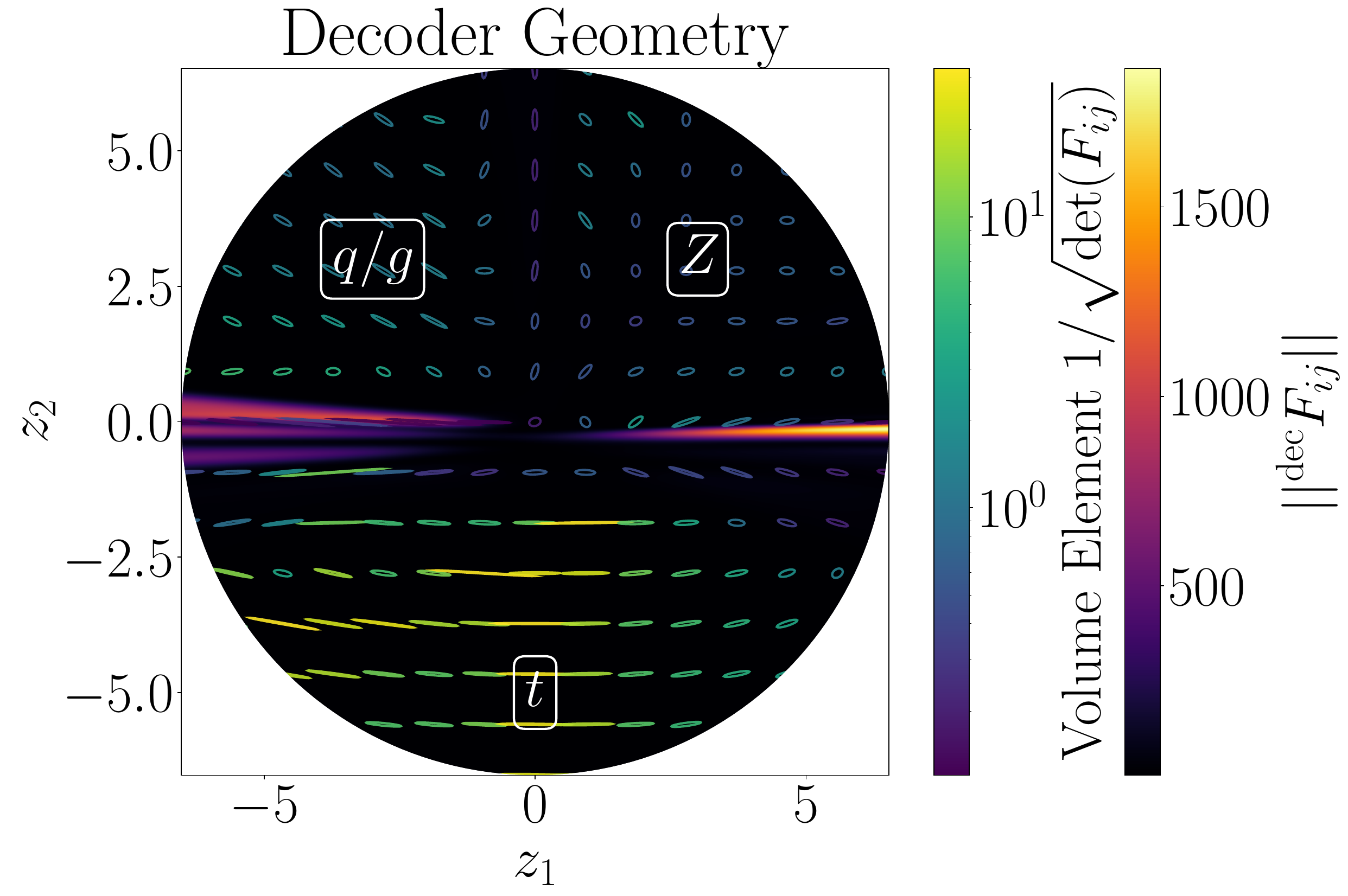}  
  \includegraphics[width=0.45\textwidth]{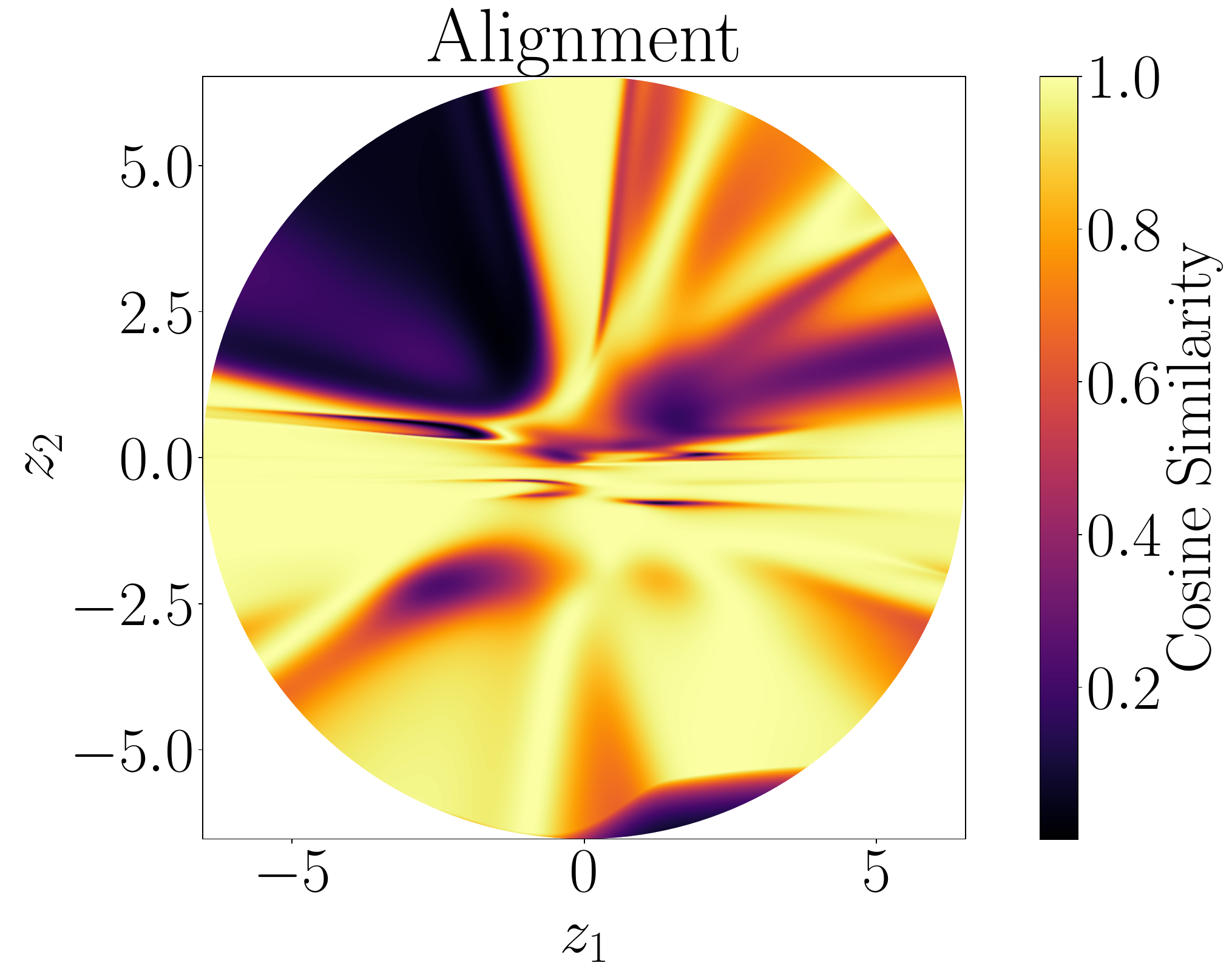}
  \caption{Left: Classifier geometry with the Frobenius norm and ellipses. Center: Cosine similarity as alignment measure between the classifier and decoder geometry. Right: Decoder geometry with the Frobenius norm and ellipses.} 
  \label{fig:jet_class_alignment}
\end{figure}

Figure~\ref{fig:jet_class_alignment} shows the Fisher ellipses for the classifier and decoder geometries. In the $q/g$ region, the classifier ellipses are strongly elongated, which means that one eigenvalue of the Fisher metric dominates locally and the classification is effectively one-dimensional. In the top vs $Z$ region sizeable minor axis indicate more than one relevant feature. The decoder geometry shows sizeable minor axes in the $q/g$ and $Z$ regions, while in the top the ellipses are extremely elongated. The bottom panel shows the alignment between the two learned metrics. It is very large along all decision boundaries and fairly large in the top and $Z$ regions and, but much smaller in the deep $q/g$ region. The effect of this misalignment on the actual classification is limited, because the misalignment only affects latent regions where the classification outcome is clear and stable.

\subsubsection*{Feature alignment}

\begin{figure}[t]
     \centering
    \includegraphics[width=\textwidth]{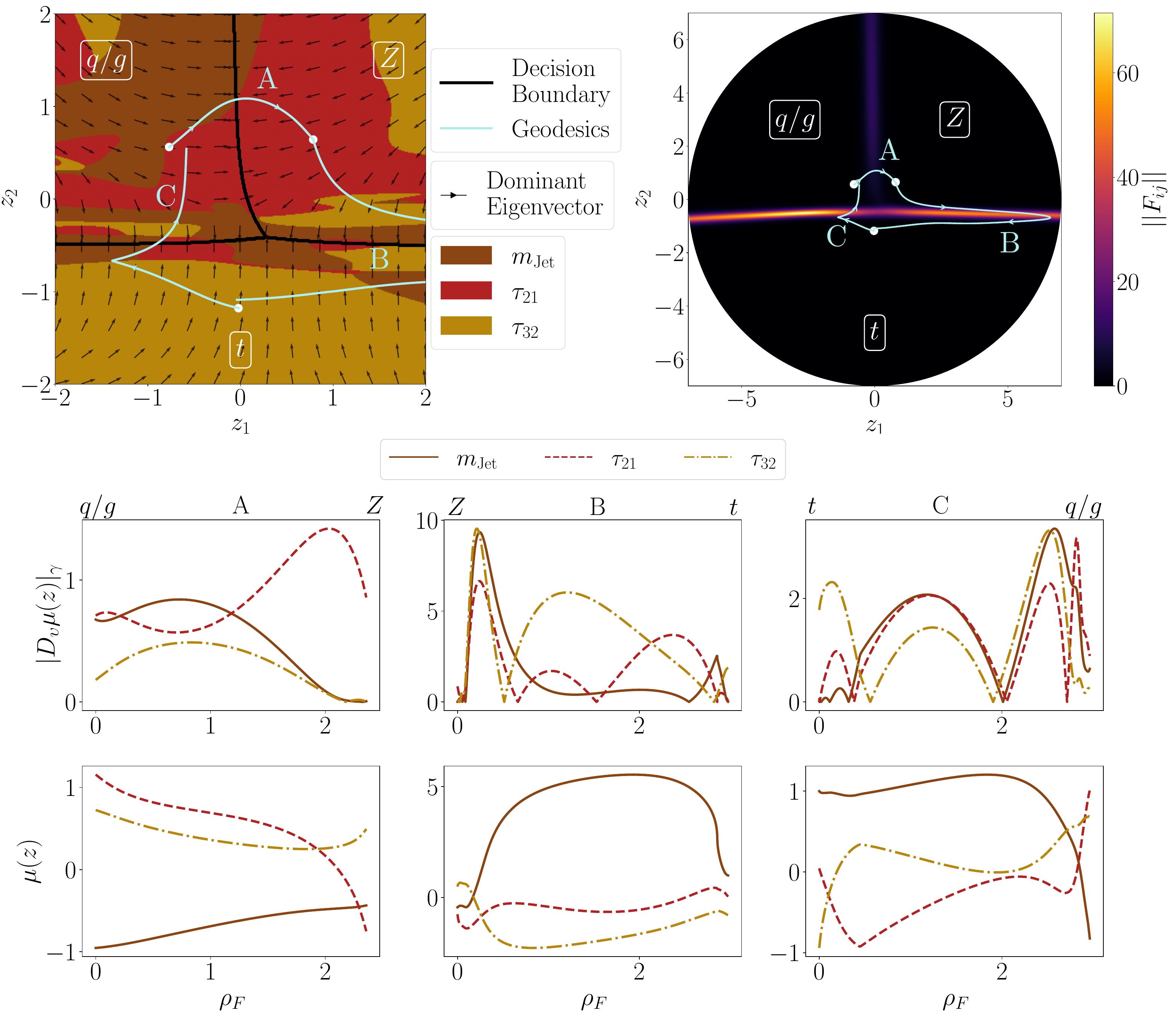}
    \caption{Latent space with average jets and geodesics in the classifier geometry connecting them. Top left: most important feature from the decoder directional derivative relevant for a class change. Top right: Frobenius norm. Bottom: normalized magnitude of the derivatives along the geodesics in the direction of the dominant eigenvector of the classifier and normalized reconstructed features.}
    \label{fig:jetclass_geodesics}
\end{figure}

Given the rich three-label latent structure, we can study which feature is important for a change of the classifier output in the direction of the other classes in Fig.~\ref{fig:jetclass_geodesics}. The region between $q/g$ and $Z$ jets with geodesic~A is dominated by $\tau_{21}$ as a measure of the number of prongs. In addition, the jet mass increases along this path. 

Geodesic B links $Z$ jets to top jets and is dominated by $\tau_{32}$, reflecting the two-step top decay. It goes out of distribution to cross the boundary, leading to potentially unphysical properties and the interesting feature that it is extremely elongated to the right to escape the pull from the $q/g$ region. A similar tendency can be seen for geodesic~A, but the path looks shorter in Euclidean metric because the Frobenius norm is smaller. 

The last geodesic~C shows the transition from a top to a $q/g$ jet. Even though the Frobenius norm is large in this region, the pull from the $Z$ allows the geodesic to cross the boundary passing through intermediate $Z$-like features.  Along the geodesic we start with a steep increase of $\tau_{32}$, indicating the loss of one prong, followed by an increase of $\tau_{21}$. In the $q/g$ region, the jet mass drops abruptly and $\tau_{21}$ rises sharply, reflecting the transition to a light, predominantly one-prong jet.
These strongly changing patterns in the dominant derivatives along geodesic~B and C have physical reasons, also reflected in sharp changes of the direction of the geodesics, where the derivative has to vanish. For instance, they raise the question if a top jet has to first transition to a $Z$-like jet to eventually turn into a $q/g$ jet, which we will discuss in the next section. 

\subsubsection*{Distances between classes}

\begin{figure}[t]
    \includegraphics[width=\textwidth]{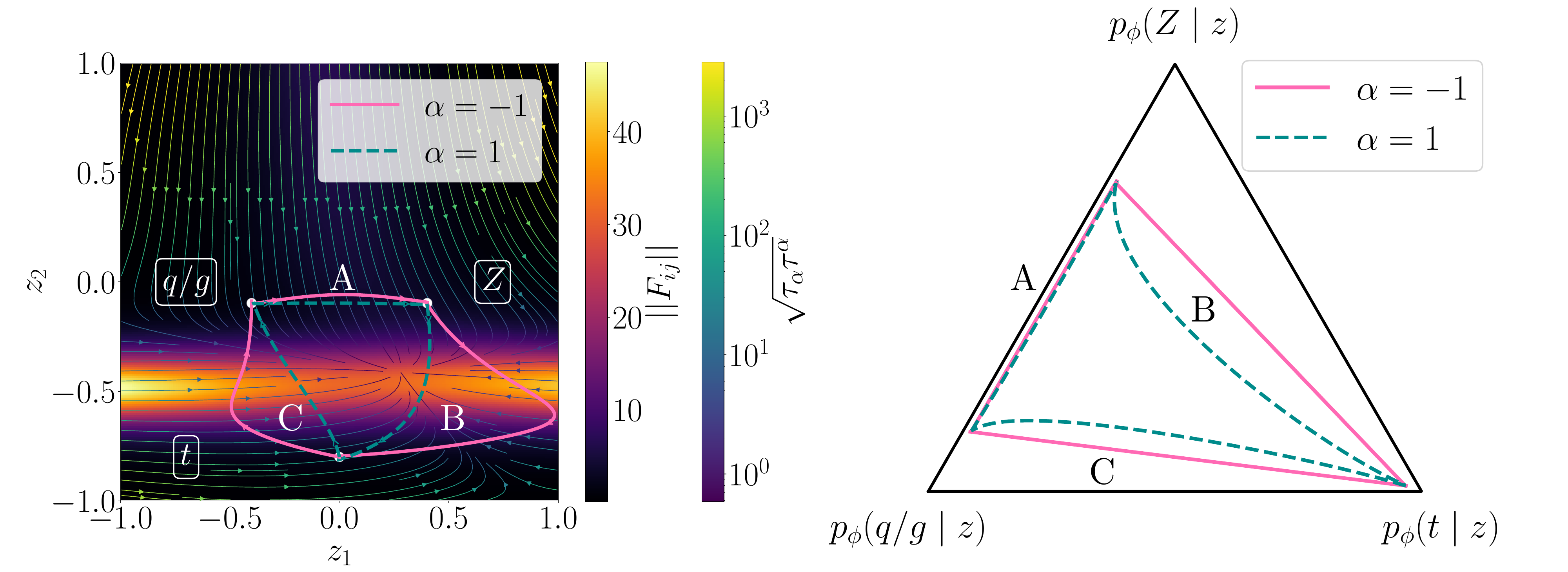}
    \caption{Left: Frobenius Norm and Chebyshev field with color coded norm. Three autoparallels connecting three test jets for the two $\alpha$ connections.  Right: Paths on the simplex along the autoparallels on the left.}
    \label{fig:jet_class_simplex}
\end{figure}

For three jet classes it is interesting to compute the Fisher–Rao distance along the geodesic, Eq.\eqref{eq:fisher_rao_distance}. It allows us to compare how far two jets are separated in the classifier-induced metric. In Tab.~\ref{tab:distances in Wasserstein} we show the Fisher-Rao and Wasserstein distances between the three classes. Both distance measures between top jets and the other two classes are large, which means the classifier can most easily distinguish top jets from other two classes. This is consistent with the Frobenius norm in  Fig.~\ref{fig:jet_class_standard} and with state-of-the-art top taggers achieving near-perfect classification performance. 

\begin{table}[b!]
\centering
\begin{small} \begin{tabular}{ccccc}
\toprule
distance measure&$p_T$ [GeV] & $d_{Z-t}$ & $d_{Z-q/g}$ & $d_{t-q/g}$ \\
\midrule
$\rho_F$ & - &$3.0$&$2.4$ & $3.0$\\
\midrule
\multirow{3}{*}{Wasserstein}&
500--650  & $0.133 \pm 0.002$ & $0.129 \pm 0.001$ & $0.171 \pm 0.003$ \\
&650--800  & $0.123 \pm 0.003$ & $0.114 \pm 0.002$ & $0.148 \pm 0.002$ \\
&800--1000 & $0.119 \pm 0.003$ & $0.101 \pm 0.002$ & $0.129 \pm 0.002$ \\
\bottomrule
\end{tabular} \end{small}
\caption{Fisher-Rao distance along a geodesic and Wasserstein distance between quark/gluon, $Z$, and top jets, computed from 15 repeated random sub-samples of 500 jets per class in each $p_T$ bin. Uncertainties denote the standard deviation across independent sub-samples.}
\label{tab:distances in Wasserstein}
\end{table}

In the last section we arrived at the question if the transition of the top jet to a $q/g$ jet passes $Z$-like features, interpreting the three classes as a sequence
\begin{align}
t \; \rightarrow \; Z \; \rightarrow \; q/g \; .
\label{eq:class_sequence}
\end{align}
Independent of network tasks, the Fisher-Rao and Wasserstein distances between $q/g$ and top jets are indeed large, but the three distance ratios are democratic rather than hierarchical. The Wasserstein distance remains the largest between $q/g$ and top jets across all $p_T$ bins. An interesting questions beyond the scope of this paper would be to derive how the Wasserstein distance and its induced metric differs from the latent classifier metric in view of the Neyman Pearson lemma.

For the three-label classification we can benefit from the dually flat geometry, \ie there are two interpretable coordinates systems in which the autoparallels are straight lines. The (mixture) $(-1)$-autoparallels are linear interpolations in probability space and therefore straight lines in the simplex in Fig.~\ref{fig:jet_class_simplex}, while the (exponential) $(+1)$-autoparallels correspond to log-linear interpolation between probabilities. The Chebyshev vector field pushes the Levi–Civita geodesics to the $(+1)$-autoparallels.

The three reference jets in Fig.~\ref{fig:jet_class_simplex} lie close to the decision boundaries, but represent typical physics features. In the $Z$-$q/g$ region the two  autoparallels coincide, which shows that this sector is largely uninfluenced by the presence of tops in the latent space. Along the straight line, in the latent space, is no region of non-zero probability for a top jet and the paths on the simplex coincide.  In contrast, regions closer to the top jet exhibit a strong dual structure and the autoparallels deviate from each other. The difference between the $(\pm1)$-autoparallels is more pronounced when going from $Z$ jets to top jets, which is in line with the intuition about the number of prongs for the jets. A two-pronged jet should transition to the three-pronged regime without passing through features characteristic of one-pronged jets, whereas a three-pronged jet can evolve toward a one-pronged configuration while still touching the two-pronged region along the way.

\clearpage
\section{Outlook}
\label{sec:outlook}

Locality, similarity, closeness or correlation lengths are the fundamental notions behind learned representations. Viewing closeness as topological allows us to employ information geometry to describe and interpret the learned network representations. While Riemannian geometry is curved, but metric-compatible and torsion-free, our information geometry is dual, but not Levi-Civita flat in general, misses metric-compatibility, but is still torsion-free. This way, low-dimensional latent spaces will drive neural networks to encode information in curvature and in non-metricity.

Based on the Fisher metric and the Amari-Chentsov tensor, and in addition to the Fisher Frobenius norm, we have proposed a set of new scalars to study the geometry of the latent representation. They trace decision boundaries and show how additional information is encoded. Moreover, geodesics and autoparallels allow us to relate continuous changes in features to the latent representation and to study the latent representation in terms of relevant distances.

We have applied this new information geometry methodology to latent models, which learn independent decoder and a classification geometries. First, we have illustrated the main aspects for a simplified version of the MNIST dataset, with controlled features. Next, we have shown how this new methodology allows us to understand the latent structure of binary quark-gluon tagging, expanding beyond our earlier study in Ref.~\cite{Vent:2025ddm}. Finally, we have applied information geometry to three-label classification, separating quark/gluon jets, $Z$-decay jets, and top-decay jets. It allowed us to understand the relation of these three classes and the physics background of the different patterns in the pairwise classification.

\section*{Acknowledgements}

This work is supported by the Deutsche Forschungsgemeinschaft (DFG, German Research Foundation) under grant 396021762 -- TRR~257 \textsl{Particle Physics Phenomenology after the Higgs Discovery}. The authors acknowledge support by the state of Baden-W\"urttemberg through bwHPC and the German Research Foundation (DFG) through grant no INST 39/963-1 FUGG (bwForCluster NEMO). SV is funded by the Carl-Zeiss-Stiftung through the project \textit{Model-Based AI: Physical Models and Deep Learning for Imaging and Cancer Treatment}. BS is supported by the Deutsche Forschungsgemeinschaft (DFG, German Research Foundation) under Germany's Excellence Strategy EXC 2181/1 - 390900948 (the Heidelberg STRUCTURES Excellence Cluster) and by the Vector-Stiftung. RMK acknowledges funding of the Stiftung der Deutschen Wirtschaft (Foundation of German Business) with funds from the Begabtenf\"orderung of the Federal Ministry of Research, Technology and Space (BMFTR).

\clearpage
\appendix

\section{Information geometry details}
\label{app:infogeo}

\subsubsection*{The dual Ricci Scalars and $C_4$}

In a local coordinate frame, the components of the (1,3)-Riemann curvature tensor $\tensor[^{(+1)}]{R}{^i_{jk\ell}}$ for the $(+1)$-connection read~\cite{Nielsen:2020} 
\begin{align}
  \tensor[^{(+1)}]{R}{^i_{jk\ell}}= \partial_k \tensor[^{(+1)}]{\Gamma}{^i_{\ell j}} - \partial_\ell  \tensor[^{(+1)}]{\Gamma}{^i_{kj}} + \tensor[^{(+1)}]{\Gamma}{^i_{k m}} \tensor[^{(+1)}]{\Gamma}{^m_{\ell  j}} - 
  \tensor[^{(+1)}]{\Gamma}{^i_{\ell m}} \tensor[^{(+1)}]{\Gamma}{^m_{kj}}\;,\label{eq: e m Riemann tensor}
\end{align}
with $\tensor[^{(+1)}]{\Gamma}{^i_{jk}}$ the corresponding connection coefficients. The $(-1)$-Riemann tensor follows in complete analogy. 

Notably, due to the failure of the $(\pm1)$-connections to preserve the metric tensor under parallel transport, the dual Riemann tensors do not inherit all symmetries of the LC-Riemann tensor. Inserting the $(\pm 1)$-connection coefficients in Eq.\eqref{eq:kl_connection} gives the dual Riemann tensors in terms of the ACT,  
\begin{align}
      \tensor[^{(\pm 1)}]{R}{^i_{j k\ell }} =  \tensor[^{\text{\text{LC}}}]{R}{^i_{j k\ell }} \pm \frac{1}{2} \left( \tensor[^{\text{LC}}]{\nabla}{_\ell} \tensor{C}{^i_{kj}} - \tensor[^{\text{LC}}]{\nabla}{_k} \tensor{C}{^i_{\ell j}} \right) + \frac{1}{4}\left(\tensor{C}{^i_{km}} \tensor{C}{^m_{\ell j}} -  \tensor{C}{^i_{\ell m}} \tensor{C}{^m_{kj}}\right) \; \label{eq: e m curvature tensor},
\end{align}
with the same index structure in the last term for $\tensor{C}{^i_{jk}}$ as in the Riemann-tensor for $\tensor{\Gamma}{^i_{jk}}$.
The Ricci tensor, which is symmetric for the geometry at hand~\cite{Matsuzoe:2006, Takeuchi:2005}, follows from this as 
\begin{align}
  \tensor[^{(\pm1)}]{R}{_{j \ell }} &= \delta_i ^k \, \tensor[^{(\pm1)}]{R}{^i _{j k \ell}} 
   = \tensor[^{\text{LC}}]{R}{_{j \ell}} \,\pm  \,\frac{1}{2} \left(\tensor[^{\text{LC}}]{\nabla}{_\ell} \tau_j  - \tensor[^{\text{LC}}]{\nabla}{_m}  \tensor{C}{^m _{\ell j}} \right) + \frac{1}{4}\left( \tau_m \tensor{C}{^m_{\ell j }} -  \tensor{C}{^n _{\ell m}} \tensor{C}{^m_{n j}}\right)\; . \label{eq:riccitensor}
\end{align}
We note the structural resemblance of the second summand to the well-known Palatini identity. Full contraction immediately gives
\begin{align}
    R_{(\pm 1)} &= F^{j\ell}\, \tensor[^{(\pm1)}]{R}{_{j \ell }} \notag \\
    &=  F^{j\ell} \,\tensor[^{\text{LC}}]{R}{_{j \ell}} \,\pm  \,\frac{1}{2} \left(\tensor[^{\text{LC}}]{\nabla}{_\ell}  F^{j\ell} \tau_j  - \tensor[^{\text{LC}}]{\nabla}{_m}   F^{j\ell} \tensor{C}{^m _{\ell j}} \right) + \frac{1}{4}  F^{j\ell} \left( \tau_m \tensor{C}{^m_{\ell j }} -  \tensor{C}{^n _{\ell m}} \tensor{C}{^m_{n j}}\right) \notag \\
    &= R_{\text{LC}} \pm \,\frac{1}{2} \left(\tensor[^{\text{LC}}]{\nabla}{_\ell}  \tau^\ell  - \tensor[^{\text{LC}}]{\nabla}{_m}   \tau^m \right) + \frac{1}{4}  \left( \tau_m \tau^m -  \tensor{C}{^n _{\ell m}} \tensor{C}{^m_{n} ^\ell}\right) \notag \\
    &= R_{\text{LC}} + \frac{1}{4}  \left( \tau_m \tau^m -  \tensor{C}{^n _{\ell m}} \tensor{C}{^m_{n} ^\ell}\right)\; .
\end{align}
The last term of the sum can be written as 
\begin{align}
    \tensor{C}{^n _{\ell m}} \tensor{C}{^m_{n} ^\ell} = F^{rn} F_{sn} \, \tensor{C}{_{ r\ell m}} \tensor{C}{^{m s \ell}} = \tensor{C}{_{r\ell m}} \tensor{C}{^{m r \ell}} = \tensor{C}{_{\ell m r}} \tensor{C}{^{\ell m r}}\; , 
\end{align}
due to the full symmetry of the ACT. With this, we find 
\begin{align}
     R_{(\pm 1)} &= R_{\text{LC}} + \frac{1}{4}  \left( \tau_m \tau^m - \tensor{C}{_{\ell m n}} \tensor{C}{^{\ell m n}} \right)\; .
\end{align}
We identify the scalar $C_4$ from 
Eq.\eqref{eq:nonmetricity_scalars} to reach the result stated in  Eq.\eqref{eq:ricciscalars},
\begin{align}
    R_{(\pm 1)} = R_{\text{LC}} - C_4\; .
\end{align}
For further information on dual curvature scalars in information geometry, see Refs.~\cite{Matsuzoe:2006, Zhang:2007}. 

\subsubsection*{Chebyshev field and conformal transformations}

Following Eq.\eqref{eq:nonmetricity_scalars} in Sec.~\ref{sec:info_geo}, we note that a suitable conformal transformation eliminates the Chebyshev field $\tau$. We show that this is indeed the case, in complete analogy to the Weyl-field in metric affine gravitation theory \cite{Cacciatori:2006, Iarley:2015}, and confirm this result using vector field divergences. 

The argument builds on a prominent result from information geometry: the Chebyshev one-form $\tau \in \Gamma(T^*\mathcal{M})$ ($\tau_i = F_{ij}\tau^j$) is closed on a statistical manifold, \ie its exterior derivative vanishes identically \cite{Matsuzoe:2006}. From the symmetry of the Ricci tensor given in Eq. \eqref{eq:riccitensor},
\begin{align}
    \tensor[^{(+1)}]{R}{_{k \ell }} - \tensor[^{(+1)}]{R}{_{ \ell k}} \overset{!}{=} 0
    \qquad \Rightarrow \qquad 
    \tensor[^{\text{LC}}]{\nabla}{_\ell}  \tau_k -\tensor[^{\text{LC}}]{\nabla}{_k} \tau_\ell  
    =0\; ,
\end{align}
this closedness follows immediately
\begin{align}
    \dd\tau = 0 
    \qquad \Leftrightarrow \qquad 
    \partial_i \tau_j - \partial_j \tau_i = \tensor[^{\text{LC}}]{\nabla}{_i} \tau_j -\tensor[^{\text{LC}}]{\nabla}{_j} \tau_i = 0 \; . \label{eq:closednesscondition}
\end{align}
On a simply connected manifold, closedness implies the exactness of a one-form. Consequently, there exists a potential $\Phi$ such that $\partial_i \Phi = \tau_i$ \cite{Matsuzoe:2006}. The fact that we can write the Chebyshev field as the gradient field of a scalar potential defines its transformation behavior under conformal transformations. For clarity, we rewrite $\tau_i$ with the ACT, 
\begin{align}
    \tau_i = F^{jk}C_{ijk} = F^{jk} \;\tensor[^{(+1)}]{\nabla}{_i} F_{jk}\; ,
\end{align}
and subsequently perform a conformal rescaling of the metric tensor, $F \mapsto e^{-\rho/2} \cdot F$, with a scalar function $\rho$. Then, the Chebychev field transforms as 
\begin{align}
    \tau_i \mapsto \tau_i' &= e^{+\rho/2}\, F^{jk} \,\left(\tensor[^{(+1)}]{\nabla}{_i} \left(e^{-\rho/2} F_{jk}\right)\right) \\
    &=  F^{jk} \;\tensor[^{(+1)}]{\nabla}{_i}  F_{jk} - \frac{F^{jk} F_{jk}}{2} \,\tensor[^{(+1)}]{\nabla}{_i}  \rho  \\
    &= F^{jk} C_{ijk} - \tensor[^{(+1)}]{\nabla}{_i}  \rho \\
    &= \tau_i - \partial_i \rho\; .
\end{align}
We see that since $\tau_i$ can be written as $\partial_i \Phi$, the choice $\rho \equiv \Phi$ (up to an additional constant) gives $\tau_i' = \tau_i - \partial_i \Phi = 0$. In conclusion, the proper conformal transformation locally erases the Chebyshev field. This argument is analogous to the treatment of the Weyl field in alternative gravity formulations \cite{Cacciatori:2006, Iarley:2015}. 

Conversely, $\tilde{C}_{ijk}$ defined in Eq.\eqref{eq:act_split} does not vanish under conformal transformations, 
\begin{align}
    \tilde{C}_{ijk}\mapsto \tilde{C}_{ijk}' &= \tensor[^{(+1)}]{\nabla}{_i} (e^{-\Phi/2} F_{jk}) - \underbrace{\left(\tau_i - \partial_i \Phi\right)}_{=0} \frac{e^{-\Phi/2}}{2}  \,F_{jk}\\ 
    &= e^{-\Phi/2} \left(\tensor[^{(+1)}]{\nabla}{_i} F_{jk} - \frac{\partial_i \Phi}{2} \cdot F_{jk} \right) \\ 
    &= e^{-\Phi/2} \left(C_{ijk}- \frac{\tau_i}{2}  F_{jk} \right)\\
    &= e^{-\Phi/2} \;\Tilde{C}_{ijk} \\ 
    &\ne 0\; .
\end{align}
The dual Ricci scalars vanish identically for the exponential family \cite{Nielsen:2020} and are thus invariant under the above conformal transformation, whilst the curvature scalar $R_{\text{LC}}$ will generally change. 

To confirm this result, we examine a vector field divergence, which only depends on $\tau_i$ and not $\tilde{C}_{ijk}$ as it is defined using a metric trace. Consider the $(-1)$-divergence of a vector field $X\in\Gamma(T\mathcal{M})$ defined with respect to the Fisher information $F$,
\begin{align}
   \text{div}_{(-1)} (X,F) =\tensor[^{(-1)}]{\nabla}{_i} X^i  &= \tensor[^{\text{LC}}]{\nabla}{_i}  X^i  + \frac{\tau_i X^i}{2} \label{eq:vectorfielddivergence}  \\ 
   &= \frac{1}{\sqrt{\det F}} \partial_i  \left( \sqrt{\det F} \;X^i \right) + \frac{\tau_i X^i}{2}  \notag  \\ 
   &= \partial_i  X^i  +\big(\partial_i  \ln \sqrt{\det F}\big) \, X^i +  \big(\partial_i \ln e^{\Phi/2} \big) \, X^i\notag \\
   &= \frac{1}{\sqrt{\det F \cdot e^\Phi}} \, \partial_i  \big(\sqrt{\det F\cdot e^\Phi} \; X^i \big) \notag \\
   &= \frac{1}{\sqrt{\det (F\cdot e^{\Phi/2})}}\partial_i  \big(\sqrt{\det (F\cdot e^{\Phi/2})} \cdot X^i \big)\\
   &= \text{div}_{\text{LC}} \,(X,e^{\Phi/2} \cdot F)\; .
\end{align}
We find that the $(-1)$-divergence of a vector field coincides with the LC-divergence of a vector field up to a conformal factor. The same is of course true for the $(+1)$-divergence, up to a sign. Hence, in expressions which only see the trace part $\tau$ of the ACT, the skewness can be eliminated by a conformal rescaling. 

Lastly, we relate the Jeffreys prior to this conformal transformation. In information geometry, Jeffreys prior appears as the Riemannian covolume for the Fisher information metric \cite{Jeffreys:1946}. Crucially, \cite{Takeuchi:2005} note that $\Phi = \ln \det F$. Also, \cite{Zhang:2014} and \cite{Jiang:2020} relate covolumes to the Chebyshev potential. From these two statements, the possibility of removing the Chebyshev field via a conformal transformation using covolumes is an immediate consequence. 

First, we confirm $\Phi =\ln \det F$ again using vector field divergences before showing that the Riemannian covolume is the proper conformal factor to eliminate the skewness in our 2d latent space. From now on, we omit the specification of the metric $F$ for simplicity. 

Eq.\eqref{eq:vectorfielddivergence} gives the divergence of $\tau$,  
\begin{align}
   \text{div}_{(+1)}(\tau) =  \text{div}_{\text{LC}}(\tau)  - \frac{ \tau^i \tau_i}{2} = \text{div}_{\text{LC}}(\tau) - \frac{ \tau^i \partial_i\Phi}{2}\; . \label{eq:divergence_Chebyshev}
\end{align}
From Proposition 9.8.7 in \cite{Udriste:2014}, it directly follows that 
\begin{align}
     \text{div}_{(+1)}\, (X) = \text{div}_{\text{LC}}(X) + X^i  \partial_i \big(\ln \sqrt{\det F}\big) \; , \label{eq:udriste_result}
\end{align}
for any vector field $X$ and a dually flat exponential family and. Combining Eq.\eqref{eq:divergence_Chebyshev} and Eq.\eqref{eq:udriste_result} or $X \equiv \tau$ gives, 
\begin{align}
 - \frac{\partial_i\Phi}{2} = \partial_i \big(\ln \sqrt{\det F} \big) \; \rightarrow \; - \Phi = \ln \det F + C\; \rightarrow \; e^{-\Phi/2} = e^{+(\ln\det F)/2} = \sqrt{\det F}\; , 
\end{align}
the Riemannian covolume, up to an uninformative constant $C$, which is omitted in the last step. Please note that the Eq.~\eqref{eq:udriste_result} requires the statistical manifold to be oriented, which is an assumption commonly made in information geometry \cite{Udriste:2014}.

\subsubsection*{Classifier geometry}

During training, the network selects the latent coordinates $z$ based on the interplay between classification and reconstruction. This coordinate frame is physically meaningful, but might have coordinate artifacts, such as degeneracies, which the network selected. Since the construction of geometric invariants, such as statistical distances, requires the use of an inverse metric tensor, we must ensure that possible degeneracies in the coordinates are treated properly. 

Because the classifier likelihoods form an exponential family, we can check for such artifacts using a coordinate change $\varphi:\lambda \mapsto z$ from the natural frame $\lambda$ with $\text{dim}\,\lambda = K-1$ for $K$ classes, in which the geometry is well represented, to the latent $z \in \mathbb{R}^2$. If this coordinate map $\varphi$  is one-to-one in a certain region, then the latent frame $z$ certainly represents the classifier geometry well.

For the three-label classifier, $\varphi$ can be one-to-one in principle, as $\text{dim}\,\lambda = 2$ in this case, \ie latent dimension coincides with the natural dimension of the statistical manifold. We can test where in the latent $\varphi$ is actually invertible, by testing where its Jacobian $J_\lambda(z)$ has full rank. Because this Jacobian defines the Fisher information in $z$, 
\begin{align}
    g_{ij}(z) = J_\lambda(z)^k_i \;g_{k\ell}(\lambda) \;J_\lambda(z)^\ell_j = \left(\frac{\partial \lambda^k}{\partial z^i}\right) g_{k\ell}(\lambda) \left(\frac{\partial \lambda^\ell}{\partial z^j}\right) \,,
\end{align}
this is equivalent to checking whether our metric becomes singular or equivalently where its determinant vanishes. 

Empirically, for our MNIST application, Fig.~\ref{fig:logdet_Chebyshev_jacobian} confirms that the classifier geometry is well-represented in $z$ almost everywhere in our regions of interest. The only degenerate points lie on a line crossing the boundary region, which we noted in Sec.~\ref{sec:info_geo}. On this line, the metric determinant vanishes abruptly. Importantly, we see that the line traverses the undecided region between two classes and is thus not caused by being close to the edges of the simplex (i.e. one probability being either 1 or 0). Rather, as explained in Sec.~\ref{sec:info_geo}, the line indicates an over-parametrization of the problem in the $z$ coordinates, \ie $\varphi$ is not one-to-one there. As a geometric consequence, any quantity which involves the inverse metric must be treated with caution when it is evaluated across this line. Numerically, the pseudo-inverse used in our computations provides a smooth continuation to all maps, hence why the line does not appear \eg in Fig.~\ref{fig:mnist_directional_derivative}. However, there is no guarantee that this continuation is faithful to the true geometry. 

The natural dimension of a binary classifier is $\text{dim} \, \lambda =1$, since one class probability fully specifies the other, $p_0 = 1- p_1$. Hence, a two-dimensional latent representation of the classifier geometry is always redundant and $\varphi$ can never be one-to-one. In this case, all relevant geometry for the classification happens in the direction of the natural parameter gradient $\nabla_\theta$. In $z$, this corresponds to paths that hit the decision boundary orthogonally. Clearly, the reconstruction depends on both latent directions, which is why we keep a two-dimensional representation of this effectively one-dimensional manifold. 

\section{Data generation MNIST 1-7}
\label{sec:datagenerationmnist}

\begin{figure}[t!]
    \includegraphics[width=\textwidth]{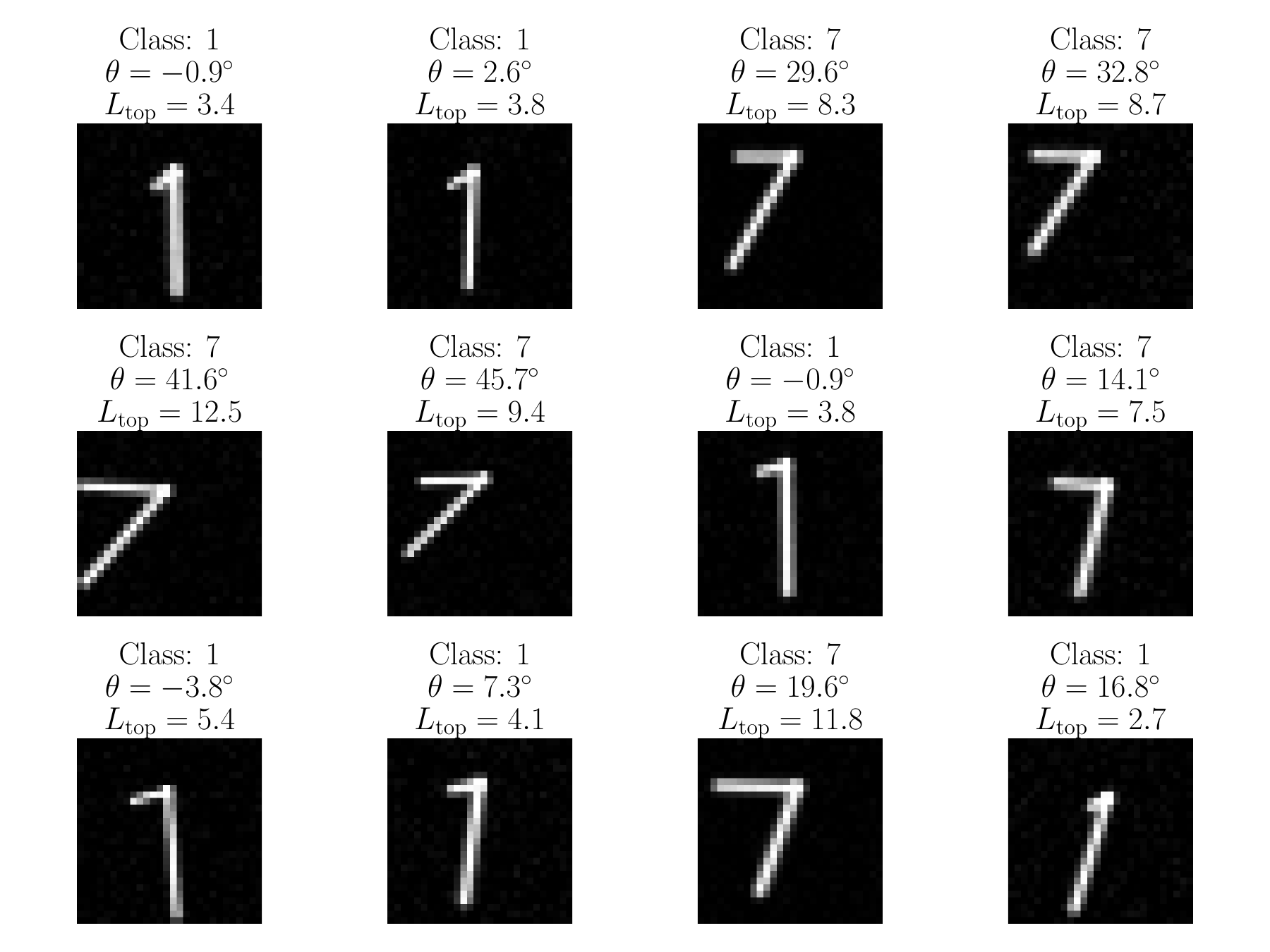}
    \caption{Example images from the 1 vs. 7 toy dataset and their class and features.}
    \label{fig:mnist_examples}
\end{figure}

We generate toy data in the style of MNIST digits. We constrain ourselves to the digits 1 and 7 as they are closely related in typical handwriting. Two features describe their difference for us, the length of the horizontal bar $L_\text{top}$ and the angle with which the vertical bar is tilted $\theta$. Additionally, the horizontal bar, or cap, is tilted with an angle, too, but we do not use this as a feature. It is only for the generation and not given to the network. It can be viewed as a nuisance parameter. While one could infer these features from the MNIST dataset by fitting templates, for this example it is much easier to generate the images. Fig.~\ref{fig:mnist_examples} shows a set of examples.

\begin{figure}[b!]
    \centering
    \includegraphics[width=.5\textwidth]{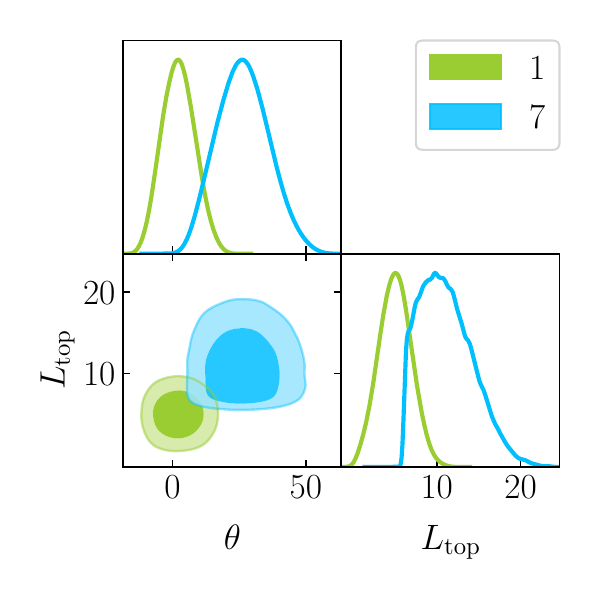}
    \caption{Joint distribution of the two features for the 1 vs. 7 toy dataset. }
    \label{fig:mnist17_feature_distribution}
\end{figure}

The data generation process goes as follows: the class is sampled from a uniform distribution, but it is made sure that there is an equal number of classes in the end. In total we produce 100\,000 images. The features are sampled from the distributions defined in Tab.~\ref{tab:toy_mnist_sampling}. We chose them based on how human-like the final images looked, but made sure to have some overlap. Fig~\ref{fig:mnist17_feature_distribution} shows the joint distribution of the two features. Note that there is some ambiguity how some of the images are classified. 

Additionally, the overall rotation, the stroke thickness, and the translational shift are sampled and applied to the digit. The output image has the same 28x28 pixel as the MNIST data. 

\begin{table}[t]
    \centering    
    \begin{small} \begin{tabular}{lccc}
    \toprule
    Class & $\theta$ (deg) & $L_{\text{top}}$ (px) & Nuisance cap rotation (deg) \\
    \midrule
    1 &
    $\mathcal{N}_{[-10,18]}(2,6)$ &
    $\mathcal{N}_{[0,12]}(5,2)$ &
    $\text{Unif}(-60,-6)$\\
    7 &
    $\mathcal{N}_{[6,55]}(26,10)$ &
    $\mathcal{N}_{[6,24]}(10,4)$ &
    $\mathcal{N}_{[-10,10]}(0,3)$ \\
    \bottomrule
    \end{tabular} \end{small}
    \caption{Sampling process for toy MNIST digits 1 and 7. $\mathcal{N}_{[a,b]}(\mu,\sigma)$ denotes a truncated normal with mean $\mu$, std $\sigma$, truncated to $[a,b]$.}
    \label{tab:toy_mnist_sampling}
\end{table}

\section{Energy flow polynomials}
\label{EFP_appendix}

Energy flow polynomials (EFPs) summarize a jet by combining momentum fractions $z_i$ with angular separations $\Delta R_{ij}$ according to multigraph patterns. For a given multigraph, one multiplies one $z_i$ per vertex and one $\Delta R_{ij}$ per edge, summing over all jet constituents. For example,
\begin{align}
\begin{gathered}
\includegraphics[scale=.2]{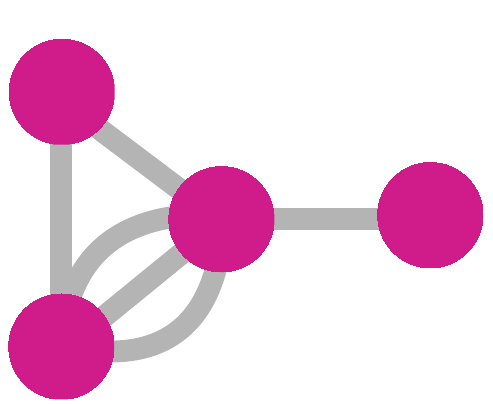}
\end{gathered}
= \sum_{i} \sum_{j} \sum_{k} \sum_{l} 
z_i z_j z_k z_l 
\Delta R_{ij}\Delta R_{ik}\Delta R_{jk}^3\Delta R_{kl}
\qquad \text{with} \qquad 
z_i = \frac{p_{T,i}}{\sum_i p_{T,i}} \, .
\end{align}
Simple EFPs that are fully specified by the number of vertices $v$ and edges $d$ are denoted by $\text{EFP}_{v,d}$. We also consider disconnected multigraphs, defining composite EFPs as
\begin{align}
\text{EFP}_G = \prod_{g \in C(G)} \text{EFP}_g \, ,
\end{align}
where $C(G)$ denotes the set of connected components of the multigraph $G$.

We include all multigraphs with up to seven edges, resulting in an overcomplete set of approximately 1000 EFPs. Because simulated data has an implicit infrared cutoff, energy flow polynomials form a linear basis for all IRC-safe observables, as well as for all permutation-invariant observables on hadronized simulated data that depend only on particle momenta. From this perspective, EFPs provide an excellent basis for probing which structures are encoded by a neural network.

\begin{figure}[t]
\includegraphics[width=0.49\linewidth]{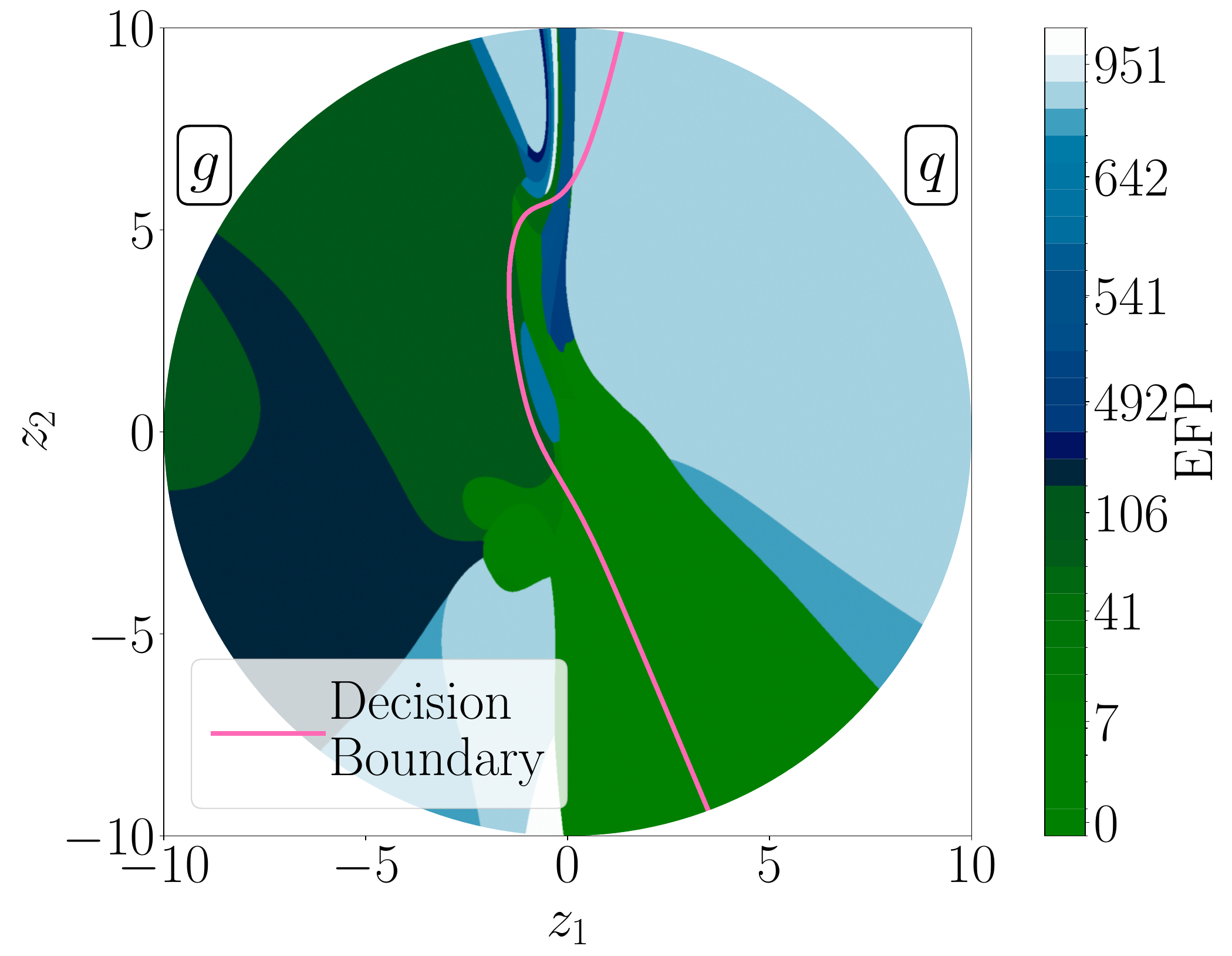}
  \includegraphics[width=0.49\linewidth]{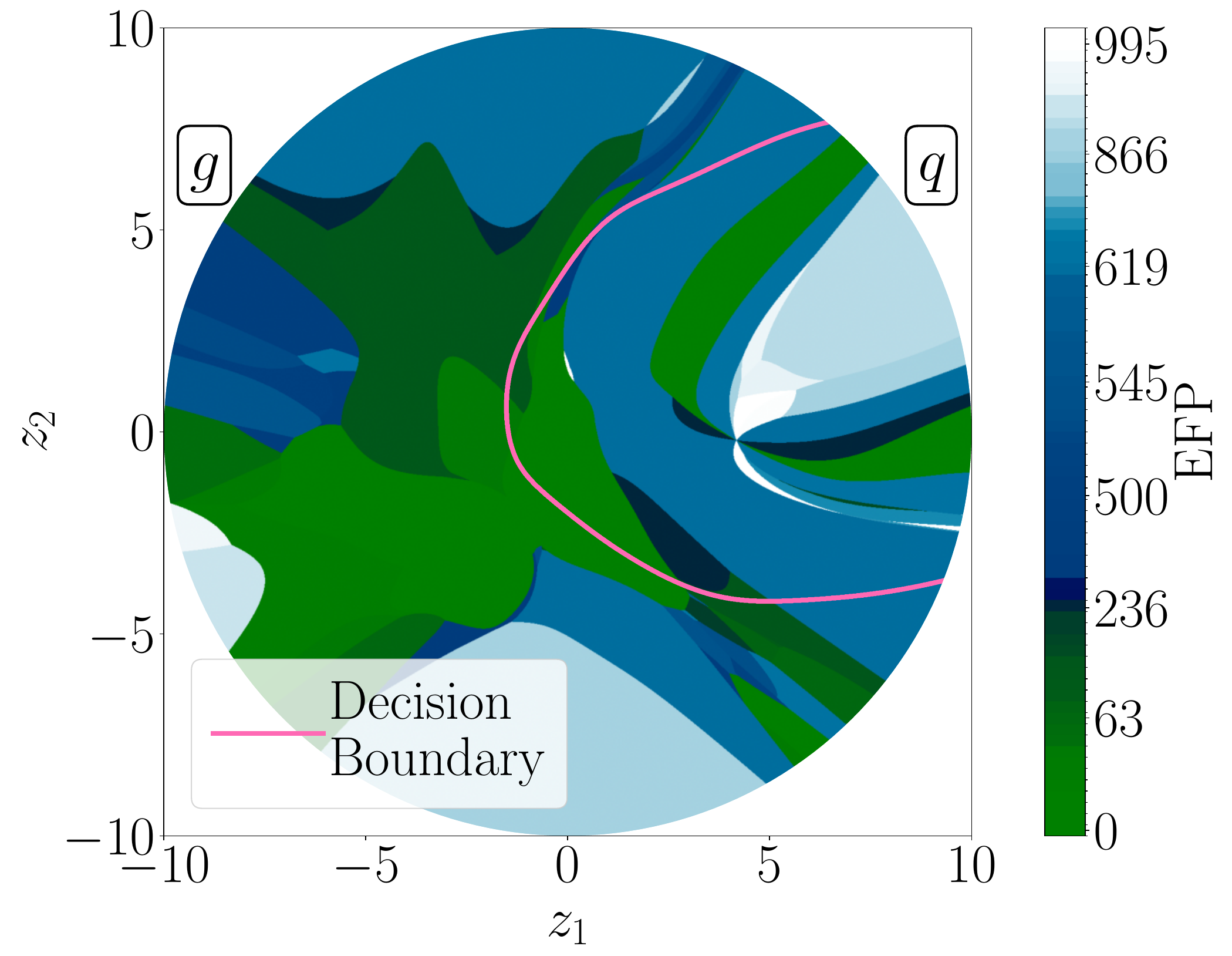}
  \caption{Dominant derivatives for quark-gluon classification with reconstructed EFP basis for two independent trainings with different random seed.}
  \label{fig:EFPS}
\end{figure}

In practice, however, it is beneficial to work with a minimal effective subset. When computing the dominant derivative with respect to the EFP basis (see Fig.~\ref{fig:EFPS}), we observe that the simplest EFP ($\sum_i z_i$) dominates a large area. Close to the discussion boundary, many complex EFPs dominate in small fractions which makes it difficult to interpret individually. This instability is partly due to the strong correlations among EFPs, which make the basis highly redundant. From an explainability standpoint, it is also challenging to assign clear physical meaning to specific EFPs. The outcome varies drastically between different runs as seen in Fig.~\ref{fig:EFPS}. Noting that the left panel used 30 EFPs and the right one 72. We refrain from giving conrete definitions for all of the EFPs just noting that an index above 500 denotes a multigraph with increasing complexity and up to 500 the EFP complexity grows approximately with the index.

\section{Network implementation}
\label{sec:implementation_details}

The VAE+classifier model is trained by minimizing a weighted sum of the standard VAE objective (reconstruction + KL regularization) and a cross-entropy term~\cite{kingma2022autoencodingvariationalbayes}
\begin{align}
    \loss 
    = \lambda & \left( \loss_{\text{recon}} + \loss_{\text{KL}} \right) + \loss_{\text{CE}}\, \notag \\
    \text{with} \qquad 
    \loss_{\text{recon}}
    &= \Langle -\log p_\phi(x' | z)\Rangle_{z\sim p_\phi(z|x)} \notag \\
    \loss_{\text{KL}}
    &= D_{\text{KL}}\left[p_\phi(z | x),p(z)\right] \notag \\
    \loss_{\text{CE}}
    &= \Langle -\log p_\phi(k | z)\Rangle_{z\sim p_\phi(z|x)} \,.
\end{align}
$\lambda$ establishes a balance between the VAE task and the classification. We optimize $\lambda$ such that we always get the best possible classification performance. We use $p(z) = \mathcal{N}(0,1)$ in the regularization term. 

All datasets provide an input, a reconstruction target, and a class label (Tab.~\ref{tab:dataset_overview}). Depending on the input modality and reconstruction target, we use the architectures in Tab.~\ref{tab:architecture}. All networks are trained for 100 epochs using AdamW. The initial learning rate $10^{-3}$ is reduced with a CosineAnnealing scheduler to $10^{-6}$. We consistently use a batch size of 256. The only dataset-dependent hyperparameter is $\lambda$: $\lambda=1$ for image reconstruction, $\lambda=0.1$ for feature reconstruction (except EFP, $\lambda=0.01$).

\begin{table}[b]
    \centering
    \begin{small} \begin{tabular}{lccccc}
    \toprule
    Dataset & Number of classes & Input dimension & Reconstruction (dim) & Number of samples \\
    \midrule
     MNIST & 2 / 3 / 4 & $28\times28$ & Image ($28\times28$) & 7\,000 per class \\
     Toy MNIST (1 vs. 7) & 2 & $28\times28$ & Features (2) & 100\,000 \\
     Quark / Gluon & 2 & 64 & Features (8) & 100\,000 \\
     JetClass & 3 & 256 & Features (3) / EFP (1000) & 2\,000\,000 per class \\
     \bottomrule
    \end{tabular} \end{small}
    \caption{Dataset overview.}
    \label{tab:dataset_overview}
\end{table}

\begin{table}[t]
\centering
\begin{small} \begin{tabular}{p{2.8cm}p{1.8cm}p{2.8cm}p{6.0cm}}
\toprule
Data & Module & Layer & Hyperparameters \\
\midrule
\multirow{6}{*}{MNIST images}
& \multirow{6}{*}{Encoder}
& Input & -- \\
&  & Conv2d + GELU & $1\!\to\!64,\;k=4,s=2,p=1$ \\
&  & Conv2d + GELU & $64\!\to\!128,\;k=4,s=2,p=1$ \\
&  & Conv2d + GELU & $128\!\to\!256,\;k=3,s=2,p=1$ \\
&  & Flatten & -- \\
&  & FC heads & $\mu:4096\!\to\!2,\;\log\sigma^2:4096\!\to\!2$ \\
\midrule
\multirow{2}{*}{Quark / Gluon}
& \multirow{4}{*}{Encoder}
& Input & -- \\
&  & Linear + GELU & $D\!\to\!128$ \\
\multirow{2}{*}{JetClass}  &  & Linear + GELU & $128\!\to\!64$ \\
&  & FC heads & $\mu:64\!\to\!L,\;\log\sigma^2:64\!\to\!L$ \\
\midrule
\multirow{6}{*}{MNIST images}
& \multirow{6}{*}{Decoder}
& Input & -- \\
&  & FC & $2\!\to\!4096$ \\
&  & Reshape & $4096\!\to\!256\times 4\times 4$ \\
&  & ConvT2d + GELU & $256\!\to\!128,\;k=3,s=2,p=1,\text{op}=1$ \\
&  & ConvT2d + GELU & $128\!\to\!64,\;k=4,s=2,p=1$ \\
&  & ConvT2d & $64\!\to\!1,\;k=4,s=2,p=3$ \\
\midrule
\multirow{4}{*}{Features}
& \multirow{4}{*}{Decoder}
& Input & -- \\
&  & Linear + GELU & $L\!\to\!32$ \\
&  & Linear + GELU & $32\!\to\!32$ \\
&  & Linear & $32\!\to\!R$ \\
\midrule
\multirow{4}{*}{Probabilities}
& \multirow{4}{*}{Classifier}
& Input & -- \\
&  & Linear + GELU & $2\!\to\!64$ \\
&  & Linear + GELU & $64\!\to\!64$ \\
&  & Linear & $64\!\to\!C$ (logits), $C=\#$classes \\
\bottomrule
\end{tabular} \end{small}
\caption{Architecture summary.}
\label{tab:architecture}
\end{table}

\clearpage
\bibliography{tilman,refs}
\end{document}